\documentclass[12pt]{article}
\usepackage{amsmath,amsfonts,amsthm, amssymb}
\usepackage{enumitem}
\usepackage{bm}
\usepackage{bbm}
\usepackage{graphicx}
\usepackage{xcolor}
\usepackage{natbib}
\usepackage{url} 
\usepackage[colorlinks=true, citecolor=blue]{hyperref}
\usepackage{multirow}
\usepackage{array}
\usepackage{caption}
\newcolumntype{C}[1]{>{\centering\arraybackslash}p{#1}}

\newtheorem{theorem}{Theorem} 

\newtheorem{assumption}{Assumption}
\newtheorem{remark}{Remark}
\newtheorem*{remark*}{Remark}
\usepackage{appendix}
\usepackage{algorithm}  
\usepackage{algpseudocode}

\usepackage{lineno}
\newcommand{\blind}{1}
\newcommand{\I}{\mathcal{I}}
\newcommand{\J}{\mathcal{J}}

\newcommand{\pr}{\mathbb{P}}
\newcommand{\R}{\mathbb{R}}
\DeclareMathOperator*{\argmin}{arg\,min}



\begin{document}

\def\spacingset#1{\renewcommand{\baselinestretch}%
{#1}\small\normalsize} \spacingset{1}


\if1\blind
{
  \title{\vspace{-0.75in}\LARGE\bf Joint Registration and Conformal Prediction for Partially Observed Functional Data}
  \author{
  Fangyi Wang, Sebastian Kurtek and Yuan Zhang\\
  Department of Statistics, The Ohio State University
  }
    \date{}
  \maketitle
} \fi

\if0\blind
{
  \bigskip
  \bigskip
  \bigskip
  \begin{center}
    {\LARGE\bf Joint Registration and Conformal Prediction for Partially Observed Functional Data}
\end{center}
  \medskip
} \fi

\begin{abstract}

Predicting missing segments in partially observed functions is challenging due to infinite-dimensionality, complex dependence within and across observations, and irregular noise. These challenges are further exacerbated by the existence of two distinct sources of variation in functional data, termed amplitude (variation along the $y$-axis) and phase (variation along the $x$-axis). While registration can disentangle them from complete functional data, the process is more difficult for partial observations. Thus, existing methods for functional data prediction often treat phase variation as negligible. Furthermore, they typically require precise model specifications and/or rely on computationally intensive tools, such as bootstrapping, to construct prediction intervals. We propose a unified registration and prediction approach for partially observed functions using conformal prediction. Our method integrates registration and prediction in one algorithm while ensuring exchangeability through carefully constructed predictor-response pairs. Using a neighborhood smoothing algorithm, the framework produces pointwise prediction bands with finite-sample marginal coverage guarantees under weak assumptions. The method is easy to implement, computationally efficient, and permits simple parallelization. Numerical studies and real-world data examples demonstrate the effectiveness and practical utility of our method.
\end{abstract}

\noindent%
{\it Keywords:} conformal prediction, elastic functional data analysis, neighborhood smoothing, phase variation
\vfill
\newpage
\spacingset{1.75} 
\vspace{-0.2in}


\section{Introduction}
\label{sec::intro}
\vspace{-0.1in}
Accurate prediction of future trajectories given historical functional data is an important question in many applications. 
For instance, given previous (complete) daily observations of traffic flow rate, and a partial observation (up to some time) for a new day, prediction of traffic flow rate for the rest of that day can help optimize transportation networks and reduce congestion during rush hour \citep{chiou2012dynamical}. 
As another example, given complete historical daily maximum temperatures at a particular location, one may be interested in a daily maximum temperature forecast for the rest of the current year, which can help mitigate societal risks related to extremely cold weather. 
See Section \ref{sec::real-data} for more detailed analyses. 
Other examples include growth rate prediction for people or other natural objects, forecasting of pollutant density, and so on. 
These predictions can be used to anticipate future trends and help inform biomedical, environmental and social decisions.

Functional data is challenging to analyze, because it is inherently infinite-dimensional, exhibits complex dependencies within and potentially across observations, and is often observed at discrete time points with heteroscedastic noise. 
Functional observations encompass two sources of variation, one along the $x$-axis called \emph{phase} and one along the $y$-axis called \emph{amplitude}. 
A common practice is to first separate amplitude and phase variations.
This is done via a process called \emph{registration}, which ``warps'' the functions' domains to align their key features with respect to time \citep{kneip1992statistical, RamsaySilverman2005, srivastava2011registration}. 
However, registration is often treated as a pre-processing step, with subsequent statistical modeling and inference applied to ``well-aligned'' data. 
Meanwhile, registration uncertainty is often ignored in downstream tasks. Furthermore, when the data contains partial observations, registration is much more challenging since the domains of partial and complete functions are fundamentally different \citep{bauer2021registration, bryner2021shape}.

To tackle the aforementioned challenges, we propose a framework for prediction of partially observed functions wherein registration is directly incorporated. 
Comparing to a sequential approach of ``registration then prediction'', our approach bypasses the need for registration of partial observations to complete ones, and propagates registration uncertainty to the prediction step. 
We adapt \emph{conformal prediction}, a distribution-free uncertainty quantification method that provides valid finite-sample coverage validity \citep{vovk2005algorithmic, lei2018distribution}. 
This requires a careful construction of exchangeable predictor-response data pairs, which are not naturally defined in our context. 
In addition, we employ a nonparametric, neighborhood smoothing prediction algorithm \citep{zhang2017estimating}, which offers flexibility without strong modeling assumptions on the relationship between predictors and responses. Thus, our main contributions are as follows.
\begin{enumerate}[leftmargin=*]
    \item We propose a novel framework that integrates registration into conformal prediction of the amplitude component for partial functional data. 
    This results in more accurate prediction intervals, as compared to procedures that do not utilize registration, when functional data contains phase variation. 
    We additionally define a separate conformal prediction procedure for the phase component, which incorporates a monotonicity constraint.
    Our approach builds on the elastic functional data analysis (EFDA) framework for registration \citep{srivastava2011registration, srivastava2016functional}.
    \item Our method offers distribution-free uncertainty quantification with a finite-sample guarantee. 
    For conformal prediction of partially observed functions, we construct predictors and responses from raw data that maintain exchangeability throughout, and leverage ideas from split conformal methods to integrate registration. 
    Unlike traditional parametric approaches, our method imposes minimal assumptions on the data generating process. 
    We employ a nonparametric neighborhood smoothing prediction algorithm that provides more flexibility than competitors such as functional linear regression.
    \item Compared to nonparametric bootstrap or Bayesian uncertainty quantification, our approach is much faster, making it well-suited for big data.
\end{enumerate}

\noindent\textbf{Related work.} There exist many approaches for functional data prediction, but most of them do not consider the challenges posed by phase variation. 
They include functional regression \citep{chiou2012dynamical, maity2017nonparametric}, time series models \citep{leroux2018dynamic, jiao2023functional} and neural networks \citep{yin2021deep}. 
For uncertainty quantification, these approaches generally use closed-form confidence intervals based on variance estimates \citep{leroux2018dynamic}, or nonparametric bootstrap residual sampling \citep{chiou2012dynamical, jiao2023functional}. 
However, these methods require strong parametric assumptions, and the resulting prediction intervals' coverage rate converges to the desired nominal level asymptotically with no finite-sample guarantee. \citet{matuk2022bayesian} and \citet{earls2017variational} define Bayesian models for joint registration and prediction, where prediction intervals are constructed from posterior samples. Nonetheless, the high computational cost of these approaches hinders their application to large datasets.

Conformal prediction provides prediction intervals that have a finite-sample coverage guarantee without imposing assumptions on the data generating process \citep{vovk2005algorithmic}. 
It is generally computationally more efficient when compared to Bayesian methods, which utilize Markov chain Monte Carlo to sample from the posterior distribution. 
Since being introduced, conformal prediction has been widely applied in a variety of scenarios, ranging from regression and classification for multivariate data \citep{lei2018distribution, cauchois2021knowing} to prediction tasks involving more complex data structures, e.g., survival analysis \citep{candes2023conformalized, gui2024conformalized}, time series \citep{zaffran2022adaptive, angelopoulos2023PID}, and matrices \citep{gui2023conformalized, shao2023distribution}. 
However, relatively few works have focused on the application of conformal prediction to functional data. 
\citet{lei2015conformal} used a basis projection for functional observations to obtain prediction sets for basis coefficients; they did not consider partially observed functions. \citet{diquigiovanni2021importance, diquigiovanni2022conformal} and \citet{ajroldi2023conformal} applied conformal prediction to functional data and functional time series using a domain-adaptive nonconformity score based on modulation functions.
Recently, \citet{diana2023distribution} and \citet{de2024conformal} studied conformalized spatial functional inference. All of the aforementioned methods treated phase variation as negligible and did not consider registration.

The rest of the article is organized as follows. 
Section \ref{sec::problem-set-up} gives a brief introduction of the elastic functional data analysis (EFDA) framework. Section \ref{subsec::FFCP} defines full conformal prediction for partial functional data without registration, while Section \ref{sec::our-method} introduces a joint registration and prediction method. Simulations and real-world data examples are presented in Sections \ref{sec::simulations} and \ref{sec::real-data} to demonstrate the validity and efficiency of the proposed method. 
We conclude with a brief discussion in Section \ref{sec::discussion}.

\vspace{-0.3in}
\section{Preliminaries - registration via EFDA}
\label{sec::problem-set-up}
\vspace{-0.1 in}

A summary of key notation used throughout Sections \ref{sec::problem-set-up}-\ref{sec::our-method} is provided in Table \ref{table::notations}. Let ${\cal F} = \{f: [0,1] \to \R | f \textrm{ is absolutely continuous}\}$ be the representation space for functional data. Phase variation in functional data is characterized through \emph{warping functions}, which are elements of the group of orientation and boundary preserving diffeomorphisms, denoted as $\Gamma := \{\gamma: [0,1] \to [0,1] | \gamma(0) = 0, \gamma(1) = 1, \dot{\gamma} > 0\}$ ($\dot{\gamma}$ is the derivative of $\gamma$).
For a $\gamma\in\Gamma$, the domain warping of a function $f\in\mathcal{F}$ is given by composition, $f \circ \gamma$, which locally stretches or compresses the domain of $f$ without changing its values. Registering a function $f_2$ with respect to another function $f_1$ involves finding $\gamma \in \Gamma$ that minimizes the distance between $f_2 \circ \gamma$ and $f_1$. To ensure that the solution does not depend on arbitrary domain warping of $f_1$ or $f_2$, we require a distance $d(\cdot,\cdot)$ that is preserved under simultaneous warping: $d(f_1,f_2)=d(f_1\circ\gamma,f_2\circ\gamma)$. 
It is well-known that the standard $\mathbb{L}^2$ distance, given by $d_2(f_1,f_2) := \Big(\int_{0}^{1}(f_1(t) - f_2(t))^2 dt\Big)^{1/2}$, does not satisfy this property (\citet{srivastava2016functional}, Chapter 4).

\begin{table}[!t]
    \begin{center}
    \captionsetup{width=.9\textwidth}
    {\small 
    \caption{Summary of key notation.}
    \vspace{-5pt}
    \label{table::notations}
    \begin{tabular}{|c|c|}
        \hline
        ${\cal F}$ & space of complete absolutely continuous functions\\
        \hline
        ${\cal Q} \subset\mathbb{L}^2 ([0,1],\mathbb{R})$ & corresponding space of square-root slope functions (SRSFs)\\
        \hline
        ${\cal F} \ni f_i:[0,1]\to\mathbb{R}$ & complete functional observation\\
        \hline
        ${\cal Q} \ni q_i:[0,1]\to\mathbb{R}$ & SRSF of $f_i$\\
        \hline
        $\bar{f}\in\mathcal{F}$ & sample Karcher mean\\
        \hline
        $\bar{q}\in\mathcal{Q}$ & sample Karcher mean SRSF\\
        \hline
        $f_i \circ\gamma_i^*\in\mathcal{F}$ & amplitude of $f_i$ with respect to $\bar{f}$\\
        \hline
        $\gamma_i^*$ & relative phase of $f_i$ with respect to $\bar{f}$\\
        \hline
        ${\cal T}$ & uniform grid of time points on $[0,1]$\\
        \hline
        ${\cal J}\subset [0,1]$ & subinterval on which partial function was observed\\
        \hline
        $f_i^{\cal J}$ & $f_i$ restricted to domain ${\cal J}$ \\
        \hline
        $(X_i,Y_i(t))$ & predictor-response pair\\
        \hline
        $\widehat{Y_i(t)}$ & estimate of $Y_i(t)$\\
        \hline
        $S_i(t) = |Y_i(t) - \widehat{Y_i(t)}|$ & nonconformity score\\
        \hline
        ${\cal I}_t$ & prediction interval at $t \in {\cal T}$\\
        \hline
    \end{tabular}
    }
    \end{center}
     \vspace{-10pt}
\end{table}

Instead, one can use (the extension of) the Fisher-Rao (FR) Riemannian metric that is preserved under simultaneous warping. However, the resulting Riemannian distance cannot be computed in closed form. 
To address this computational bottleneck, \citet{srivastava2011registration} proposed the square-root slope function (SRSF) that reduces the FR metric to the $\mathbb{L}^2$ metric. 
For a function $f\in\mathcal{F}$, its SRSF $q:[0,1]\to\mathbb{R}$ is given by $Q(f) = q := \textrm{sign}(\dot{f}) \sqrt{|\dot{f}|}$, where $Q:\mathcal{F}\to {\cal Q}$ and ${\cal Q}\subset\mathbb{L}^2([0,1],\mathbb{R})$. 
Given $f(0)$, the inverse of $Q$ is $Q^{-1}(f(0),q)(t) = f(0) + \int_{0}^{t}q(s)|q(s)|ds$.  
Finally, the domain warping of a function $f$ by $\gamma$, $f\circ\gamma$, is given by the following transformation of its SRSF: $(q\circ\gamma)\sqrt{\dot\gamma}$. 
Given two functions $f_1,\ f_2\in\mathcal{F}$ and their SRSFs $q_1,\ q_2\in\mathcal{Q}$, $d_{FR}(f_1,f_2)=d_2(q_1,q_2)=d_2((q_1\circ\gamma)\sqrt{\dot\gamma},(q_2\circ\gamma)\sqrt{\dot\gamma})$. 
Thus, the SRSF simplifies calculation of the FR Riemannian distance while maintaining its invariance under simultaneous warping (\citet{srivastava2016functional}, Chapter 4).
Pairwise registration of $f_2$ to $f_1$, with SRSFs $q_2$ and $q_1$, is given by the optimization problem
\begin{align}
\label{eq::registration-SRSF}
    \gamma^*
    =
    \argmin_{\gamma \in \Gamma} 
    d_2\big(
        q_1,
        (q_2\circ\gamma) \sqrt{\dot{\gamma}}
    \big),
\end{align}
which can be solved using dynamic programming \citep{DynPro}. 
Then, $f_2 \circ \gamma^*$ is registered to $f_1$, and the amplitude distance between them is given by $d_a(f_1,f_2)=d_2(q_1, (q_2 \circ \gamma^*)\sqrt{\dot{\gamma}^*})$; here, $\gamma^*$ is the relative phase of $f_2$ with respect to $f_1$. 
Multiple registration of $f_1,\dots,f_n$ is performed pairwise using \eqref{eq::registration-SRSF} with respect to a template function. 
Let $q_1,\dots,q_n$ be the SRSFs of $f_1,\dots,f_n$, and define the \emph{sample Karcher mean} as the template, which is given by
\begin{align}
\label{eq::def-Karcher-mean}
    \bar{q} :=
    \argmin_{q \in {\cal Q}} \sum_{i=1}^{n} \min_{\gamma_i \in \Gamma} 
    d_2^2(q,(q_i\circ\gamma_i)\sqrt{\dot{\gamma}_i}).
\end{align}
The corresponding mean in $\cal F$ is $\bar{f} := Q^{-1}(\bar{f}(0), \bar{q})$, where $\bar{f}(0) = \frac{1}{n}\sum_{i=1}^{n} f_i(0)$. 
Multiple registration produces (i) $\{\gamma_i^*\}$, relative phases with respect to the Karcher mean, and (ii) $\{f_i\circ\gamma_i^*\}$, amplitudes of $\{f_i\}$. 
See Algorithms 2 and 3, and Section 3.4 in \citet{srivastava2011registration} for implementation details; we use the \texttt{fdasrvf} package in R \citep{tucker2017fdasrvf}.

\vspace{-0.3in}
\section{Functional conformal prediction without registration}
\label{subsec::FFCP}
\vspace{-0.1 in}
Our data are functions, $f_1,\dots,f_{n+1} \stackrel{i.i.d.}{\sim} \pr_{\cal F}$, where $\pr_{\cal F}$ is a probability distribution on ${\cal F}$.
The observed portion of the $(n+1)^{\rm th}$ sample is denoted by $f_{n+1}^{\cal J}$, where ${\cal J} \subseteq [0,1]$. For simplicity, we focus on ${\cal J} = [0,U]$, where $U \sim \pi_u$ for some distribution $\pi_u$ on $[0,1]$. 
This can be generalized to other observation patterns (Section \ref{subsec::other-obs-pattern}). 
Given $f_1,\dots,f_n$ and $f_{n+1}^{\cal J}$, our goal is to construct pointwise prediction intervals for $f_{n+1}$ on a fixed, uniform grid of time points,
${\cal T} := \{t_1,\dots,t_T\}$, where $t_1 = 0,\ t_T = 1$ and $t_{k+1} - t_k = 1/(T-1),\ k=1,\dots,T-1$. Given $\alpha \in (0,1)$, we want the prediction interval ${\cal I}_t$ for $f_{n+1}(t)$ to satisfy
\begin{align}
\label{eq::coverage-validity-FFCP}
    \pr(f_{n+1}(t) \in {\cal I}_t) \geq 1-\alpha
\end{align}
for each time point $t \in {\cal T}$, which defines finite-sample coverage validity.
To achieve \eqref{eq::coverage-validity-FFCP}, we appeal to conformal prediction, which usually has the following three key components:

\begin{enumerate}[leftmargin=*]
    \item \textbf{exchangeable} (or i.i.d.) predictor-response data pairs $\big((X_1,Y_1),\dots,(X_{n+1},Y_{n+1})\big) \sim \mathbb{P}_{{\cal X} \times {\cal Y}}$, where $\cal X$ and $\cal Y$ represent the predictor and response sample spaces, respectively, and $\mathbb{P}_{{\cal X} \times {\cal Y}}$ is a joint probability distribution defined on ${\cal X} \times {\cal Y}$. 
    Exchangeability means that $\big((X_1,Y_1),\dots,(X_{n+1},Y_{n+1})\big) \stackrel{d}{=} \big((X_{\pi(1)},Y_{\pi(1)}),\dots,(X_{\pi(n+1)},Y_{\pi(n+1)})\big)$ for all permutations $\pi:[1:n+1] \rightarrow [1:n+1]$;
    \item a \textbf{permutation symmetric algorithm} that uses augmented data $\big((X_1,Y_1),\dots,$ $(X_n,Y_n),$ $(X_{n+1},y)\big)$ ($y$ is a trial value for $Y_{n+1}$ that may be included in the prediction set) to fit model $\hat{\mu}_y: {\cal X} \to {\cal Y}$. Permutation symmetry means that $\hat{\mu}_y$ remains unchanged under permutations, $\pi: [1:n+1] \rightarrow [1:n+1]$, of the data pairs $\big((X_1,Y_1),\dots,(X_{n+1},y)\big)$.
    \item a \textbf{nonconformity score} that measures the discrepancy of each observation relative to the fit, e.g., the absolute residual $S_i = |Y_i - \hat{\mu}_y(X_i)|$.
\end{enumerate}

In the classical settings of conformal prediction for regression or classification, the predictor-response pairs are well-defined. \emph{In contrast, the main challenge in applying conformal prediction to partially observed functional data is to construct meaningful $\{(X_i,Y_i)\}$ while ensuring exchangeability, as there are no natural predictor and response variables.}
In this setting, the prediction target $f_{n+1}(t)$ is analogous to the response $Y_{n+1}$, and thus, we define $Y_i(t) := f_i(t),\ i=1,\dots,n+1$, i.e., we are predicting $f_{n+1}$ at time $t$, $\forall$ $t \in {\cal T}$.
Since $f_1,\dots,f_{n+1}$ are i.i.d. samples from $\pr_{\cal F}$, $Y_1(t),\dots,Y_{n+1}(t)$ are exchangeable.
Further, since we observe $f_{n+1}^{\J}$, it can be viewed as the new feature $X_{n+1}$. 
To define $X_1,\dots,X_n$ that are exchangeable with $X_{n+1}$, we cut $f_1,\dots,f_n$ at $t = U$, and set $X_i := f_i^{\J},\ i = 1,\dots,n+1$.
The construction of $X_i$ depends on both $f_{i}$ and $U \sim \pi_u$. Thus, to ensure exchangeability of the predictors, we require
Assumption \ref{assumption::MCAR}.
\vspace{-0.07 in}
\begin{assumption}
\label{assumption::MCAR}
$U$ is independent of $f_i$, $\forall$ $i=1,\dots,n+1$.
\end{assumption}
\vspace{-0.07 in}

Assumption \ref{assumption::MCAR} 
is fairly mild and holds in many real-world data scenarios where the truncation time point does not depend on the observation process, e.g., monitoring of environmental measurements.
With exchangeable $\{(X_i,Y_i(t))\}$ and a properly chosen symmetric algorithm and nonconformity score, we can implement Full Functional Conformal Prediction ({\tt FFCP}) as follows. 
Let ${\cal Y}_{trial}$ be a set of trial values for $Y_{n+1}(t)$.
The set ${\cal Y}_{trial}$ is a user-specified uniform grid of points along the $y$-axis with a sufficiently large range and resolution.
For each $y \in {\cal Y}_{trial}$, we set $Y_{n+1}(t) = y$ and fit $\hat{\mu}_{y}$ to the augmented data $((X_1,Y_1(t)),\dots,(X_n,Y_n(t)),(X_{n+1},y))$. 
We then compute $S_i(t) = |Y_i(t) - \hat{\mu}_y(X_i)|$ for $i = 1,\dots,n+1$ and include $y$ in the prediction interval ${\cal I}_t$ if $S_{n+1}(t)$ is within the $1-\alpha$ quantile of the empirical distribution of $\{S_1(t),\dots,S_{n+1}(t)\}$, i.e., ${\cal I}_t := \{y: |y - \hat{\mu}_y(X_{n+1})| = S_{n+1}(t) \leq {\cal Q}_{1-\alpha}(\{S_1(t),\dots,S_{n+1}(t)\})\}$, where ${\cal Q}_{\beta}(\cdot)$ denotes the lower $\beta$ quantile function. 
We repeat this procedure for each $t \in {\cal T}$, and get the set of prediction intervals ${\cal I}_{t_1},\dots,{\cal I}_{t_T}$ for $f_{n+1}(t_1),\dots,f_{n+1}(t_T)$, where each prediction interval ${\cal I}_{t_k}$ satisfies the coverage validity in \eqref{eq::coverage-validity-FFCP} (\citet{vovk2005algorithmic}, Proposition 2.3).
\begin{remark}
We consider prediction of the entire function $f_{n+1}$, even on the interval ${\J}$ where it was observed, for reasons that will become clear in Section \ref{sec::our-method}; however, in the current setting, which does not consider phase variation, one can simply predict $f_{n+1}$ on $[0,1] \backslash{\cal J}$.
\end{remark}

A permutation symmetric algorithm is required to fit $\hat{\mu}_y$ to the augmented data composed of functional predictors $X_i$ and scalar responses $Y_i(t)$. One choice is linear scalar-on-function regression, applied independently for each $t \in {\cal T}$. But, it imposes strong assumptions on the relationship between predictors and responses, and the error, which might not hold for complex functional data. 
Instead, we use the \emph{neighborhood smoothing} technique from \citet{zhang2017estimating}, which is an adaptation of the Nadaraya-Watson estimator \citep{nadaraya1964estimating, watson1964smooth} for non-conventional data types. This approach is similar in spirit to nonparametric functional regression \citep{ferraty2007nonparametric} and enforces minimal assumptions on the predictor-response relationship.
Specifically, the neighborhood smoothing estimator for $Y_i(t)$ is
\begin{align}
\label{eq::neighborhood-smoothing}
    \widehat{Y_i(t)} :=
    \frac{\sum_{i' \neq i} K(h^{-1}d(X_i, X_{i'})) Y_{i'}(t)}{\sum_{i' \neq i} K(h^{-1}d(X_i, X_{i'}))},\quad i = 1,\dots,n+1,
\end{align}
where $K(\cdot)$ is a kernel function, e.g., triangular or Gaussian kernel, $h$ is a bandwidth parameter, and $d(X_i,X_{i'})$ is a distance between functions $X_i$ and $X_{i'}$. Higher weights are assigned to $Y_{i'}$ for predictors $X_{i'}$ that are closer to $X_i$ as measured via distance $d(\cdot,\cdot)$; $h$ controls the weights' concentration. This approach is computationally efficient, because \eqref{eq::neighborhood-smoothing} can be computed for all $t\in{\cal T}$ using a one-pass evaluation of a distance matrix $D$ with $D_{i,j} := d(X_i, X_j)$ \citep{shao2023distribution}. In our implementation, we use the Gaussian kernel for $K(\cdot)$. In addition, we must choose the distance $d(\cdot,\cdot)$ between functional predictors and bandwidth parameter $h$ in the kernel, both of which can affect prediction accuracy.

\noindent\textbf{Choice of distance.} 
Any valid distance between functions can be used. The $\mathbb{L}^2$ distance on $\mathcal{F}$, $d_2(\cdot,\cdot)$, is most commonly used in functional data analysis. Another choice is the FR Riemannian distance on $\mathcal{F}$, $d_{FR}(\cdot,\cdot)$ (Section \ref{sec::problem-set-up}). 
These two choices are computationally efficient when evaluating the distance matrix $D$, since they do not incorporate registration. 
Alternatively, one may consider the amplitude distance $d_a(\cdot,\cdot)$ (Section \ref{sec::problem-set-up}), which involves registration of predictors. This choice can lead to better prediction accuracy, but is slower to compute due to optimization over $\Gamma$.

\noindent\textbf{Choice of bandwidth parameter.} To tune the bandwidth parameter $h$, we adapt the method of \citet{liang2024conformal}, which ensures coverage validity while selecting a model that minimizes PI length. Specifically, we consider a set of candidate bandwidth values $h \in {\cal H}$ and obtain prediction intervals ${\cal I}_{t_k}^{h}$ for $k=1,\dots,T$. We can set $\cal H$ to be a grid of fixed values, or the lower $\beta$ quantile of the empirical distribution of the distances $\{D_{i,j}\}_{i < j}$ for a grid of $\beta \in (0,1)$. We select the optimal bandwidth parameter $h^*$ (i) globally, $h^* := \argmin_{h \in {\cal H}} \frac{1}{T} \sum_{k=1}^{T} {\rm length}({\cal I}_{t_k}^{h})$, or (ii) locally, $h_{t_k}^* := \argmin_{h \in {\cal H}}{\rm length}({\cal I}_{t_k}^{h}),\ k = 1,\dots,T$.

\begin{figure}[!t]
    \centering
    \includegraphics[width=0.8\linewidth]{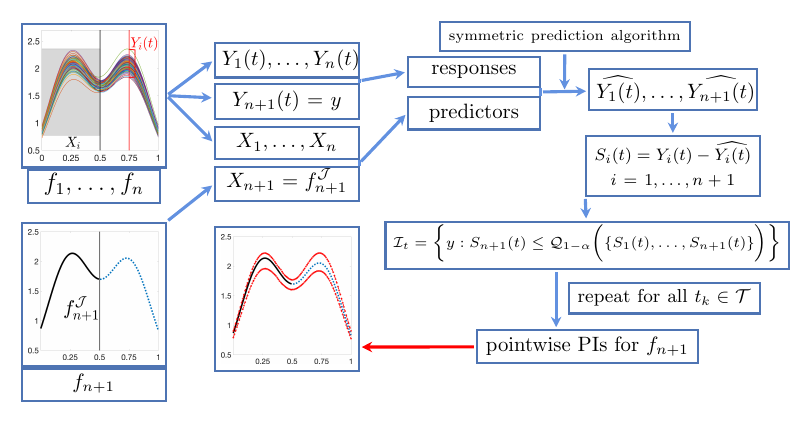}
    \vspace{-10pt}
    \caption{\small Full Functional Conformal Prediction ({\tt FFCP}) algorithm.}
    \label{fig::alg::FFCP}
    \vspace{-10pt}
\end{figure}

\begin{figure}[!t]
    \centering
    \begin{tabular}{cccc}
    (a)&(b)&(c)&(d)\\
        \includegraphics[width = 0.15\textwidth]{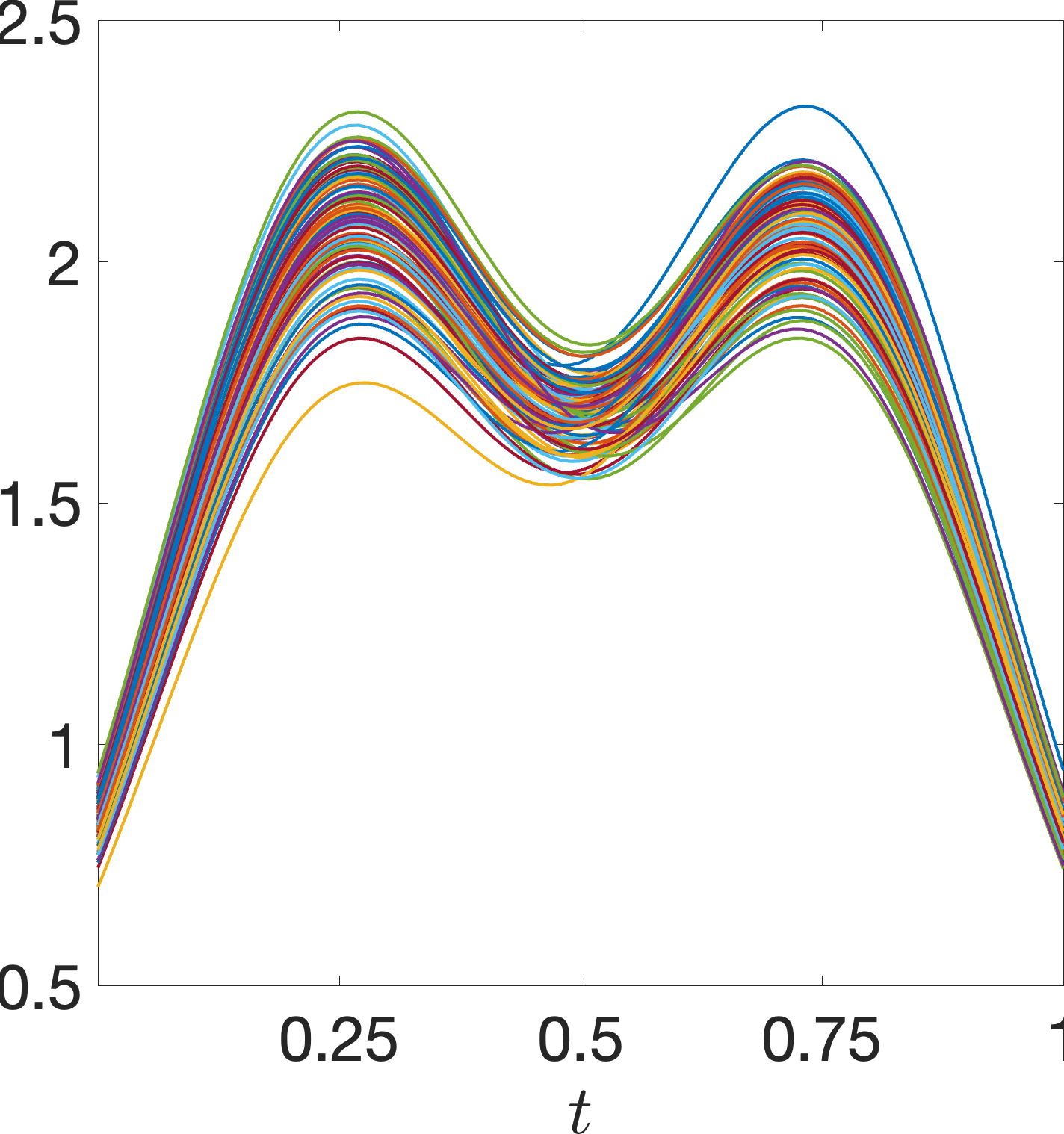} &
        \includegraphics[width = 0.17\textwidth]{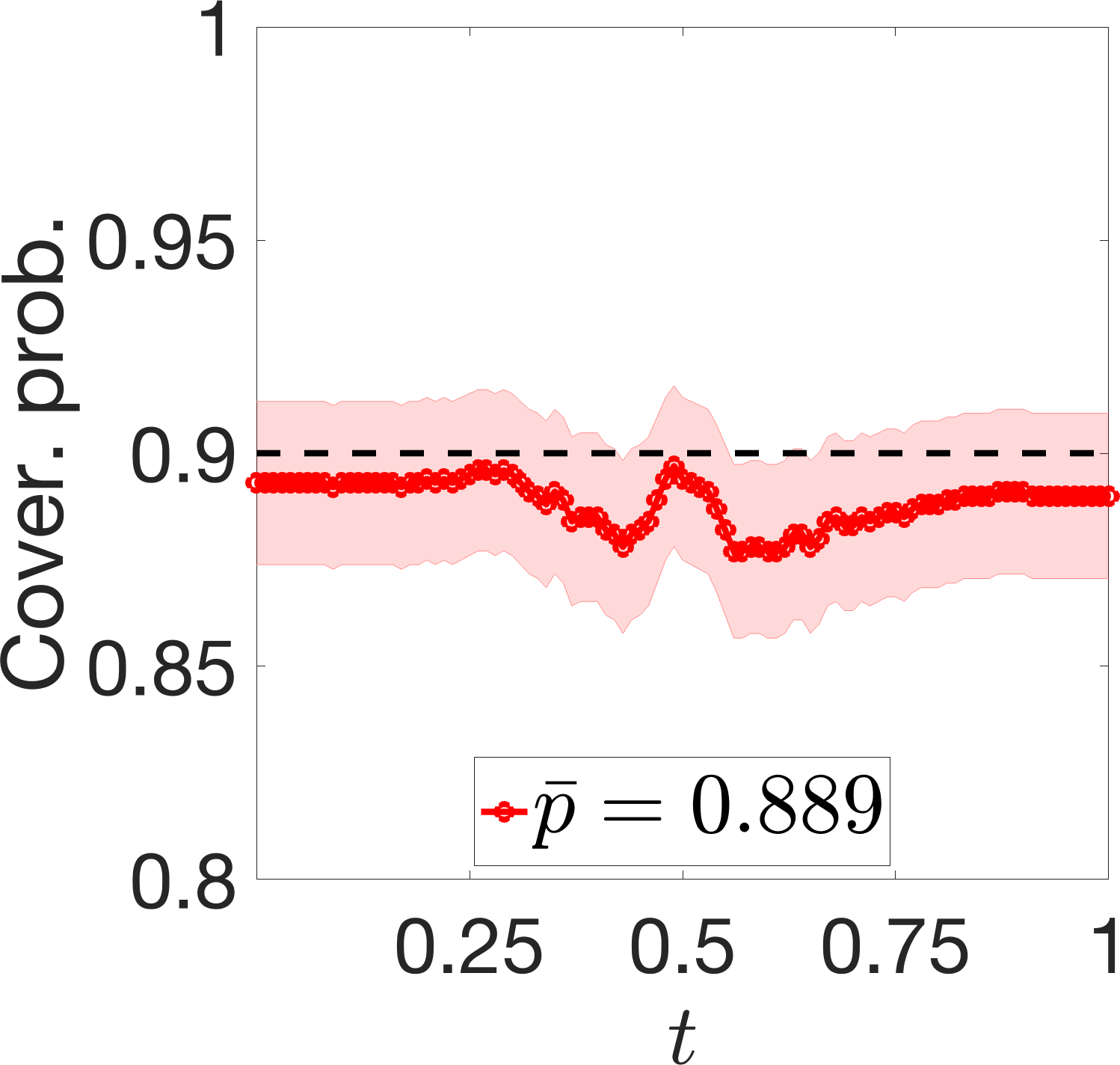} &
        \includegraphics[width = 0.17\textwidth]{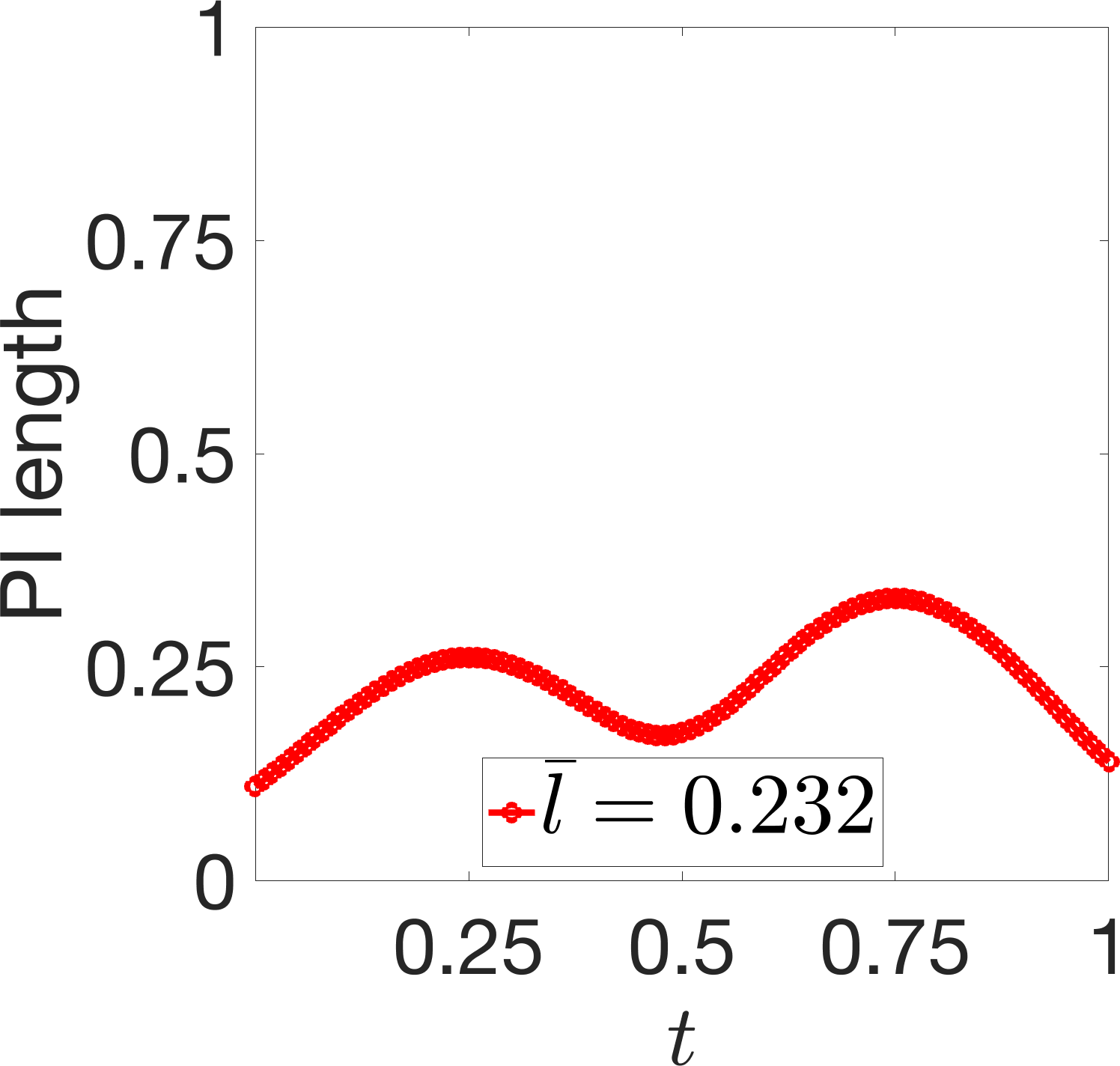} &
        \includegraphics[width = 0.15\textwidth]{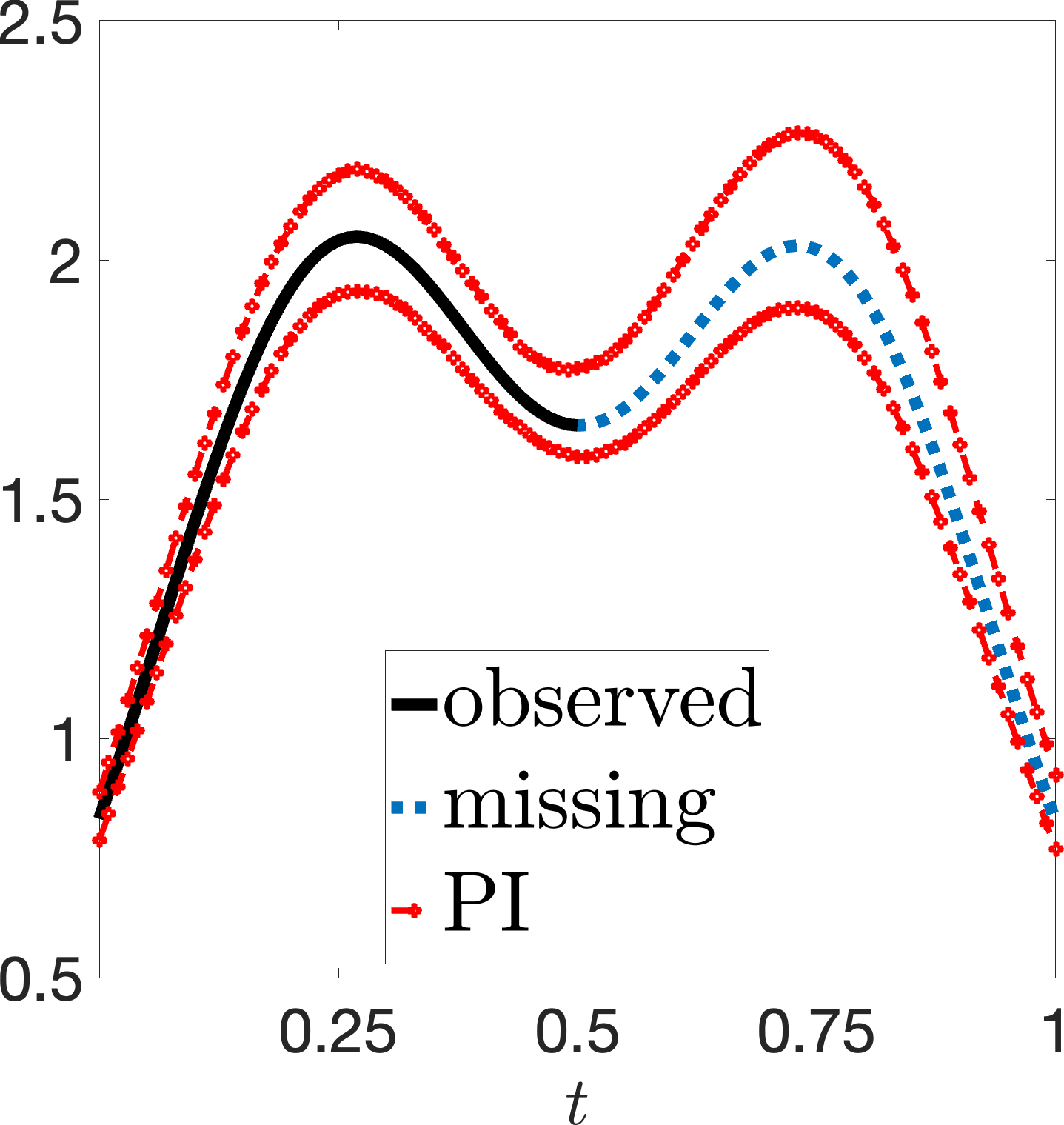}
        \\
        \includegraphics[width = 0.15\textwidth]{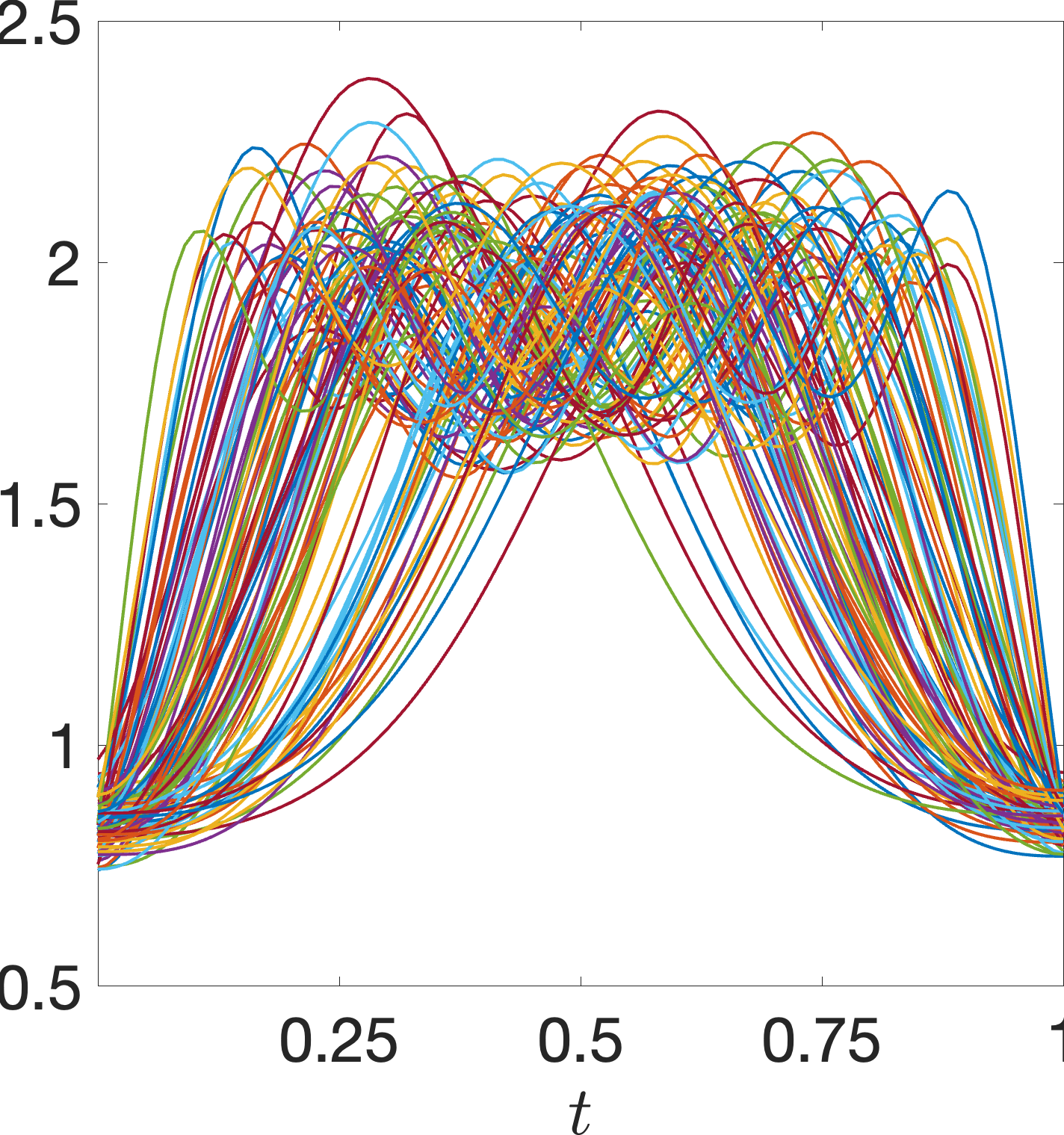} &
        \includegraphics[width = 0.17\textwidth]{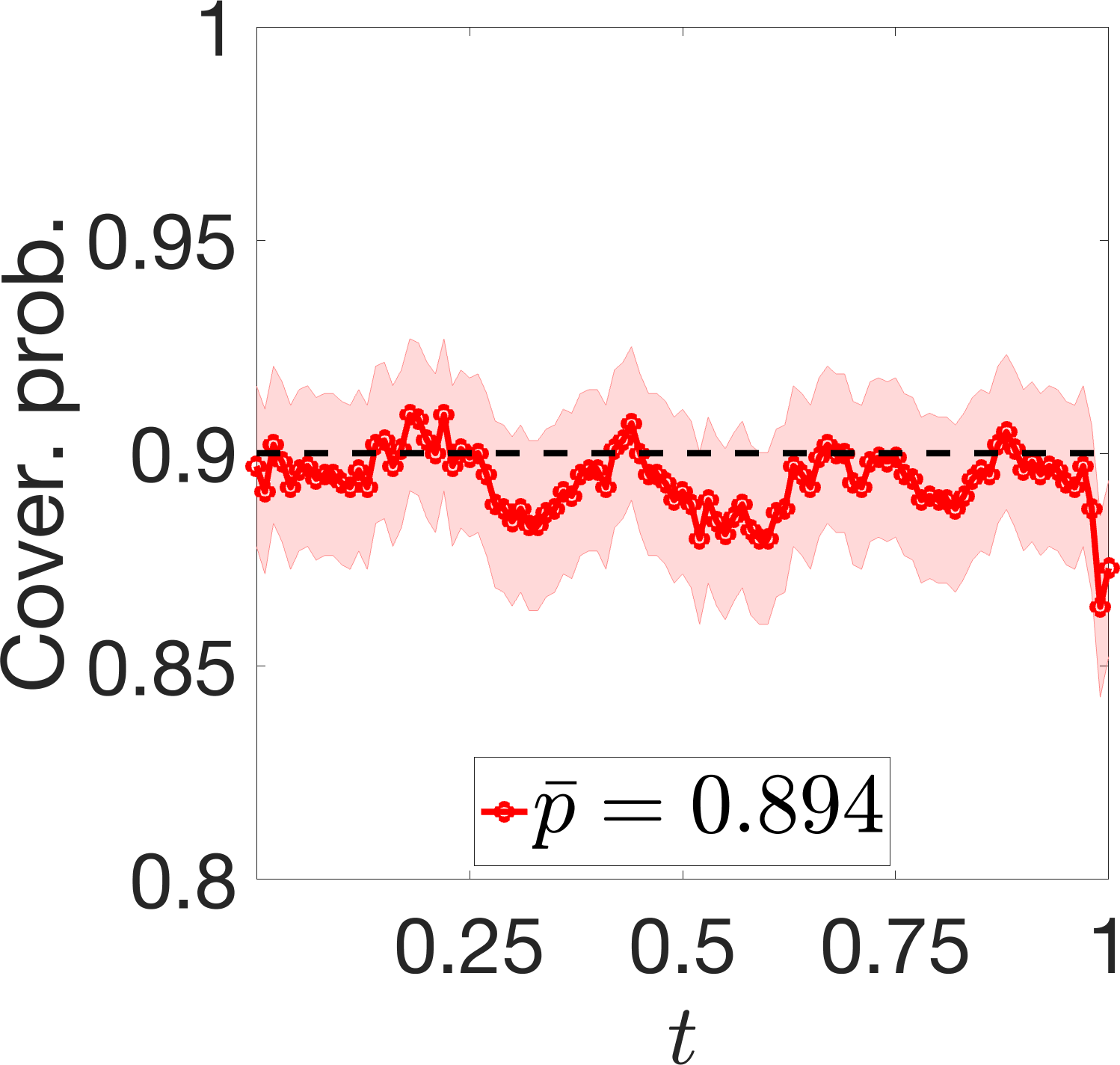} &
        \includegraphics[width = 0.17\textwidth]{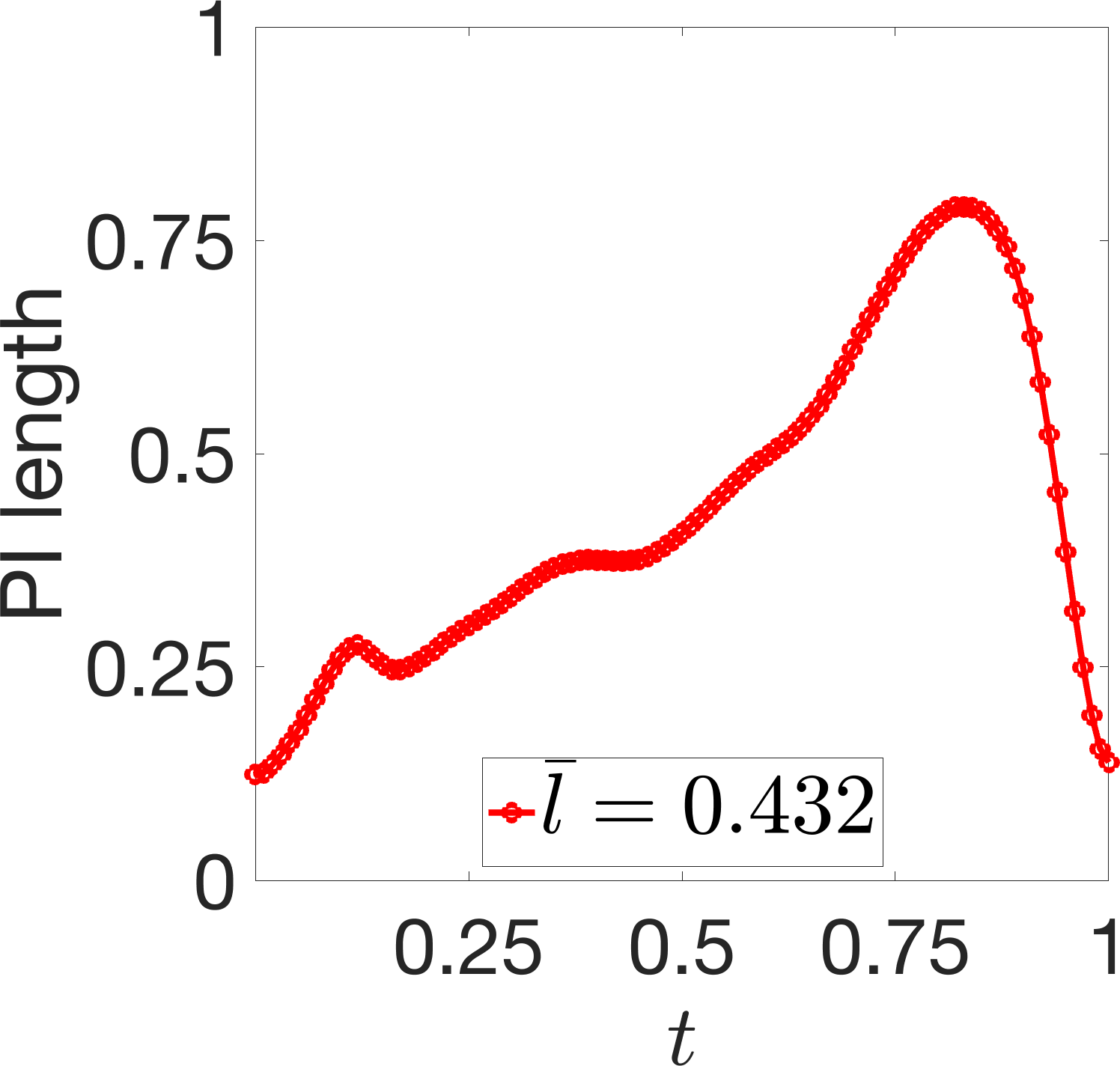} &
        \includegraphics[width = 0.15\textwidth]{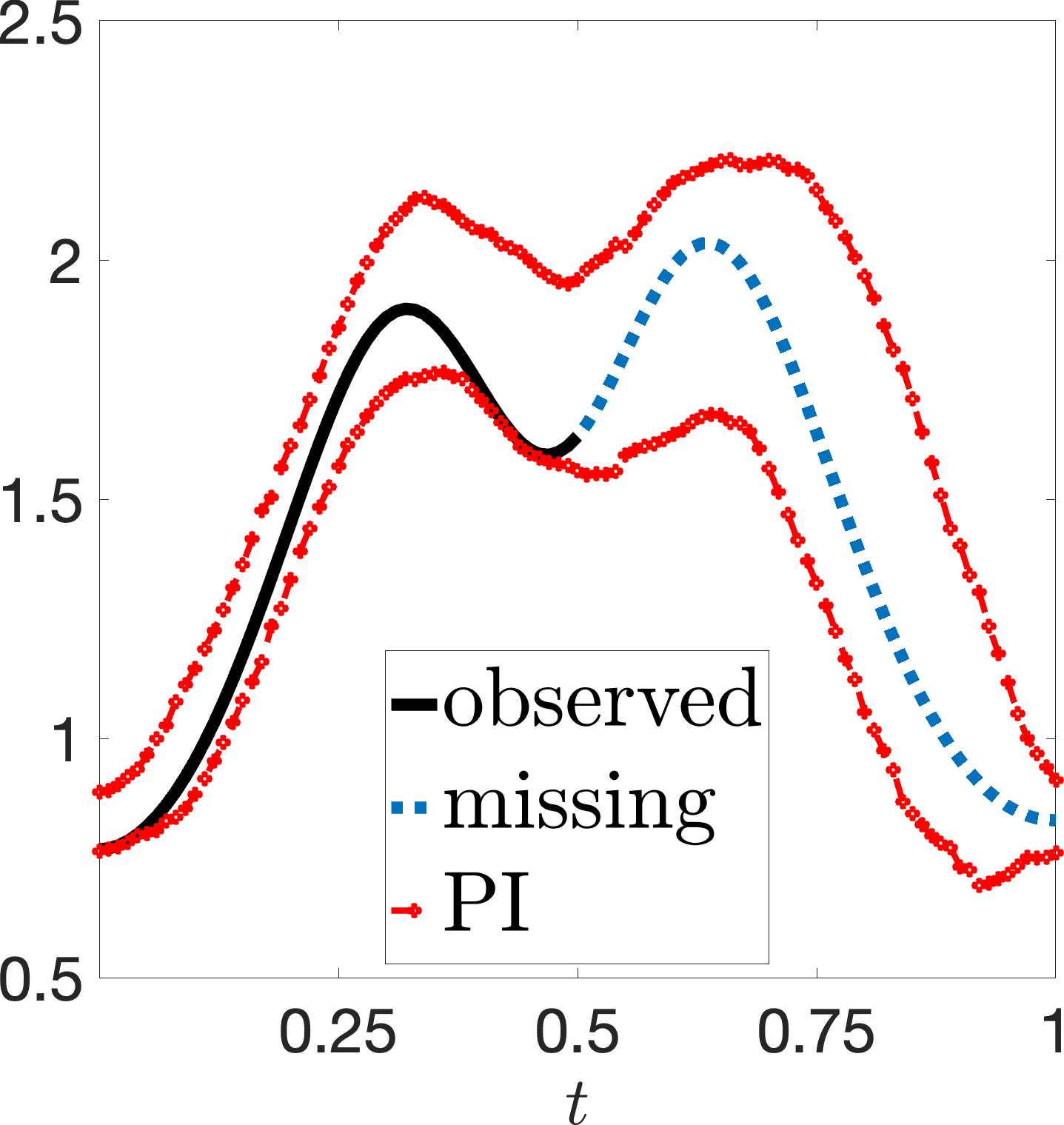}    \end{tabular}
    \vspace{-10pt}
    \caption{\small {\tt FFCP} without (row 1) and with (row 2) phase variation. (a) Data. (b) \& (c) Coverage rates $p_k$ (solid red) with 95\% CIs (shaded region) and average PI lengths $\ell_k$, respectively. (d) Partial observation $f_{n+1}^{\cal J}$ (black), ground truth $f_{n+1}^{[0,1] \backslash{\cal J}}$ (blue), and pointwise PIs (red).}
    \label{fig::full-conformal}
    \vspace{-10pt}
\end{figure}

The full {\tt FFCP} algorithm is presented in Figure \ref{fig::alg::FFCP}. To illustrate the practical performance of $\tt FFCP$, we present a quick simulation. 
Consider two-peak functions without and with phase variation (top and bottom of Figure \ref{fig::full-conformal}(a), respectively). 
We draw $B = 1000$ Monte Carlo samples with $n=100$, truncation time point $U=0.5$ for $f_{n+1}$, and $1-\alpha=0.9$. Evaluation is based on two criteria,
    (i) \textbf{coverage validity}: empirical pointwise coverage rate, 
$
        p_k = \frac{1}{B} \sum_{b=1}^{B}
        \mathbbm{1}
        \Big\{
            f^{(b)}_{n+1}(t_k) \in {\cal I}_{t_k}^{(b)}
        \Big\},\ k = 1,\dots,T 
$;
    (ii) \textbf{prediction accuracy}: pointwise average prediction interval (PI) length
    $
        \ell_k = \frac{1}{B}\sum_{b=1}^{B} {\rm length}({\cal I}_{t_k}^{(b)}),\ k=1,\dots,T
    $.
\setlength{\tabcolsep}{0pt}
 Results for functional data without (with) phase variation are given in the top (bottom) row of Figure \ref{fig::full-conformal}. 
Panels (b) and (c) show pointwise coverage rates $p_k$ (solid red) with pointwise 95\% confidence intervals (CIs) (shaded region), and pointwise average PI lengths $\ell_k$ for $k=1,\dots,T$, respectively. 
The time-averaged coverage rate $\bar{p} = \sum_{k=1}^{T} p_k/T$ and PI length  $\bar{\ell} = \sum_{k=1}^{T} \ell_k/T$ are also reported. 
We see that {\tt FFCP} guarantees finite-sample coverage validity for any $t \in {\cal T}$ for functions without and with phase variation. 
However, pointwise PI lengths tend to be much larger when phase variation is present in the data. 
Panel (d) demonstrates pointwise prediction intervals ${\cal I}_{t_1},\dots,{\cal I}_{t_T}$ for $f_{n+1}$ for a randomly chosen Monte Carlo sample. 
We show the partial observation $f_{n+1}^{\cal J}$ (black), the ground truth missing segment $f_{n+1}^{[0,1]\backslash \cal J}$ (blue), and the lower and upper boundaries of the pointwise PIs (red).  
When there is no phase variation in the data, the PIs form a prediction band that accurately captures the geometric features, i.e., two peaks and one valley, of the underlying function $f_{n+1}$. 
However, when phase variation is present, the prediction band is less effective at capturing such geometric features, since their timing varies considerably across observations. 
As a result, $Y_i(t)$ fails to provide useful information for predicting $Y_{n+1}(t)$. 
These results indicate that {\tt FFCP} is better suited for functional data without phase variation. 

\vspace{-0.3in}
\section{Joint registration and prediction for functional data}
\label{sec::our-method}
\vspace{-0.1 in}
Next, we incorporate registration into conformal prediction when functional data exhibits phase variation. 
In this context, the prediction target changes to the amplitude $f_{n+1} \circ \gamma^*_{n+1}$, where $\gamma^*_{n+1} \in \Gamma$ registers $f_{n+1}$ to a template function. Thus, we define new response variables $Y_i(t) := (f_i \circ \gamma^*_i)(t)$ for each $t \in {\cal T}$, while keeping the predictors $X_i = f_i^{\cal J}$ unchanged. However, a critical challenge now is the choice of template for registration that preserves exchangeability of $\{(X_i,Y_i(t))\}$. One simple choice is to randomly select a function from the complete observations $f_1,\dots,f_{n}$ as the template, e.g., $f_1$. However, this would break exchangeability of $\{(X_i,Y_i(t))\}$, since the registration process would treat $f_1$ differently from $f_2,\dots,f_{n+1}$. Alternatively, to preserve exchangeability, one could still use $f_1$ as the template, but only use $(f_i\circ\gamma_i^*)(t),\ i=2,\dots,n+1$ to define the responses. However, this approach may lead to unstable predictions, as the randomly selected template function may not be representative of the population, and could even be an outlier. To address this, we need a template function that is (i) representative of the population, and (ii) independent of the functions being registered. This will be accomplished using the split conformal prediction approach.

\begin{figure}[!t]
    \centering
    \includegraphics[width=0.8\linewidth]{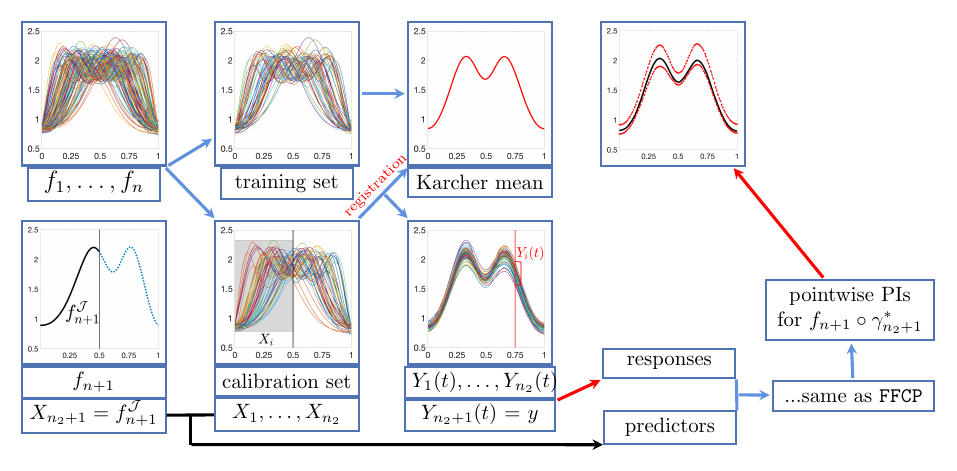}
    \vspace{-10pt}
    \caption{\small Split Functional Conformal Prediction ({\tt SFCP}) algorithm.}
    \label{fig::alg::SFCP}
    \vspace{-10pt}
\end{figure}

Specifically, we randomly split $f_1,\dots,f_n$ into two independent training and calibration sets of sizes $n_1$ and $n_2 = n - n_1$, respectively. We first use the training set $f_1,\dots,f_{n_1}$ to compute the sample Karcher mean $\bar{f}$ using \eqref{eq::def-Karcher-mean} and the inverse SRSF mapping $Q^{-1}$. We then register each function in the calibration set, $f_{n_1+1},\dots,f_n$, to $\bar{f}$ by computing $\gamma_1^*,\dots,\gamma_{n_2}^*$ via \eqref{eq::registration-SRSF} ($\gamma^*_i$ is the relative phase of $f_{n_1+i}$ with respect to $\bar{f}$). Finally, we use the amplitudes $\tilde{f}_i := f_{n_1+i} \circ \gamma_i^*$ to construct the response variables $Y_{i}(t) = \tilde{f}_i(t),\ i=1,\dots,n_2$. The response corresponding to $f_{n+1}$ is $Y_{n_2+1}(t) = (f_{n+1} \circ \gamma_{n_2+1}^*)(t)$. This procedure maintains exchangeability of $\{(X_i,Y_i(t))\}$, as permutation symmetry is preserved when the template function is computed using an independent training set. After constructing exchangeable $\{(X_i,Y_i(t))\}$, they can be used as the input of {\tt FFCP}. We call this procedure Split Functional Conformal Prediction ({\tt SFCP}) and present it in Figure \ref{fig::alg::SFCP}. Like {\tt FFCP}, {\tt SFCP} provides a marginal finite-sample coverage guarantee, which is formally stated in Theorem \ref{theorem::split-conformal::coverage-validity}.
\begin{theorem}
\label{theorem::split-conformal::coverage-validity}
    Under Assumption \ref{assumption::MCAR}, for any $T \geq 2$, any calibration set size $n_2$ such that $0 < n_2 < n$, any $\alpha \in (0,1)$ and each $t_k \in {\cal T}$, the prediction set ${\cal I}_{t_k}$ from {\tt SFCP} satisfies $1-\alpha \leq
        \pr\Big(
            (f_{n+1}\circ\gamma^*_{n_2+1})(t_k) 
            \in {\cal I}_{t_k}
        \Big) 
        \leq 1-\alpha + n_2^{-1}$.
\end{theorem}
\begin{remark}
    While the anti-conservative bound improves with larger calibration set size $n_2$, a sufficiently large training set size $n_1$ is also needed to ensure stable estimation of the sample Karcher mean. This trade-off motivates a balanced data split.
\end{remark}
\begin{remark}
    {\tt SFCP} targets the amplitude of $f_{n+1}$, $f_{n+1} \circ \gamma_{n_2+1}^*$, where the latent $\gamma_{n_2+1}^*$ registers $f_{n+1}$ to the Karcher mean of the training set. Since $\gamma_{n_2+1}^*(U)$ is unobserved, it is unknown which time interval $[0,1]\backslash\J$ corresponds to after accounting for phase variation. Thus, we must generate prediction sets over the entire domain $[0,1]$, i.e., for all $t_k\in{\cal T}$.
\end{remark}

The relationship between {\tt FFCP} and {\tt SFCP} is different from the relationship between classical full and split conformal prediction. 
The split conformal method was originally developed to improve computational efficiency of the full conformal method. 
In our case, however, the training set is used to estimate the sample Karcher mean, which acts as the template for registration of the calibration set, as well as the target function. 
The goal is not to speed up computation, but rather to construct new response variables that improve prediction accuracy in the presence of phase variation. 
Both {\tt FFCP} and {\tt SFCP} are suitable prediction methods for functional data with or without phase variation as they both guarantee finite-sample coverage. 
However, failing to account for phase variation generally inflates variance \citep{marron2015functional}, and as a result, impairs prediction (simulation in Section \ref{subsec::FFCP}). 
Thus, we advocate the use of {\tt SFCP} rather than {\tt FFCP} for functional data with phase variation.


\vspace{-0.2 in}
\subsection{Other observational regimes for partial functional data}
\label{subsec::other-obs-pattern}
\vspace{-0.1 in}
$\tt FFCP$ and $\tt SFCP$ are not limited to settings where a single continuous segment of $f_{n+1}$ on ${\cal J} = [0,U]$ is observed. 
The approaches can be generalized to the case of ${\cal J} = [U_1,U_2]$, where $U_1, U_2 \sim \pi_{\bm u},\ 0 \leq U_1 < U_2 \leq 1$ and $\pi_{\bm u}$ is a probability density on $[0,1]^2$. Two other generalizations apply to fragmented and sparse functional data.

\noindent{\bf Fragmented.} The observed portions of $f_{n+1}$ are spread across random disjoint subintervals, ${\cal J} = \cup_{j=1}^{J} {\cal J}_j$, ${\cal J}_j = [U_{j,1}, U_{j,2}]$.  This scenario is often encountered in applications involving segmented functional observations, e.g., X-ray measurements of bone mineral density \citep{bachrach1999bone}. In this case, the predictors are $X_i=\bm{f}_{i} := (f_{i}^{{\cal J}_1}, \dots, f_{i}^{{\cal J}_J}),\ i=1,\dots,n+1$. One can then define a distance for neighborhood smoothing as $d_{\rm prod}(X_i, X_{i'}) := \sum_{j=1}^{J} \lambda_j d(f_i^{{\cal J}_j}, f_{i'}^{{\cal J}_j})$, where $\lambda_j$ are weights with $\sum_{j=1}^{J} \lambda_j = 1$. The simplest choice is $\lambda_j = 1/J\ \forall\ j$. Alternatively, one can choose $\lambda_j$ to be proportional to the length of each subinterval.

\noindent{\bf Sparse.} Here, $f_{n+1}$ is observed at a set of discrete time points, ${\cal J} = \{t_1,\dots,t_{N_t}\} \subseteq [0,1]$. Such observations are common when full continuous measurements are difficult or impractical to collect, e.g., longitudinal data in clinical trials \citep{yao2005functional}. Now, the predictors are $X_i := (f_i(t_1),\dots,f_i(t_{N_t})) \in \R^{N_t}$, and we use the Euclidean distance in $\R^{N_t}$ for neighborhood smoothing: $d_e(X_i, X_{i'}) 
        :=
        \sqrt{\sum_{k=1}^{N_t} (f_i(t_k) - f_{i'}(t_k))^2}$.
\vspace{-0.2 in}
\subsection{Conformal prediction of relative phase}
\label{subsec::phase-prediction}
\vspace{-0.1 in}
{\tt SFCP} focuses on predicting the amplitude component $f_{n+1}\circ\gamma_{n_2+1}^*$, but it is also useful for prediction of the relative phase $\gamma_{n_2+1}^*$. Phase prediction allows assessment of uncertainty in the relative timing of amplitude features of $f_{n+1}$ with respect to the sample Karcher mean $\bar{f}$. 
We use the same functional predictors as before, $X_i = f_{n_1+i}^{\cal J},\ i=1,\dots,n_2+1$, which include amplitude and phase variation. However, we cannot predict the target function $\gamma_{n_2+1}^*$ pointwise at each $t \in {\cal T}$ independently, because the resulting PIs $\{{\cal I}_{t_k}\}$ would not be monotonically increasing (recall that $\dot{\gamma}(t)>0\ \forall\ t$). 
Instead, we predict $\gamma_{n_2+1}^*$ jointly for all $t \in {\cal T}$, by considering $\bm{Y}_{n_2+1} = (\gamma_{n_2+1}^*(t_1),\dots,\gamma_{n_2+1}^*(t_T)) \in \R^T$ as the prediction target. Thus, we define the other response variables as $\bm{Y}_i = (\gamma_i^*(t_1), \dots,\gamma_i^*(t_T))$ for $i=1,\dots,n_2$. This results in exchangeable $\{(X_i,\bm{Y}_i)\}$. We again employ the neighborhood smoothing estimator in \eqref{eq::neighborhood-smoothing}, which yields $\widehat{\bm{Y}}_i \in \R^T,\ i=1,\dots,n_2+1$. Note that the discretization grid ${\cal T}$ for prediction of relative phase can be different from the discretization grid used for prediction of amplitude. Due to computational considerations, we use a much coarser (equally spaced) time grid for relative phase prediction.
\begin{figure}[!t]
    \centering
    \includegraphics[width=0.8\linewidth]{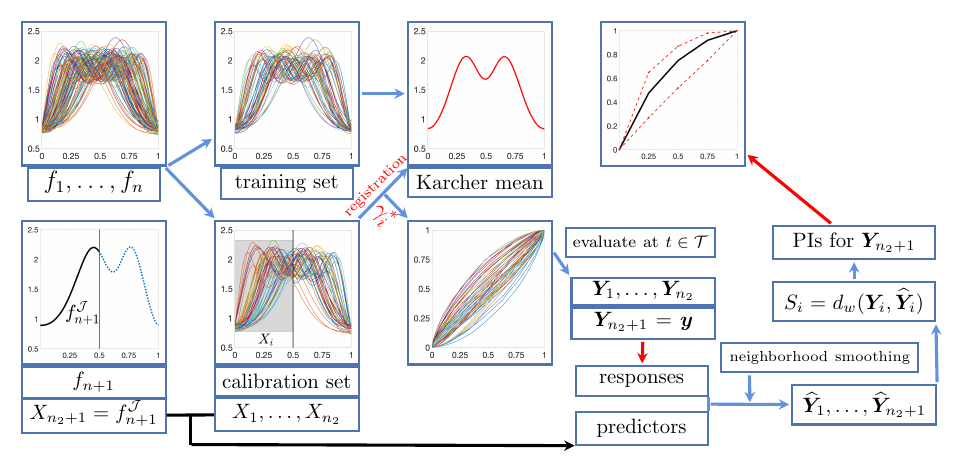}
    \vspace{-10pt}
    \caption{\small Split Functional Conformal Prediction of relative Phase ({\tt SFCPP}) algorithm.}
    \label{fig::alg::SFCPP}
    \vspace{-10pt}
\end{figure}

Since $\bm{Y}_i \in \R^T$ is a discretized version of $\gamma_i^*$, we use the FR distance on $\Gamma$ as the nonconformity score. To efficiently compute the FR distance, we again employ the SRSF representation: for a $\gamma \in \Gamma$, its SRSF is $q^{\gamma} = \sqrt{\dot{\gamma}}$. Under the SRSF representation, the space of warping functions is the positive orthant of the Hilbert sphere and the FR metric simplifies to the $\mathbb{L}^2$ metric (\citet{srivastava2016functional}, Section 4.10.2). Thus, the distance between two warping functions $\gamma_i$ and $\gamma_j$ is 
$
d_w(\gamma_i,\gamma_j) 
    :=
    \cos^{-1}(
    \int_{0}^1 q^{\gamma}_i(t)q^\gamma_j(t)dt
    ),
$ resulting in the nonconformity score $S_i = d_w(\bm{Y}_i, \widehat{\bm{Y}}_i)$ for $i=1,\dots,n_2+1$.
The prediction set is given by
$
    \I_{\gamma}
    :=
    \{
    \bm{y}: S_{n_2+1} 
    \leq
    {\cal Q}_{1-\alpha}
    (\{S_1,\dots,S_{n_2+1}\})
    \}
$,
where $\bm{y} = (y_1,\dots,y_T) \in \R^T$ are trial vectors for $\bm{Y}_{n_2+1}$ with fixed $y_1=0$ and $y_T=1$. Figure \ref{fig::alg::SFCPP} illustrates the Split Functional Conformal Prediction of relative Phase ({\tt SFCPP}) algorithm.

\vspace{-0.3in}
\section{Simulations}
\label{sec::simulations}
\vspace{-0.1 in}
We simulate i.i.d. $f_1,\dots,f_{n+1}$ without phase variation from a homogeneous population of two-peak functions: $f_i(t) = \bm{Z}_{i1}\exp \{-(t-0.25)^2/0.072 \} + \bm{Z}_{i2}\exp\{-(t-0.75)^2/0.072\}$, $\bm{Z}_i := (\bm{Z}_{i1}, \bm{Z}_{i2}) \stackrel{i.i.d.}{\sim} N(2, 0.1 {\bf I_2})$.
We induce phase variation in $\{f_i\}$ by simulating $\gamma_i = F_{a,b}$, where $F_{a,b}$ is the cumulative distribution function of a $\textrm{Beta}(a,b)$ with $a,b \stackrel{i.i.d.}{\sim}\textrm{Unif}(1,3)$, and computing $f_i \circ \gamma_i$. To evaluate coverage validity, prediction accuracy and computational efficiency, we use $B=500$ Monte Carlo samples with $n=100$, $T = 100$ and $1-\alpha=0.9$. We use $d_2(\cdot,\cdot)$ on $\mathcal{F}$ and local tuning for the bandwidth parameter in neighborhood smoothing.

\noindent\textbf{Simulation 1: comparison to other prediction methods.} We compare {\tt SFCP} and {\tt FFCP} to two state-of-the-art functional regression methods, implemented in the R package {\tt refund} \citep{R-refund}: (i) scalar-on-function regression ({\tt SoF}), with predictors $f_i^{\cal J}$ and responses $f_i(t)$ fitted independently for each $t \in {\cal T}$, i.e., $\mathbb{E}[f_i(t)] = \beta_0 + \int_{\cal J} \beta(s)f_i(s)  ds$, $i=1,\dots,n,\ t \in {\cal T}$; (ii) function-on-function regression ({\tt FoF}), with predictors $f_i^{\cal J}$ and responses $f_i$, i.e., $\mathbb{E}[f_i] = \beta_0 + \int_{\cal J} \beta(s,t)f_i(s)  ds, i=1,\dots,n$. The PIs for these two approaches are constructed using the method of \citet{marra2012coverage}, which utilizes the Bayesian posterior covariance matrix of the parameters. To ensure numerical stability of the optimization procedures for {\tt SoF} and {\tt FoF}, we apply functional principal component analysis (FPCA) to the predictors, fixing the dimension to 8 or a number that explains 90\% of the total variance, whichever is smaller. In addition to the pointwise coverage rates $p_k$, we also evaluate the empirical overall coverage rate, 
$
    p = \frac{1}{B} \sum_{b=1}^{B}
    \mathbbm{1}
    \big\{
        f^{(b)}_{n+1}(t_k) \in {\cal I}_{t_k}^{(b)}\ \forall\ k\in[1:T]
    \big\},
$
which checks whether the entire target function is within the pointwise prediction band. 
\setlength{\tabcolsep}{0pt}
\begin{figure}[!t]
    \centering
    \begin{tabular}{cccccc}
    (a)&(b)&(c)&(d)&(e)&(f)\\
    \includegraphics[width = 0.16\textwidth]{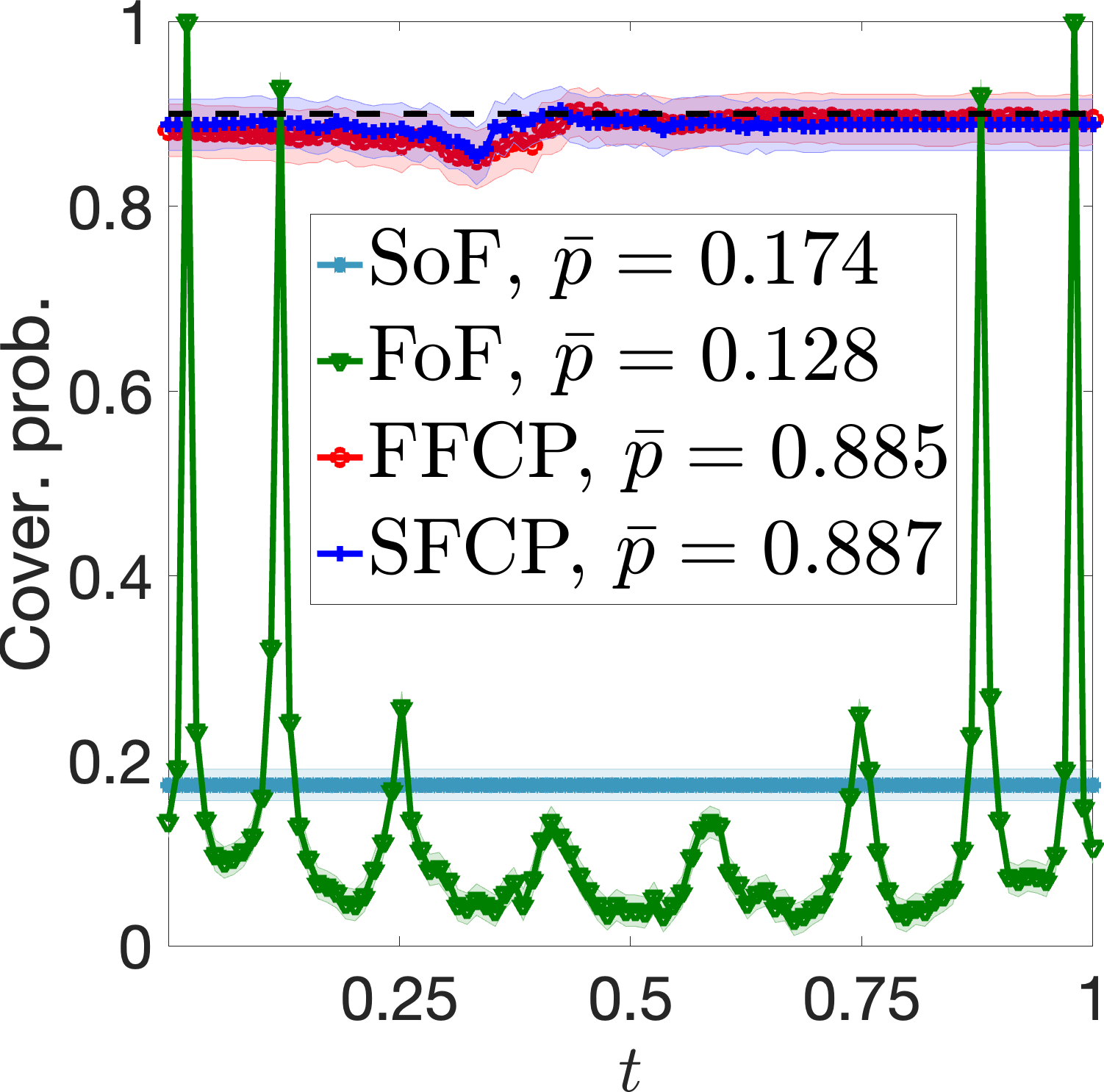} &
    \includegraphics[width = 0.16\textwidth]{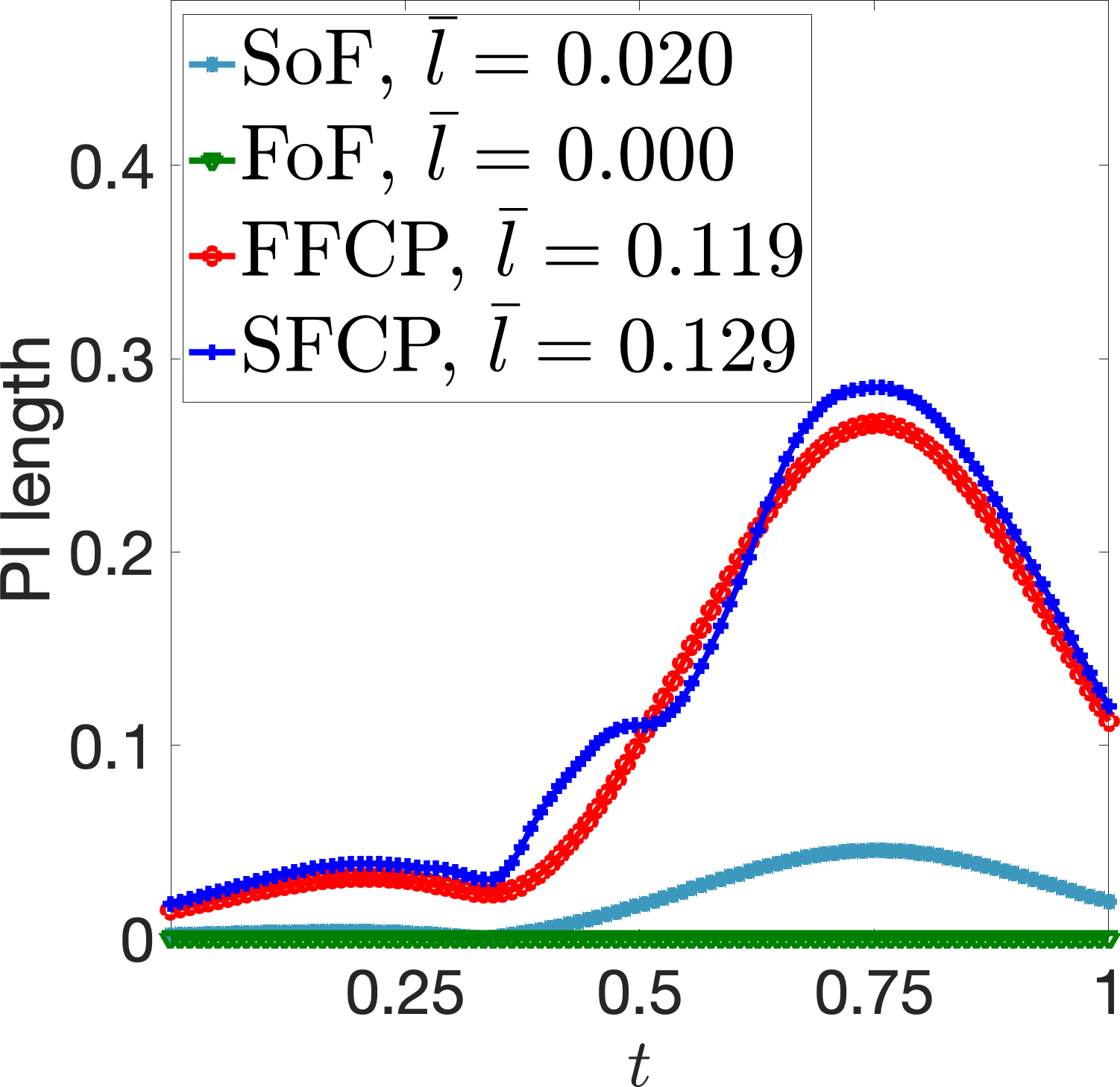} &
    \includegraphics[width = 0.16\textwidth]{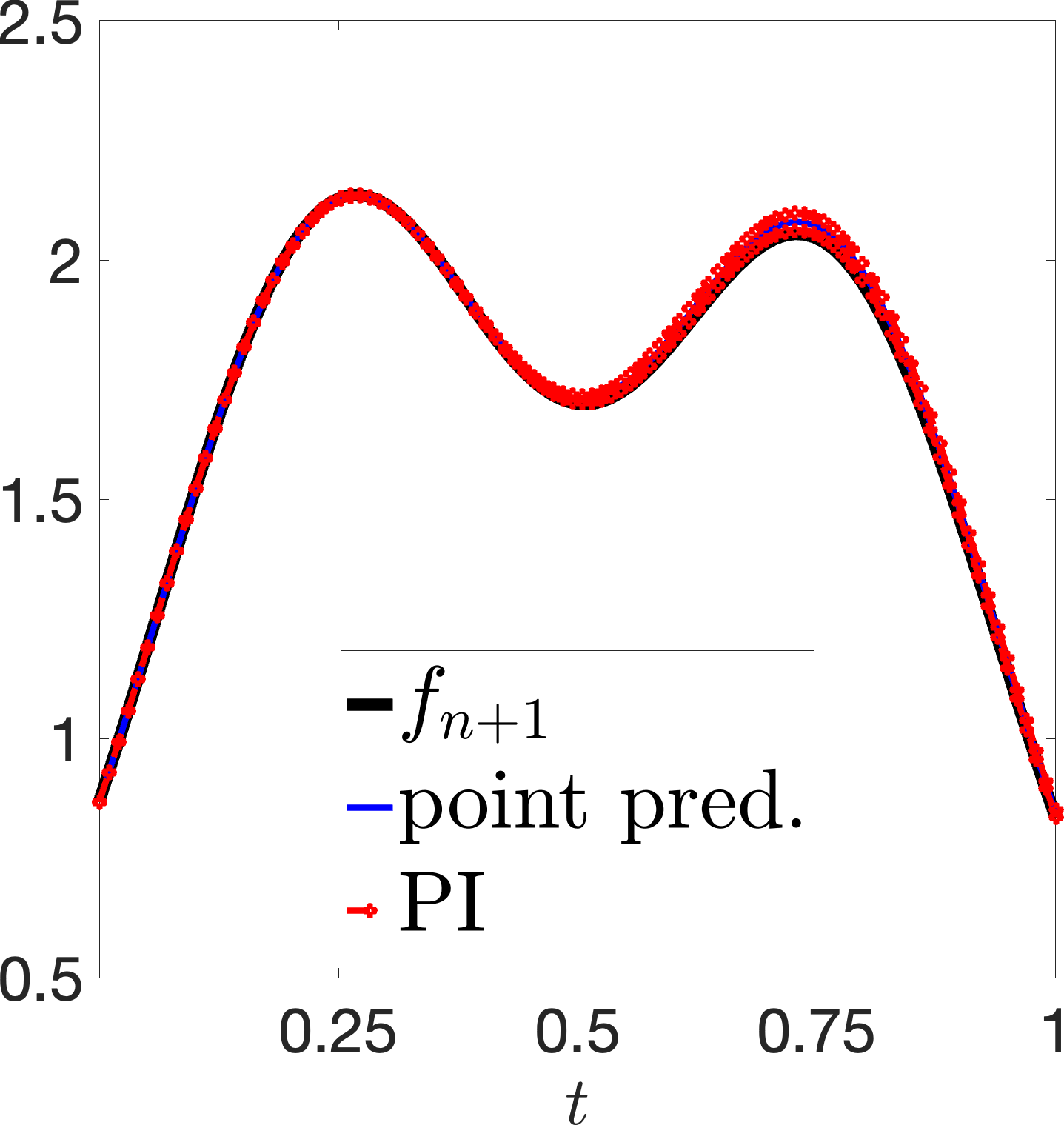} &
    \includegraphics[width = 0.16\textwidth]{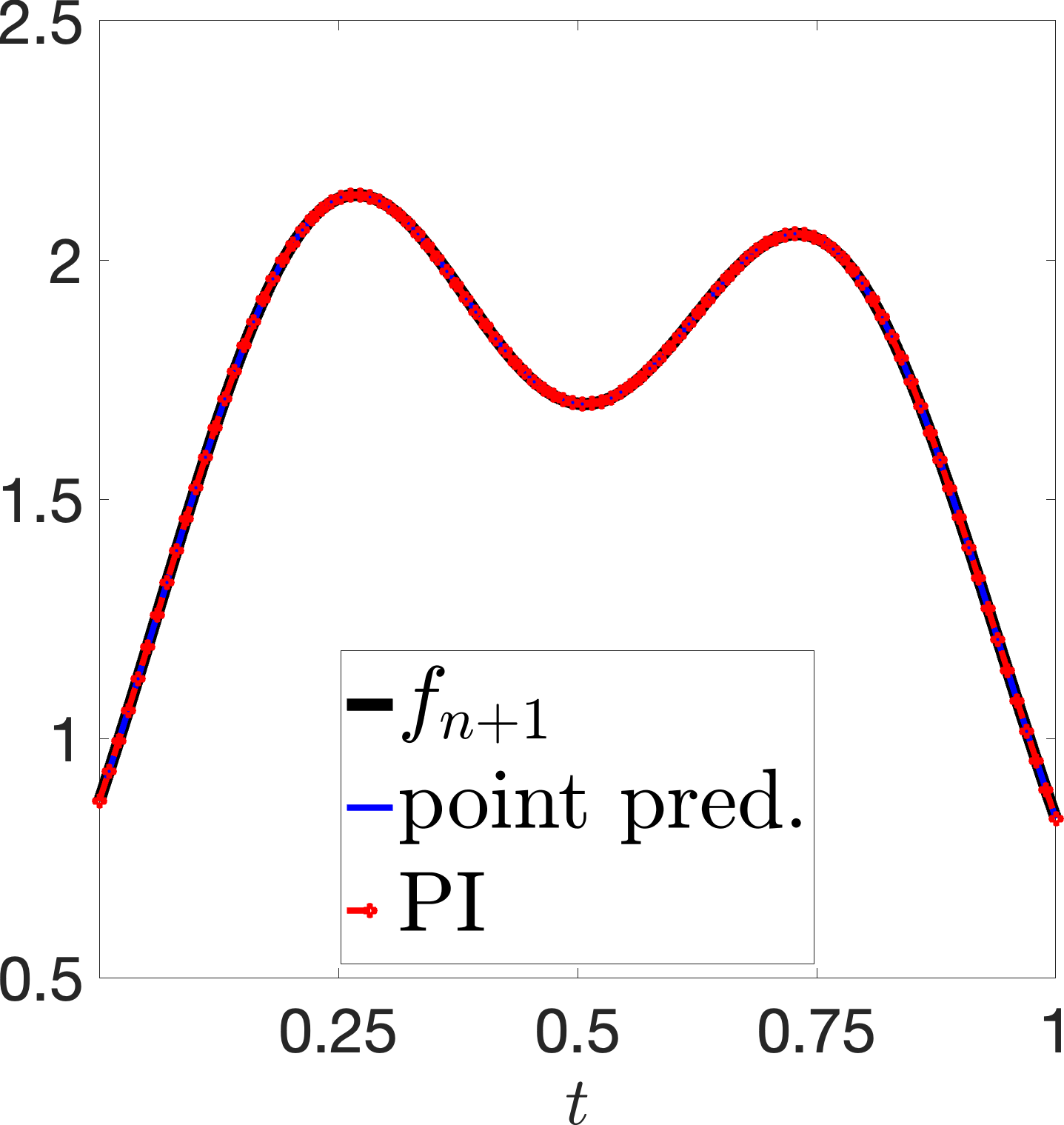} &
    \includegraphics[width = 0.16\textwidth]{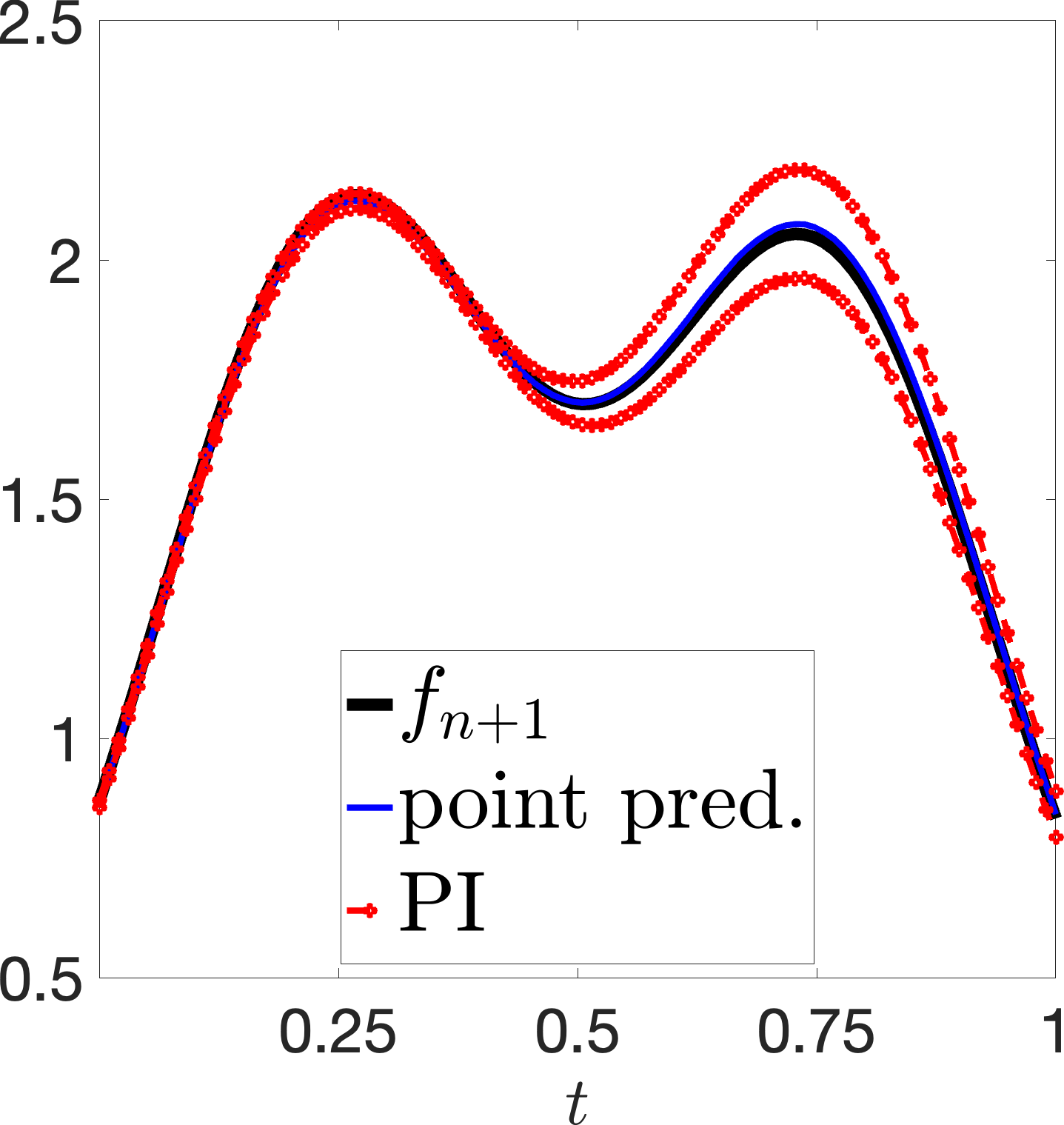} &
    \includegraphics[width = 0.16\textwidth]{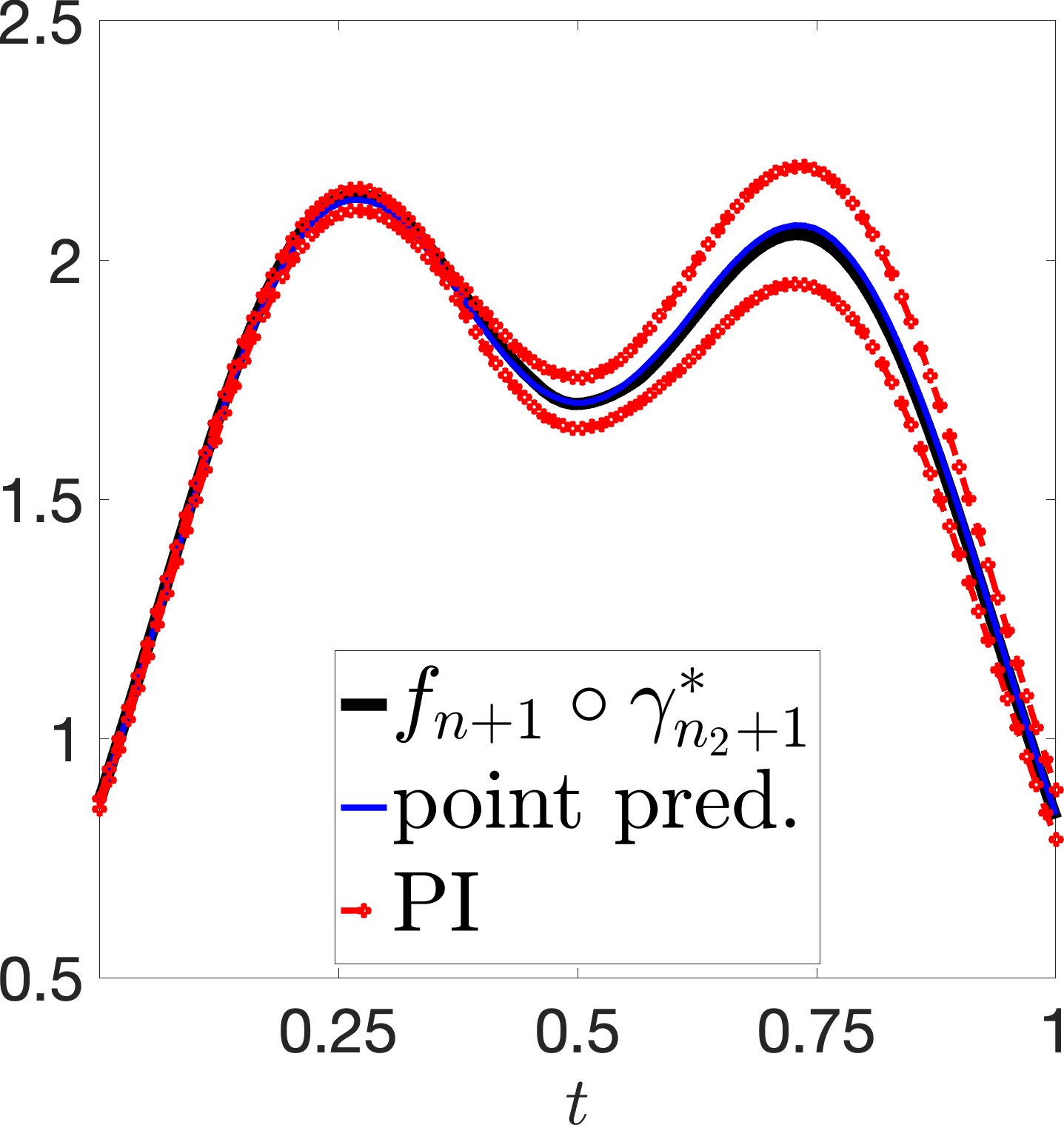}
    \\
    \includegraphics[width = 0.16\textwidth]{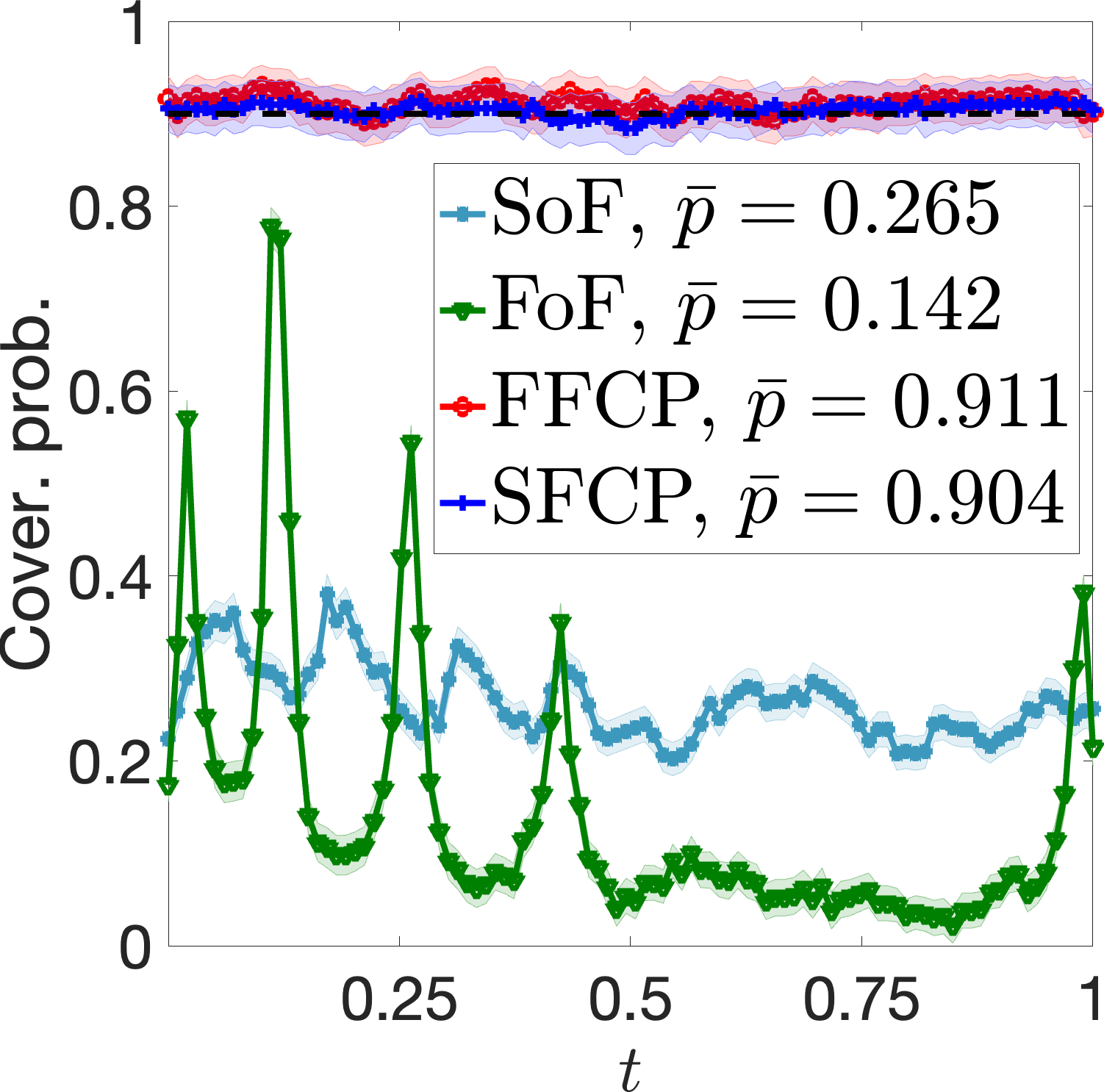} &
    \includegraphics[width = 0.16\textwidth]{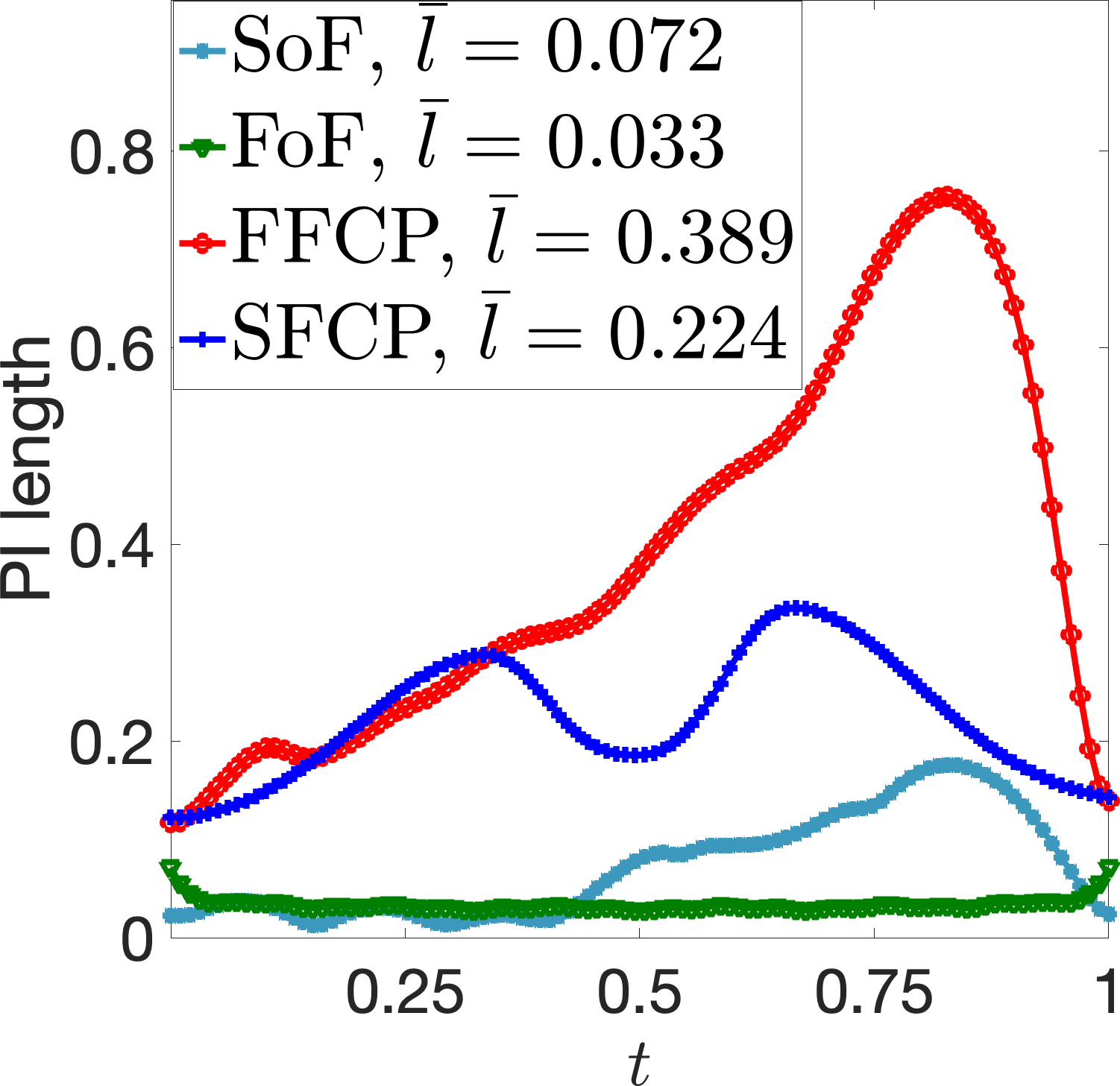} &
    \includegraphics[width = 0.16\textwidth]{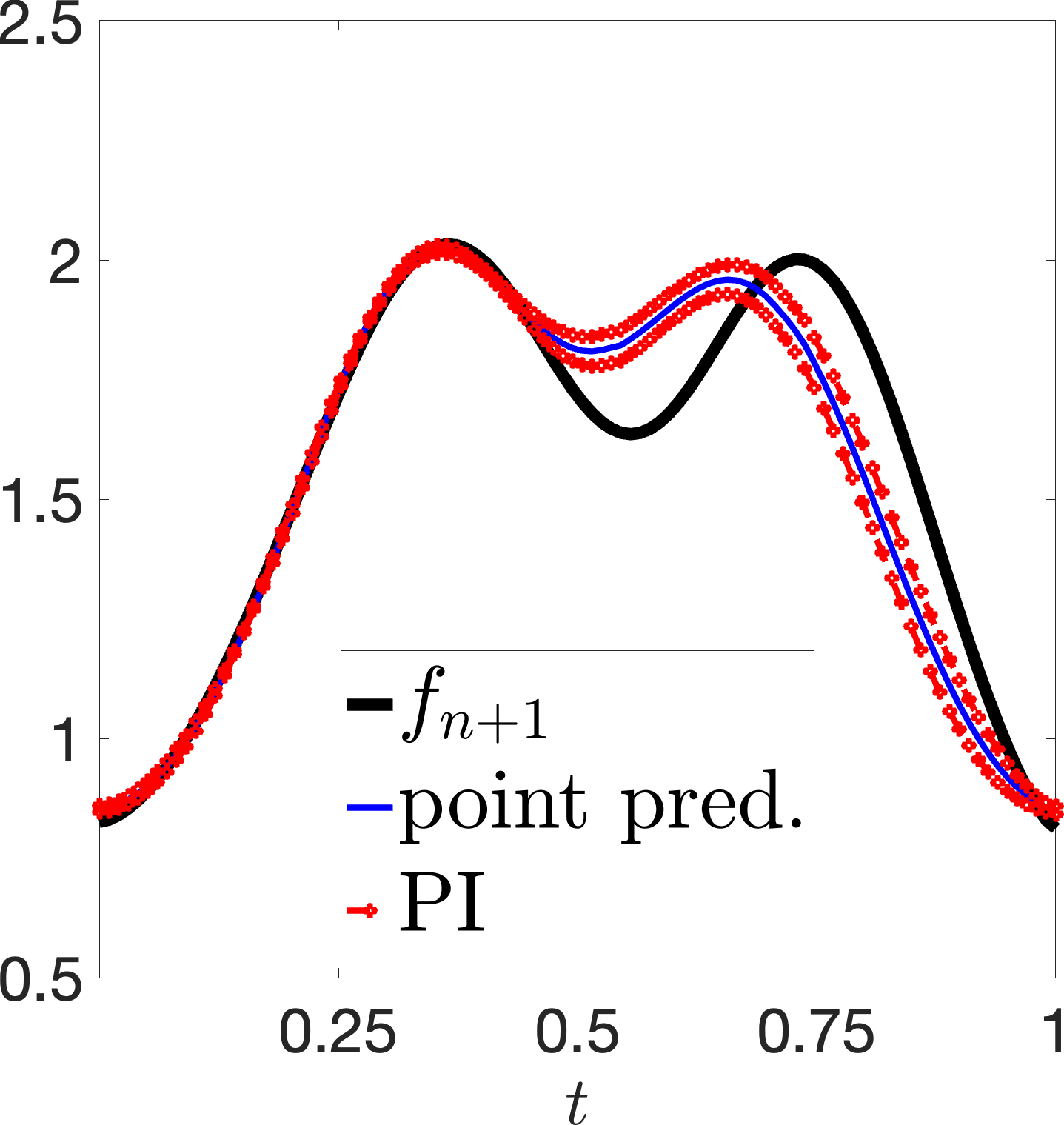} &
    \includegraphics[width = 0.16\textwidth]{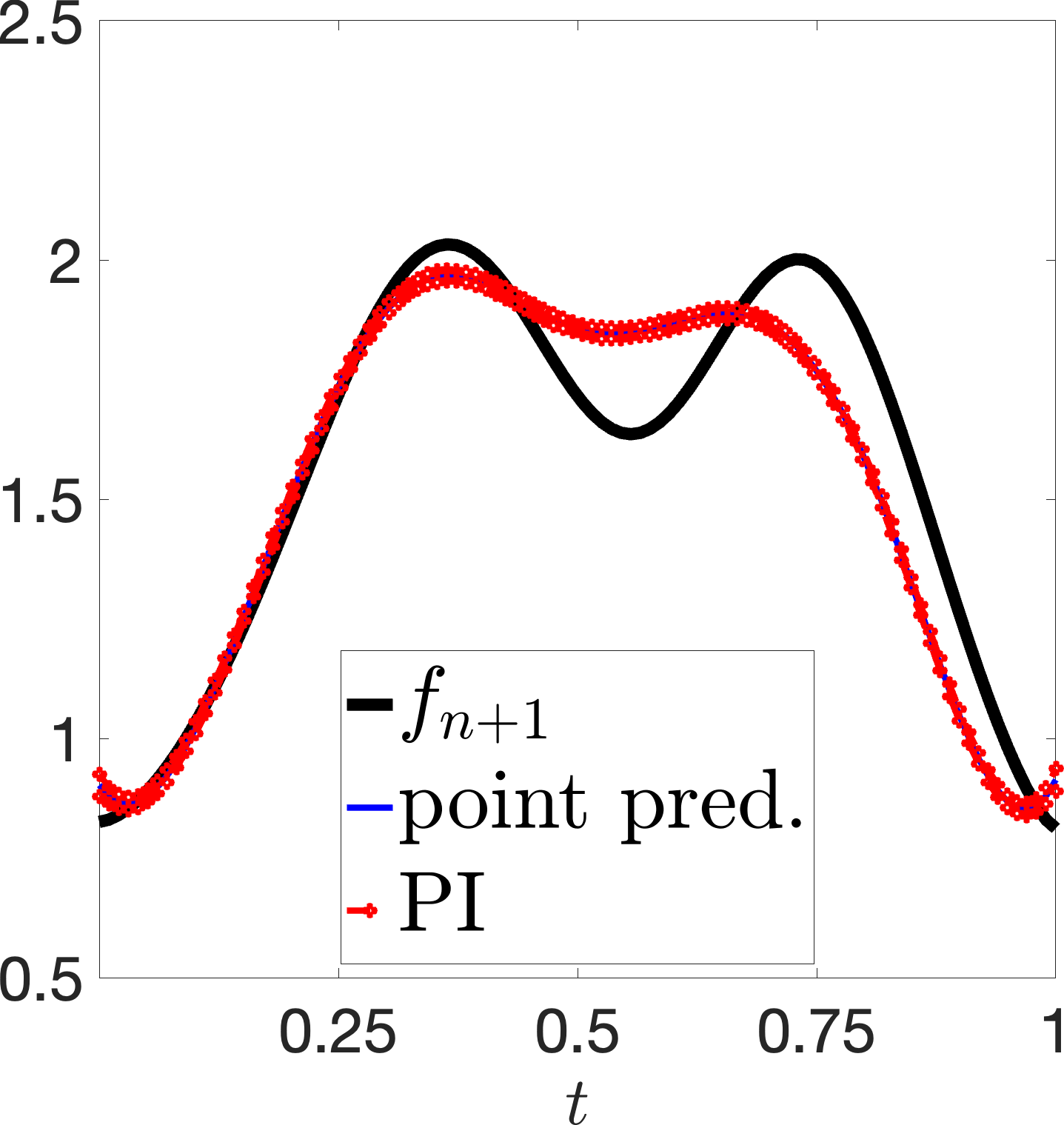} &
    \includegraphics[width = 0.16\textwidth]{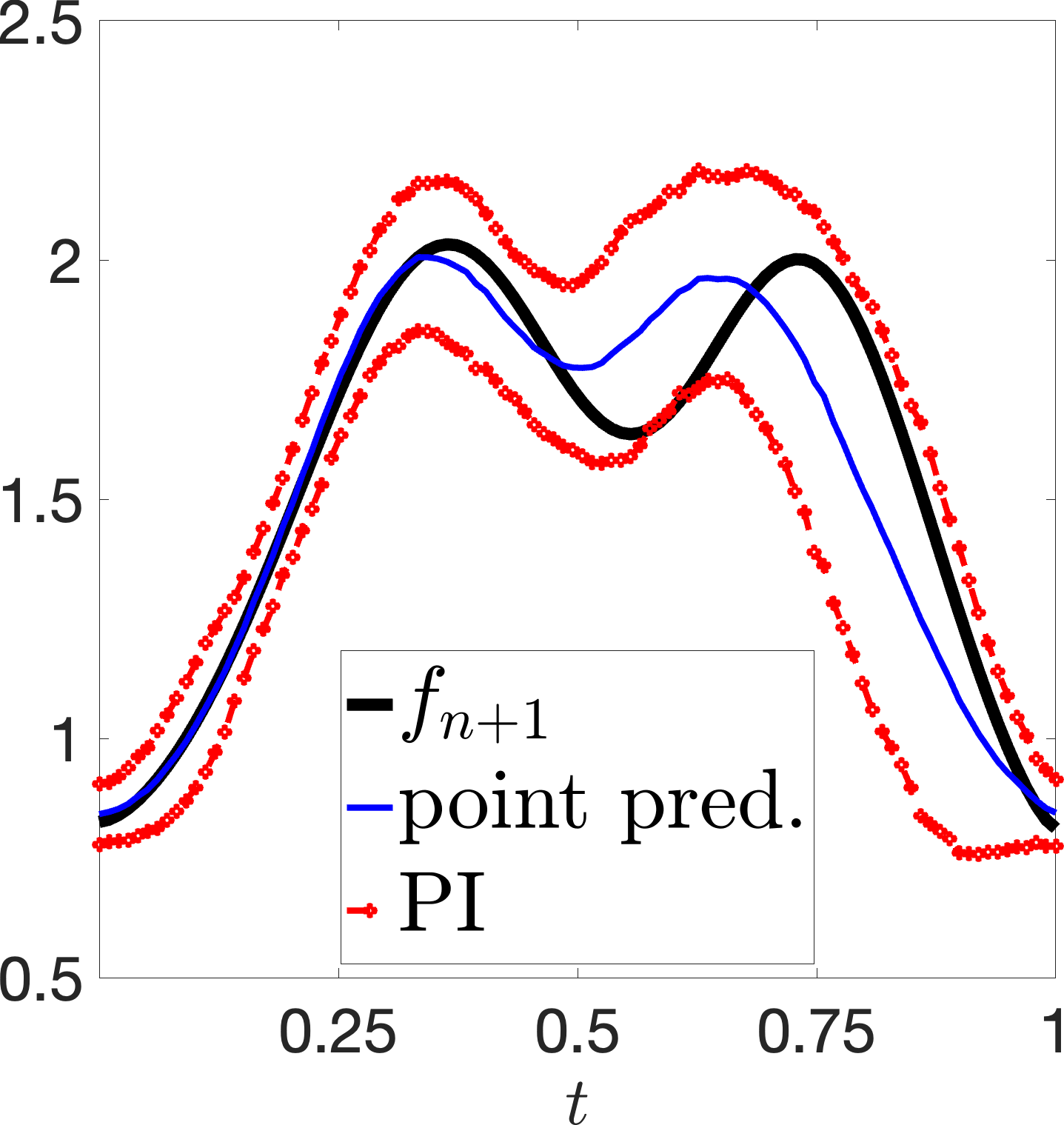} &
    \includegraphics[width = 0.16\textwidth]{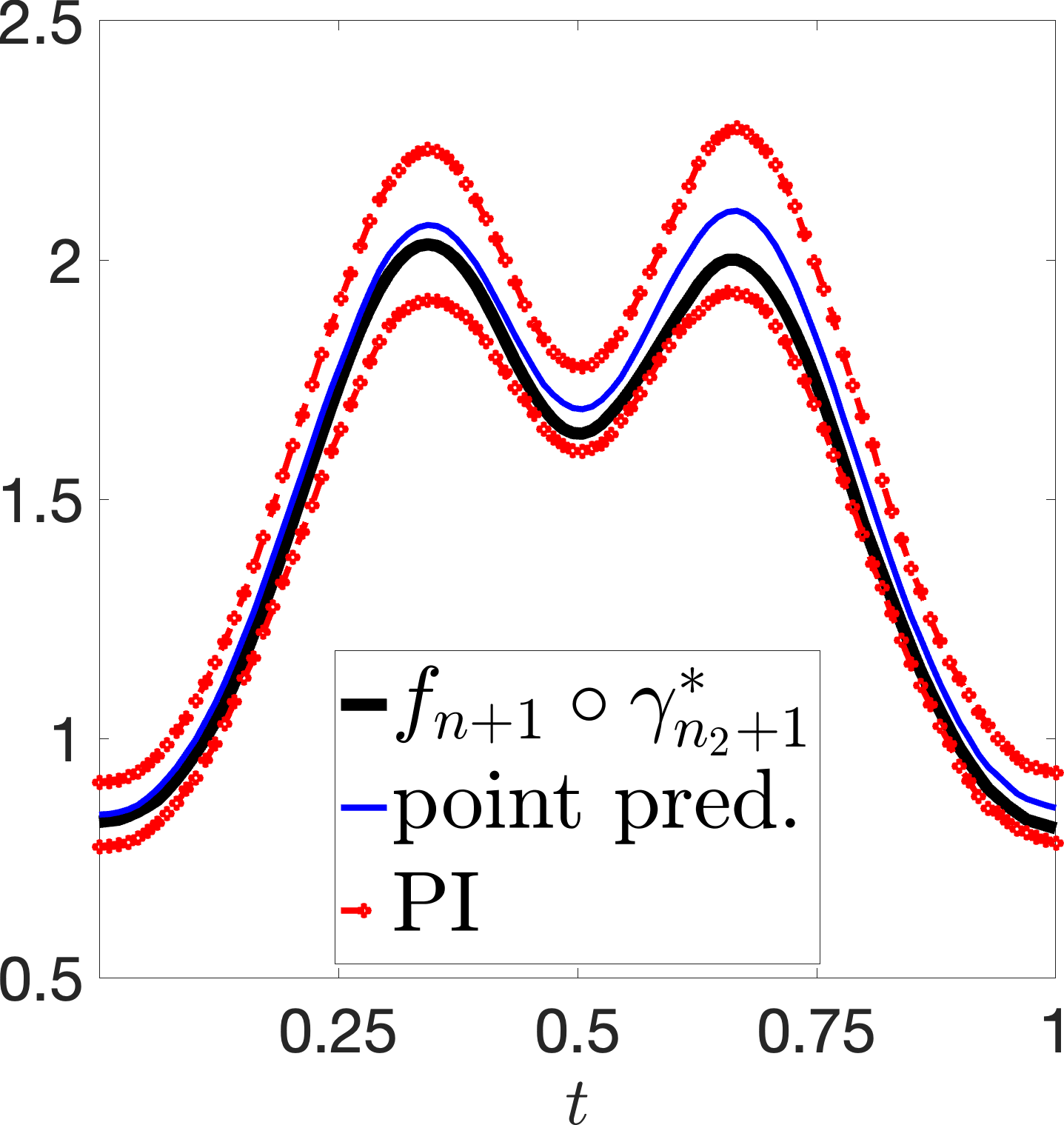}
    \end{tabular}
    \vspace{-10pt}
    \caption{\small Row 1: without phase variation. Row 2: with phase variation. (a) Pointwise coverage rates (solid) with 95\% CIs (shaded regions) and (b) average PI lengths for {\tt SoF} (light blue), {\tt FoF} (green), {\tt FFCP} (red) and {\tt SFCP} (dark blue). (c)-(f) Target function (black), point prediction (blue), pointwise PIs (red) for {\tt SoF}, {\tt FoF}, {\tt FFCP} and {\tt SFCP}, respectively.}
    \label{fig::sim-method-comparison}
    \vspace{-10pt}
\end{figure}

Figure \ref{fig::sim-method-comparison} shows results for functions without (row 1) and with (row 2) phase variation, using a truncation time point $U=0.5$ for $f_{n+1}$. Panels (a) \& (b) show pointwise coverage rates (with 95\% CIs as shaded regions) and average PI lengths, respectively. Panels (c)-(f) show examples of PIs (red) and a point prediction (blue) from {\tt SoF}, {\tt FoF}, {\tt FFCP} and {\tt SFCP}, respectively. For {\tt FFCP} and {\tt SFCP}, the point prediction is taken to be the midpoint of PIs at each time point. Note that the prediction target (black) for the first three methods is $f_{n+1}$, but for $\tt SFCP$ it is the amplitude of $f_{n+1}$. {\tt SoF} and {\tt FoF} generate PIs with very small lengths, but fail to provide valid coverage. Pointwise coverage for {\tt FoF} exhibits periodic spikes, which are related to the location of knots for the B-spline basis used in estimating the coefficient surface $\beta(s,t)$. PIs from {\tt FFCP} and {\tt SFCP} have valid coverage. However, in the presence of phase variation, {\tt SFCP} yields PIs with smaller length than {\tt FFCP}; qualitatively, {\tt SFCP} results in much better pointwise PIs in this case. Table \ref{table::coverage-comparison-overall} reports the overall coverage rate $p$. In absence of phase variation, {\tt SoF} and {\tt FoF} have $p=0.1740$ and $p=0$, respectively, whereas {\tt FFCP} and {\tt SFCP} have $p \approx 0.7$. When phase variation is present in the data, the overall coverage decreases to $0.362$ for {\tt FFCP}, but remains stable for {\tt SFCP} at $p=0.674$. Computationally, {\tt SFCP} is faster than {\tt SoF} and {\tt FFCP}, but slower than {\tt FoF}, when phase variation is not present in the data. When phase variation is present, {\tt SFCP} and {\tt SoF} have comparable speed and are both faster than {\tt FFCP}, but slower than {\tt FoF}.

\setlength{\tabcolsep}{0pt}
\begin{table}[!t]
    \centering
    \captionsetup{width=.9\textwidth}
    \caption{\small Overall coverage rates with standard errors in parentheses.}
    \vspace{-5pt}
    \label{table::coverage-comparison-overall}
    \begin{small}
    \begin{tabular}{|C{3.5cm}|C{3cm}|C{3cm}|C{3cm}|C{3cm}|}
        \hline
        Data &{\tt SoF} & {\tt FoF} & {\tt FFCP} & {\tt SFCP} \\
        \hline
        No phase variation & 0.1740 (0.0170) & 0 (0) & 0.6900 (0.0207) & 0.6800 (0.0209)\\
        \hline
        Phase variation & 0 (0) & 0 (0) & 0.3620 (0.0215) & 0.6740 (0.0210)\\
        \hline
    \end{tabular}
    \end{small}
    \vspace{-10pt}
\end{table}

\setlength{\tabcolsep}{0pt}
\begin{figure}[!t]
    \centering
    \begin{tabular}{cccc}
    (a)&(b)&(c)&(d)\\
        \includegraphics[width = 0.18\textwidth]{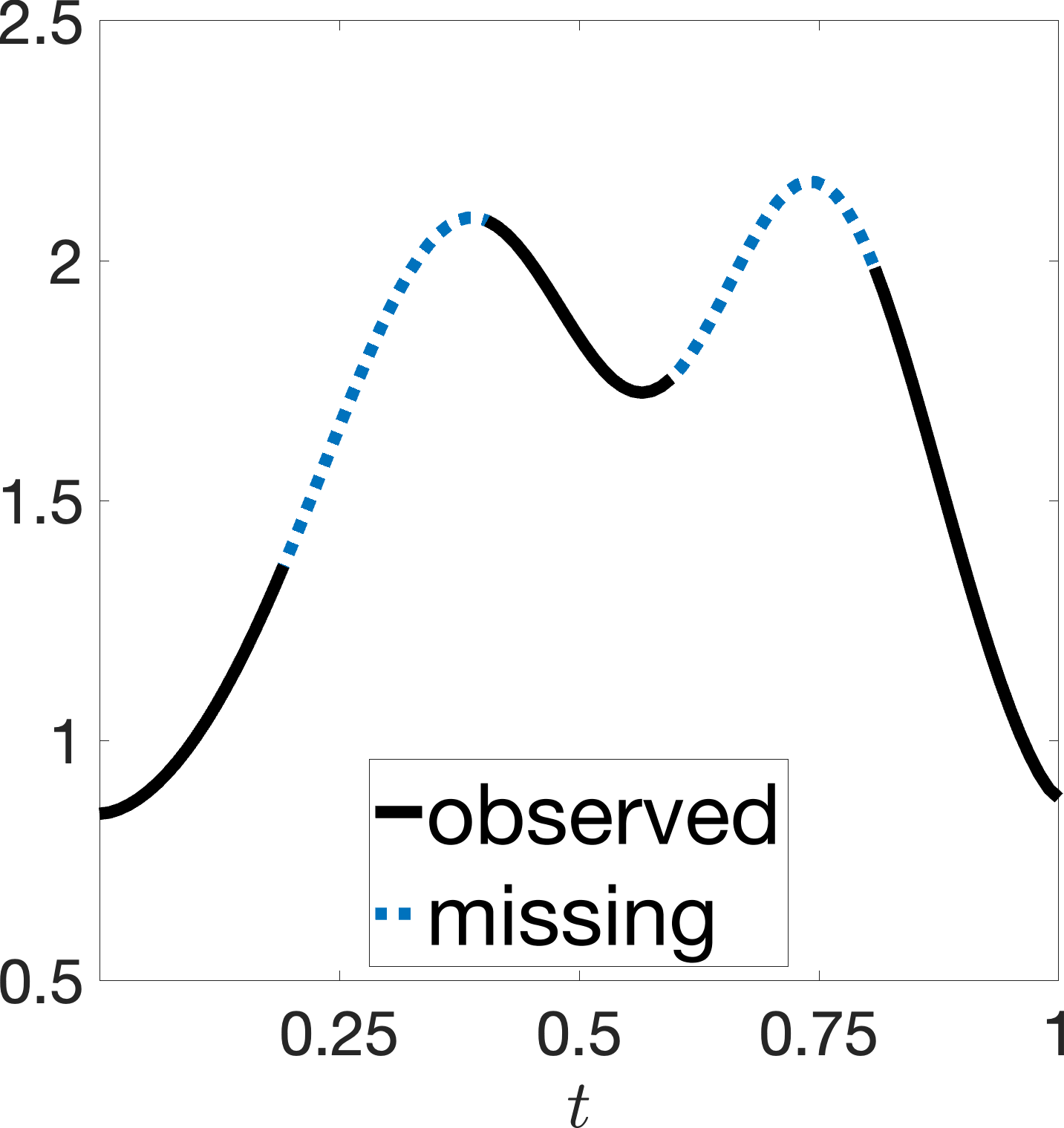} &
        \includegraphics[width = 0.2\textwidth]{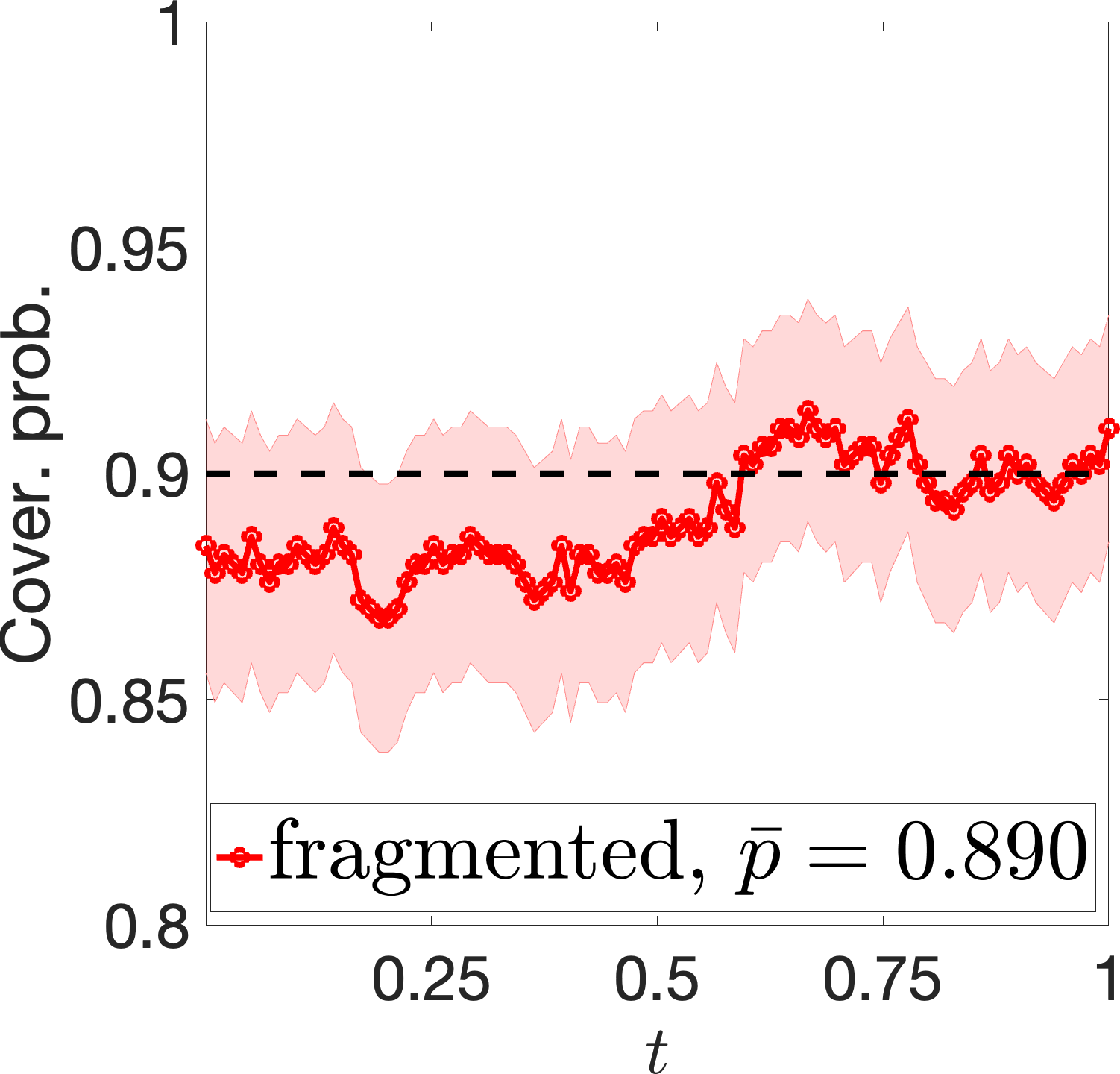} &
        \includegraphics[width = 0.2\textwidth]{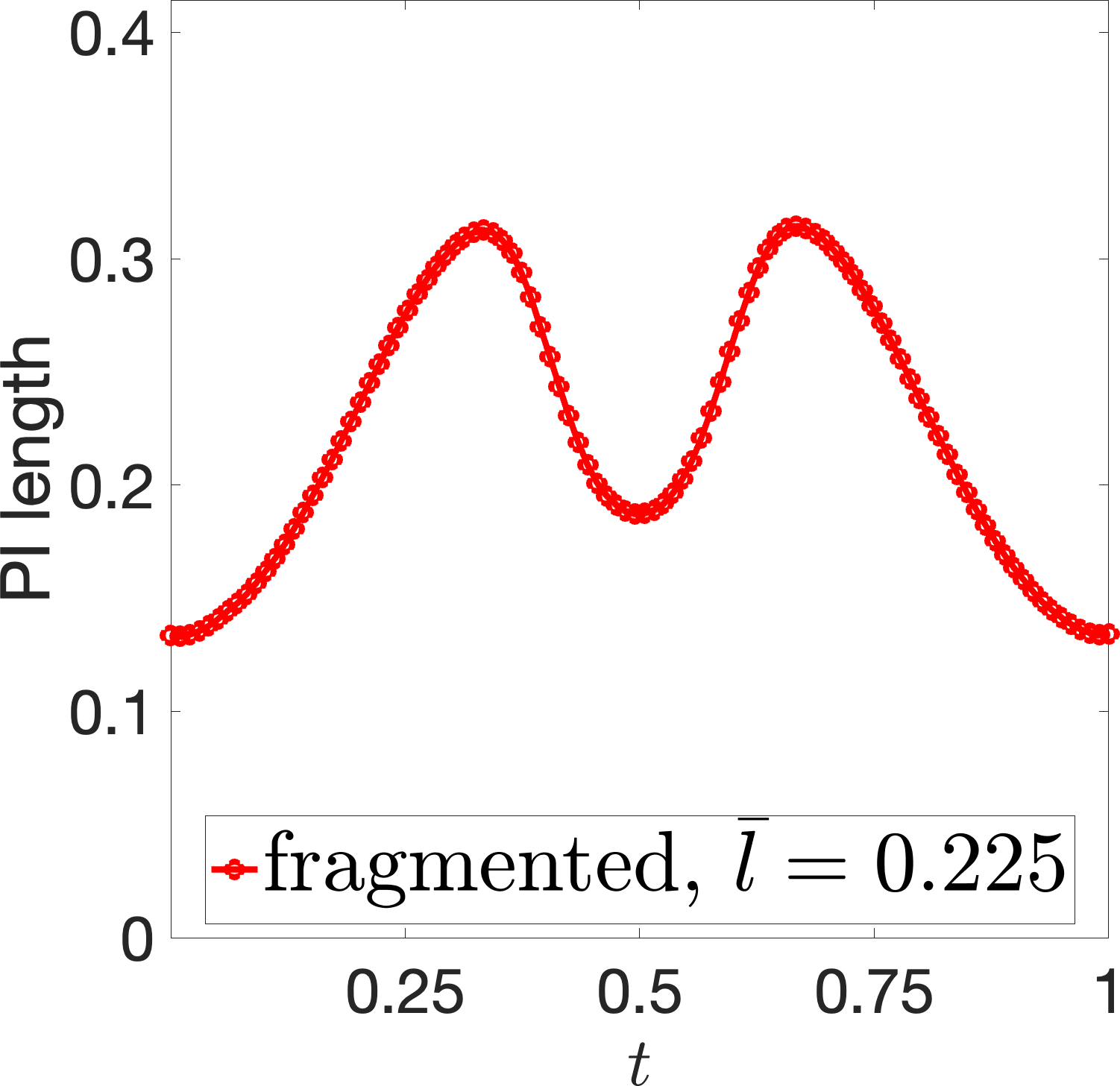} &
        \includegraphics[width = 0.18\textwidth]{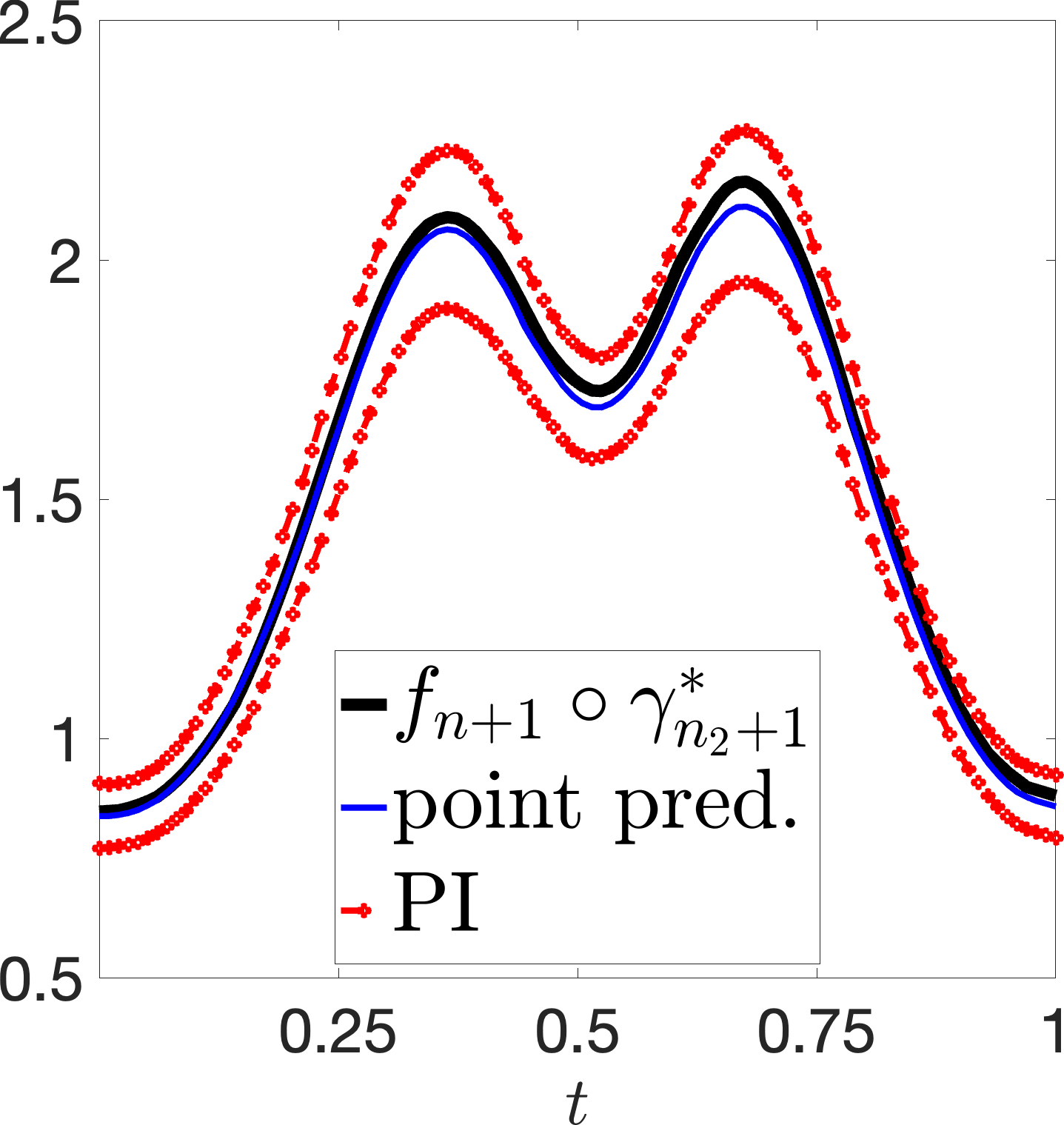}
        \\
        \includegraphics[width = 0.18\textwidth]{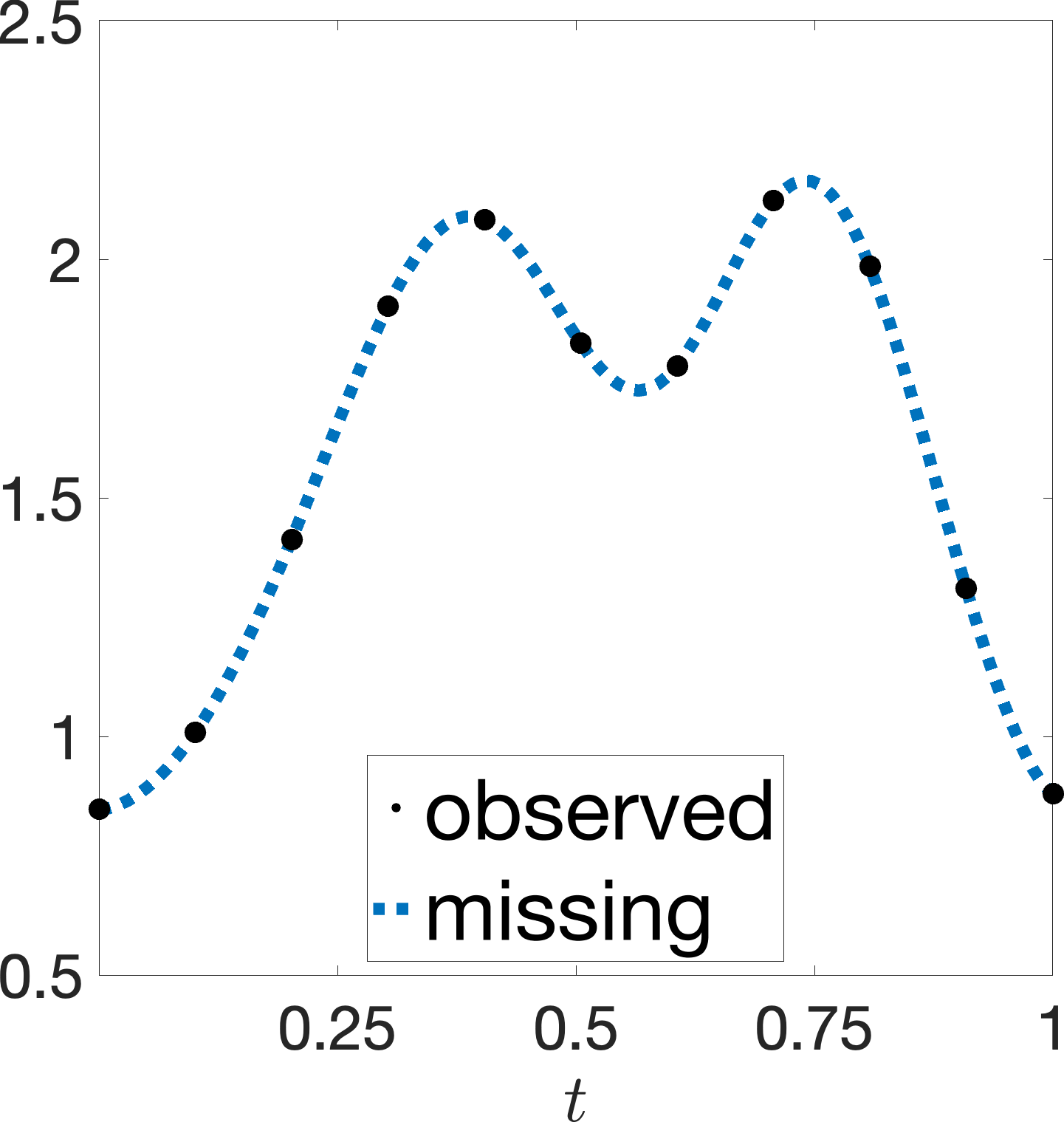} &
        \includegraphics[width = 0.2\textwidth]{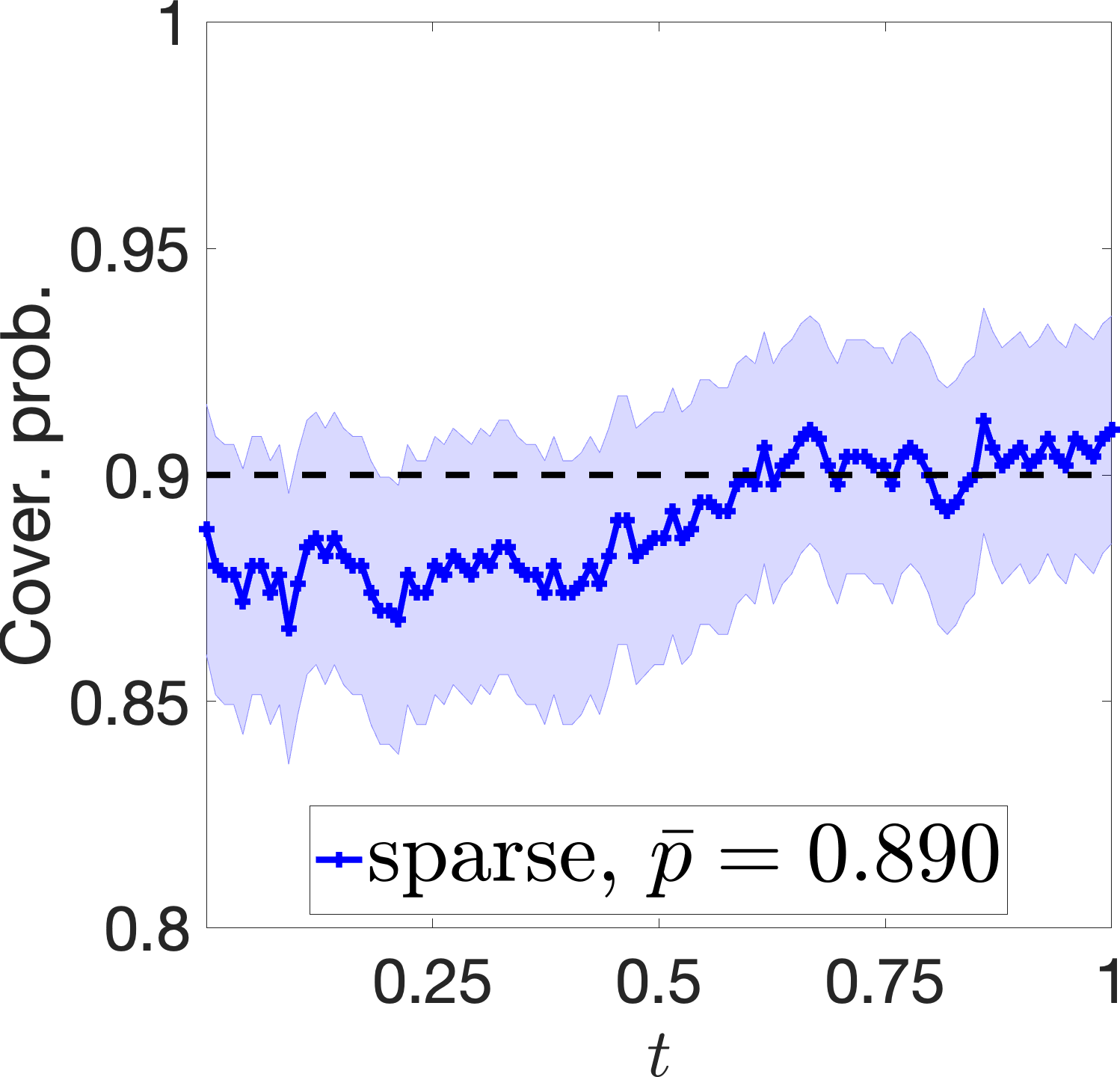} &
        \includegraphics[width = 0.2\textwidth]{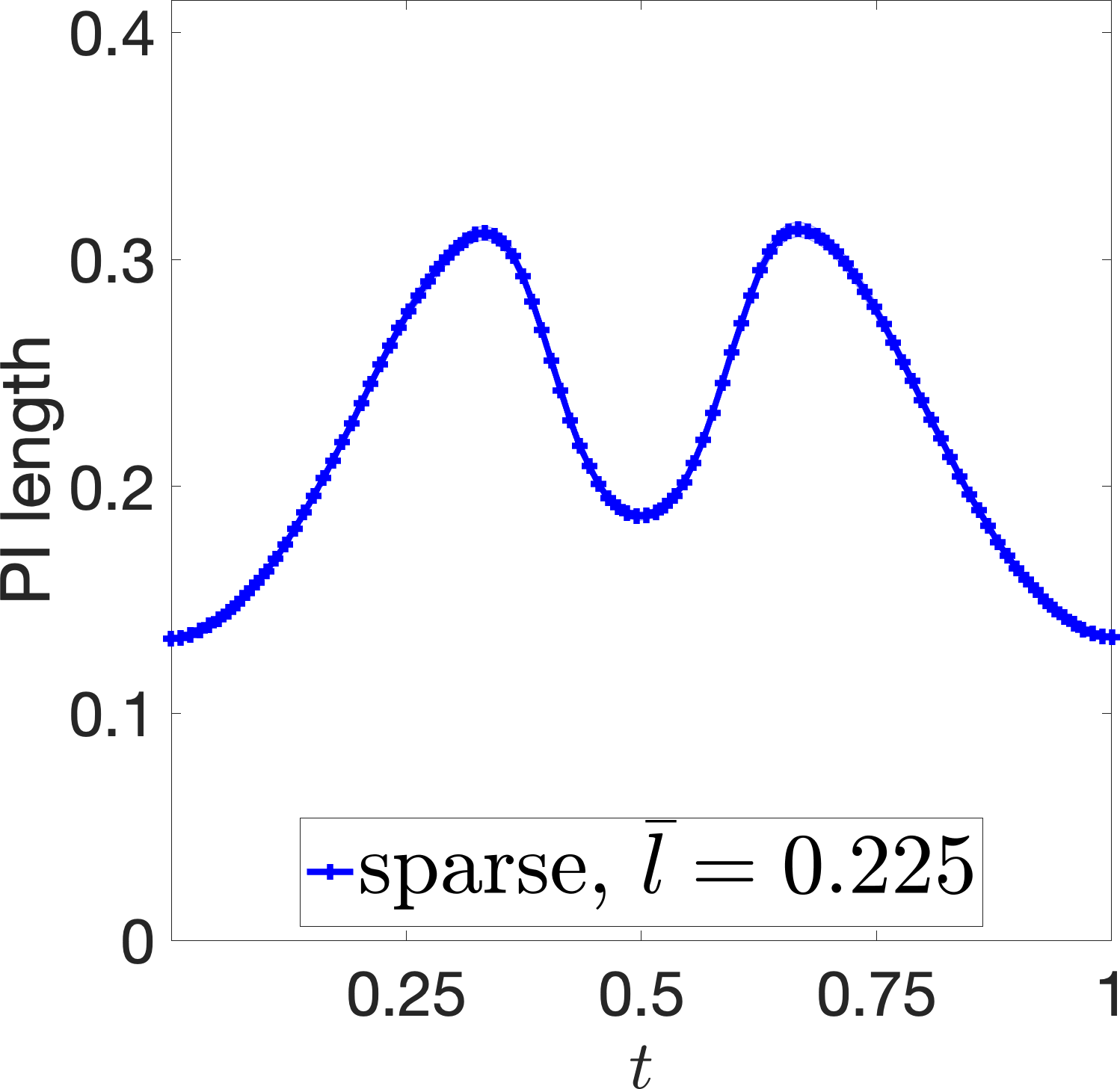} &
        \includegraphics[width = 0.18\textwidth]{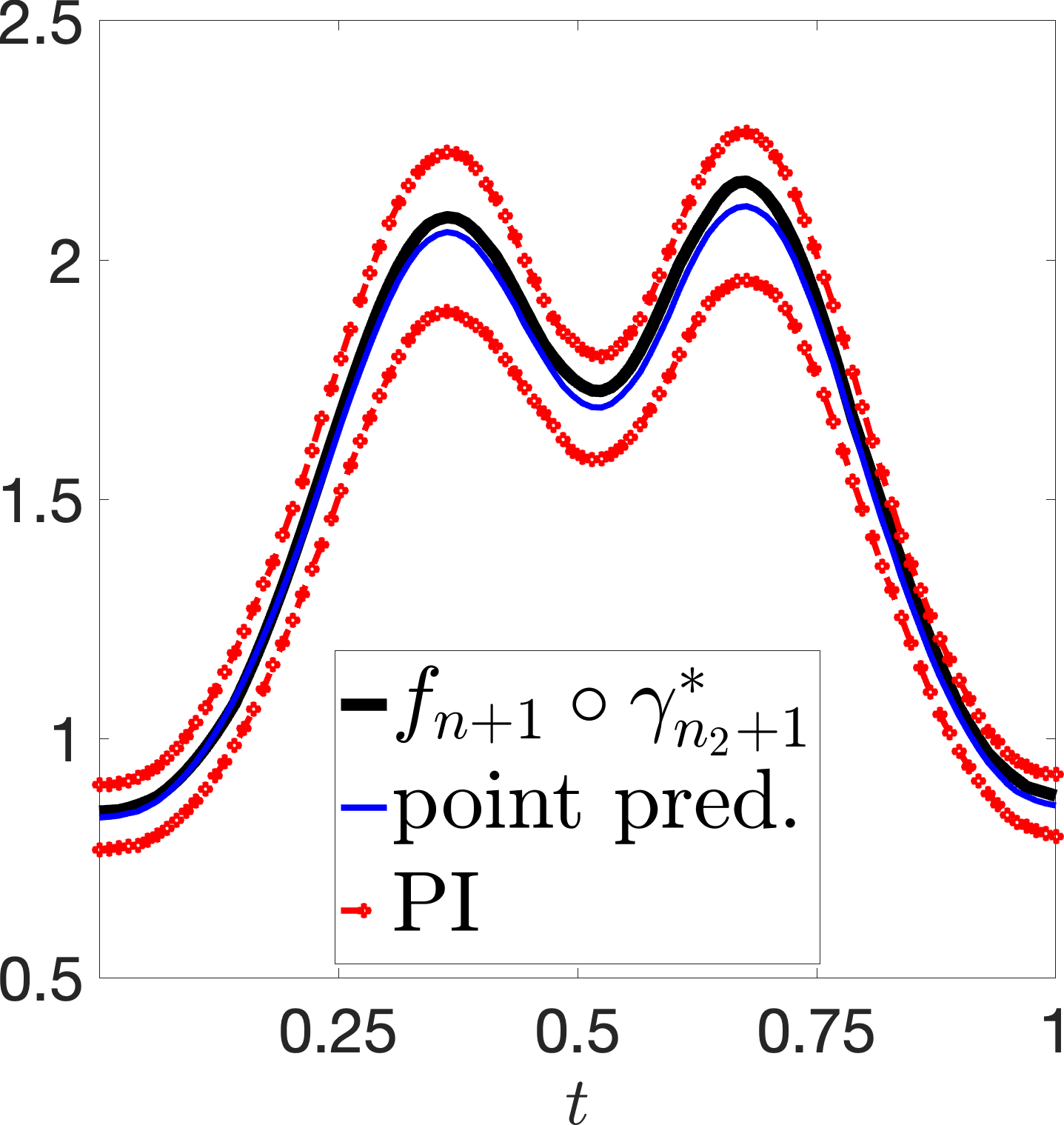}
        \end{tabular}
    \vspace{-10pt}
    \caption{\small Row 1: fragmented. Row 2: sparse. (a) Observed (black) and missing (blue) segments for $f_{n+1}$. (b) \& (c) Coverage rates (solid) with 95\% CIs (shaded regions) and average PI lengths, respectively. (d) Ground truth $f_{n+1} \circ \gamma^*_{n+1}$ (black), point prediction (blue) and pointwise PIs (red).}
    \label{fig::other-obs-pattern}
    \vspace{-10pt}
\end{figure}
\noindent\textbf{Simulation 2: other observational regimes.} We evaluate the performance of {\tt SFCP} for fragmented and sparse observations. For fragmented data, we simulate ${\cal J}_1 = [0,0.2]$, ${\cal J}_2 = [0.4,0.6]$, ${\cal J}_3 = [0.8,1]$, and use $\lambda_j = 1/3$ in the distance $d_{\rm prod}(\cdot,\cdot)$. For sparse data, we simulate ${\cal J} = \{0,0.1,\dots,0.9,1\}$. Here, we use data with phase variation. Row 1 (row 2) in Figure \ref{fig::other-obs-pattern} shows results for the fragmented (sparse) case. Panel (a) shows observed (black) and missing (blue) segments of $f_{n+1}$. Panels (b) \& (c) show pointwise coverage rates (with 95\% CIs as shaded regions) and average PI lengths, respectively. Panel (d) shows the resulting PIs (red) with a point prediction (blue) for the amplitude of $f_{n+1}$ (black). {\tt SFCP} yields PIs that preserve coverage validity in both observational regimes, with very similar PI lengths.

\begin{figure}[!t]
    \centering
    \begin{tabular}{ccc}
    (a)&(b)&(c)\\
        \includegraphics[width = 0.2\textwidth]{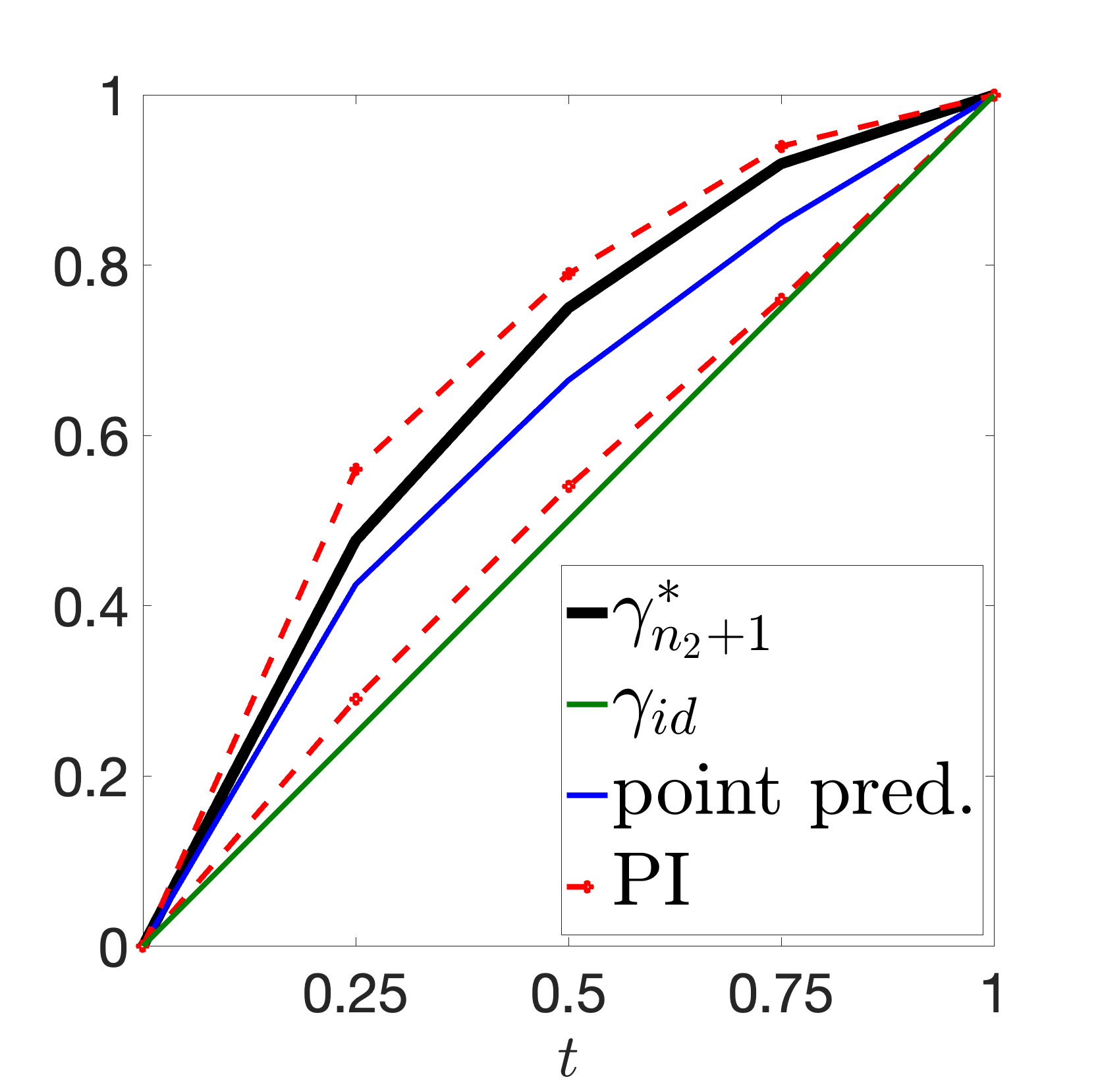} &
        \includegraphics[width = 0.2\textwidth]{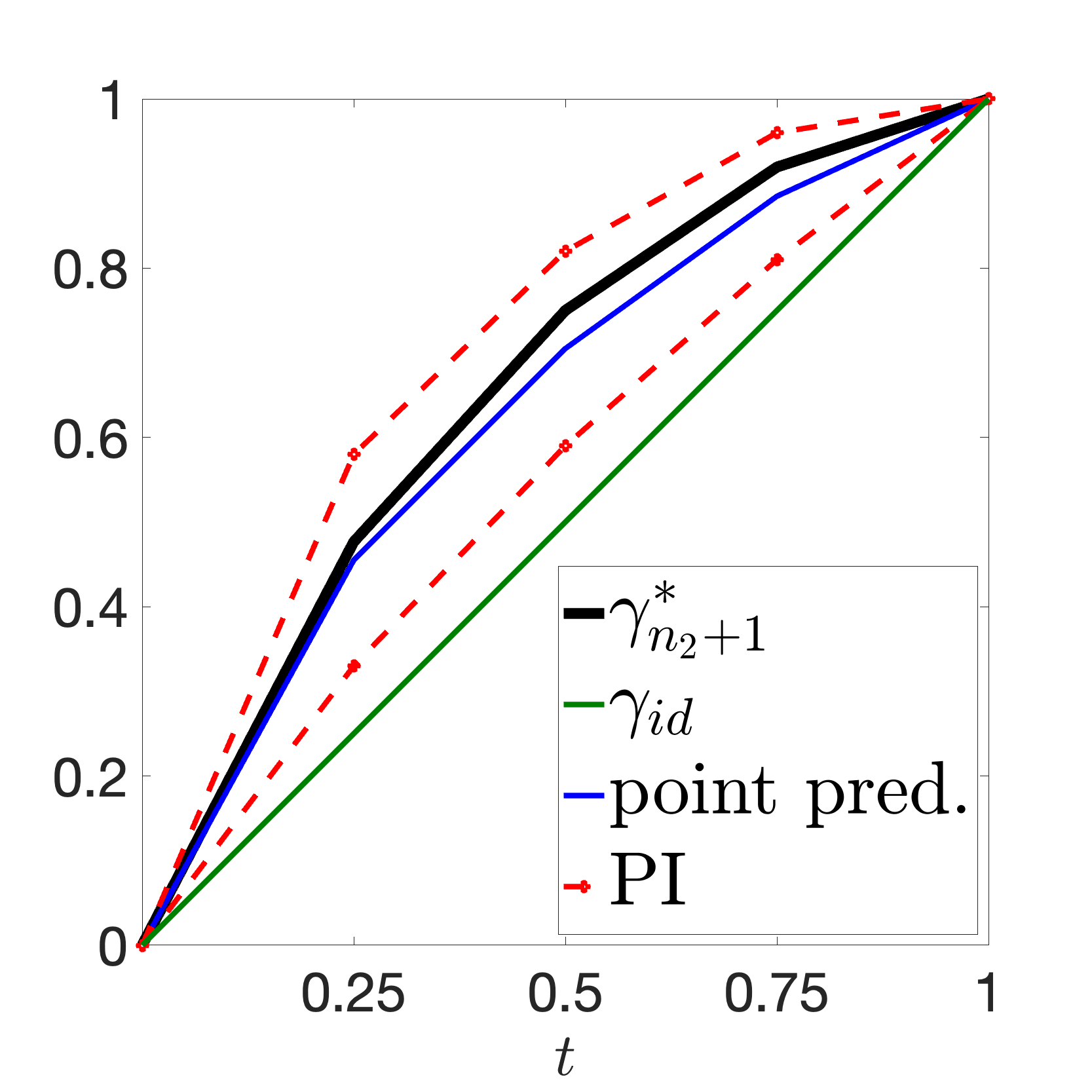} &
        \includegraphics[width = 0.2\textwidth]{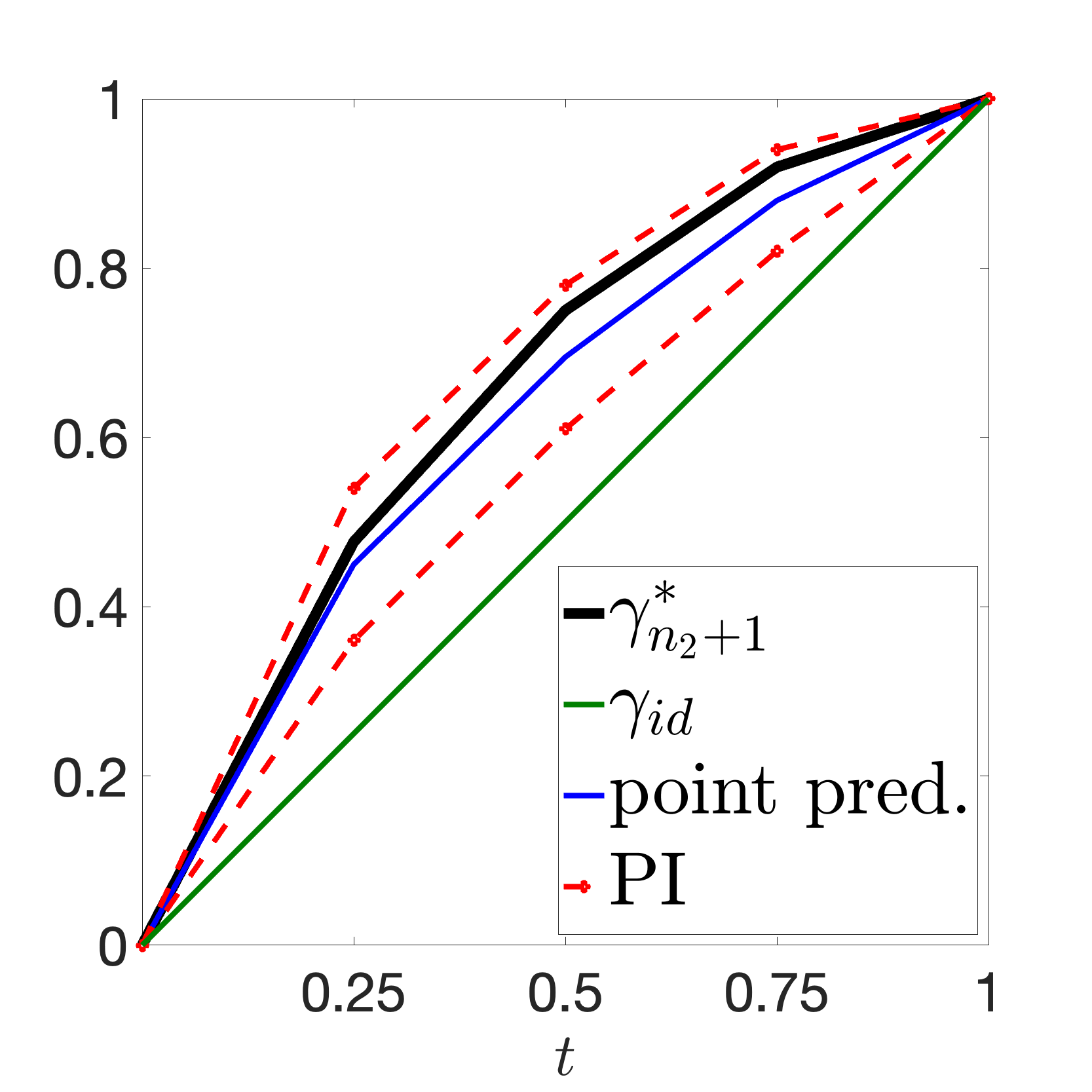}
        \end{tabular}
    \vspace{-10pt}
    \caption{\small Ground truth $\gamma_{n_2+1}^*$ (black), point prediction (blue), PIs (red) and identity warping (green). (a)-(c) Truncation time points $U = 0.25,\ 0.5$ and $0.75$, respectively.}
    \vspace{-10pt}
    \label{fig::predict-relative-phase}
\end{figure}

\noindent\textbf{Simulation 3: prediction of relative phase.} We apply {\tt SFCPP} to predict the relative phase component $\gamma_{n_2+1}^*$ using data with phase variation. Since {\tt SFCPP} predicts $\gamma_{n_2+1}^*$ for all time points simultaneously, we use global tuning for the bandwidth parameter in neighborhood smoothing. 
We use a coarse grid of time points for prediction, ${\cal T} := \{0,0.25,0.5,0.75,1\}$, and set $\alpha=0.1$ as before. Figure \ref{fig::predict-relative-phase} shows the prediction results for a randomly selected Monte Carlo replicate; panels (a)-(c) correspond to truncation of $f_{n+1}$ at $U=0.25,\ 0.5$ and $0.75$, respectively. In all cases, {\tt SFCPP} provides a reasonable point prediction (blue) and informative PIs (red). The time averaged PI lengths (average computed over the coarse time grid ${\cal T}$) for (a)-(c) are $0.233,\ 0.210$ and $0.157$, respectively. Overall, the PIs become narrower as we observe more of $f_{n+1}$ matching intuition. Across all Monte Carlo samples, for $U=0.25,\ 0.5$ and $0.75$, the coverage rates and associated $95\%$ CIs are $0.868\ (0.838,0.898)$, $0.878\ (0.849,0.907)$ and $0.894\ (0.867, 0.921)$, the average time averaged PI lengths are 0.199, 0.169 and 0.168, and the average computational costs (in seconds) are $47.73, 47.23$ and $47.33$, respectively. Thus, the proposed method has valid coverage. Computing time heavily depends on the number of trial values used at each time point; here, we used 100.

\setlength{\tabcolsep}{0pt}
\vspace{-0.3in}
\section{Data examples}
\vspace{-0.1in}
\label{sec::real-data}

\begin{figure}[!t]
    \centering
    \begin{tabular}{ccccc}
    (a)&(b)&(c)&(d)&(e)\\
        \includegraphics[width = 0.16\textwidth]{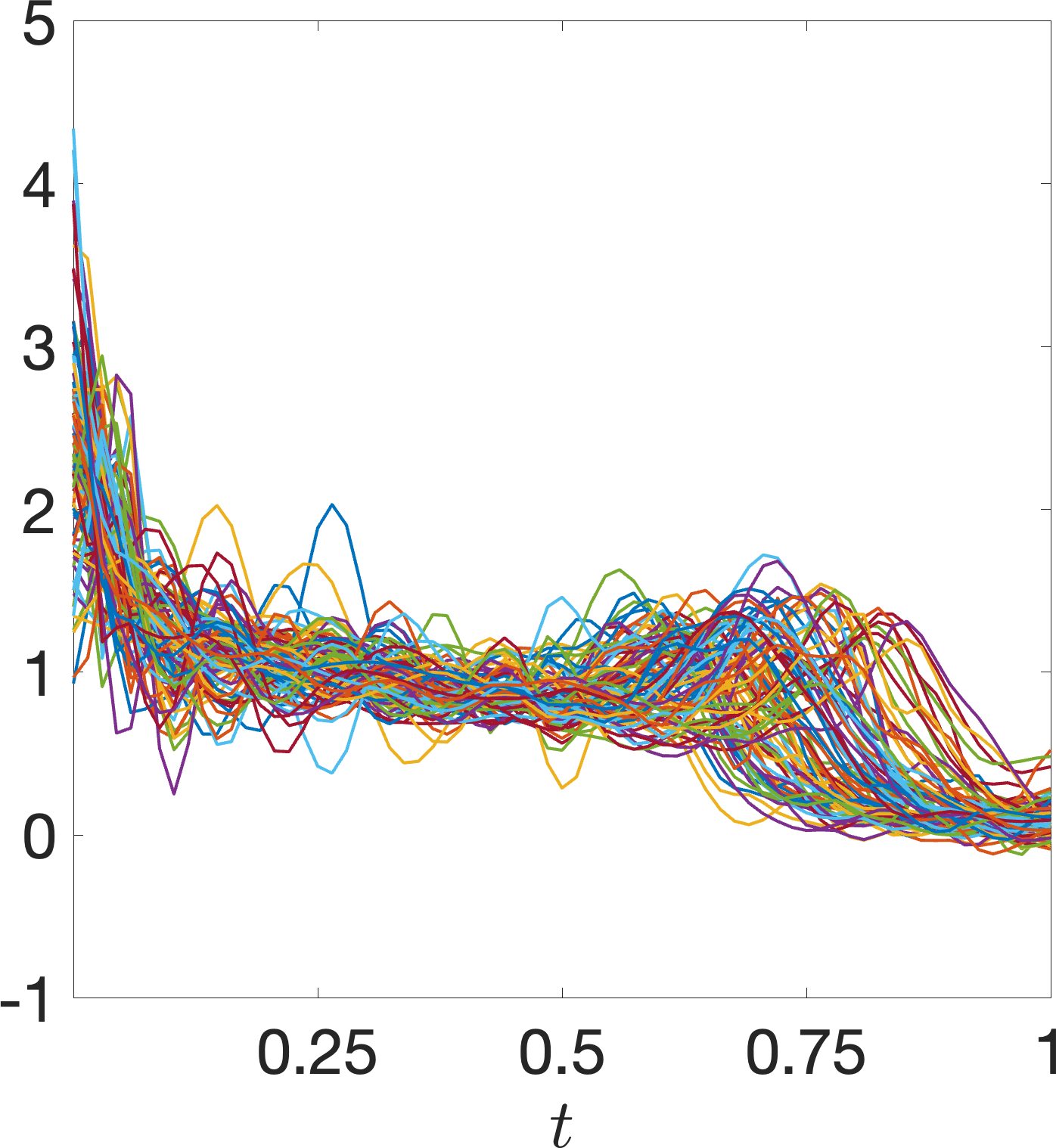} &
        \includegraphics[width = 0.16\textwidth]{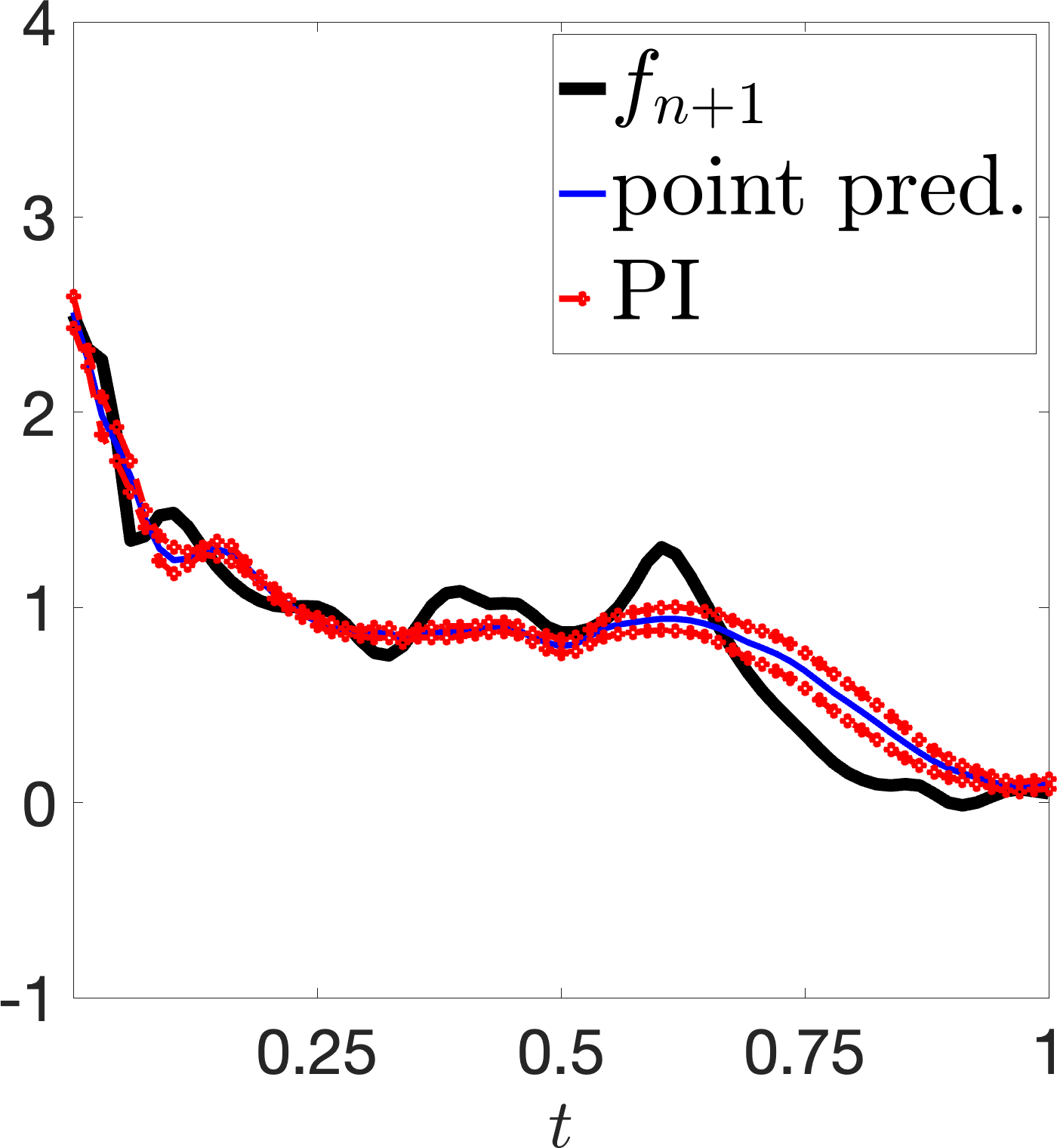} &
        \includegraphics[width = 0.16\textwidth]{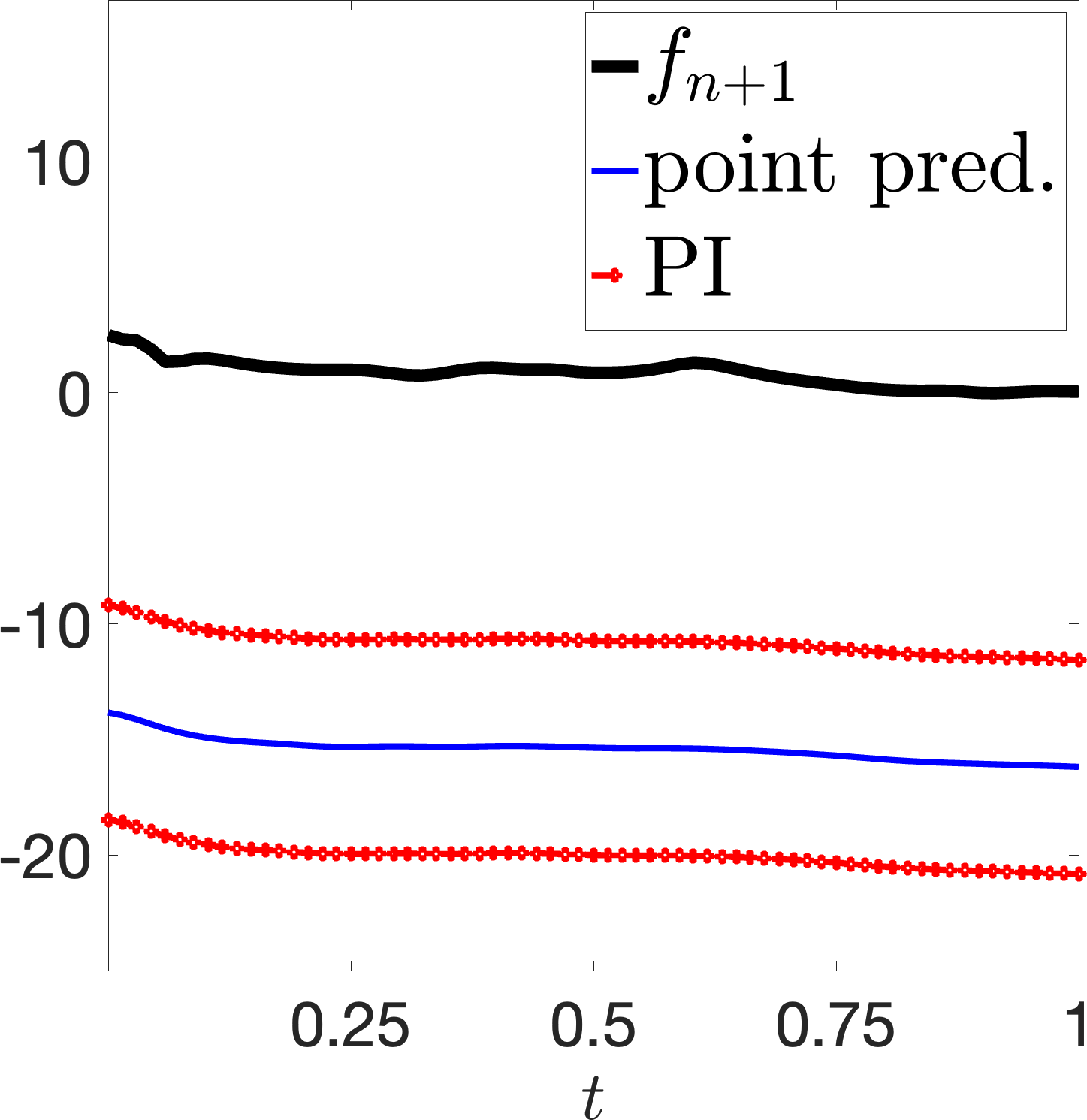} &
        \includegraphics[width = 0.16\textwidth]{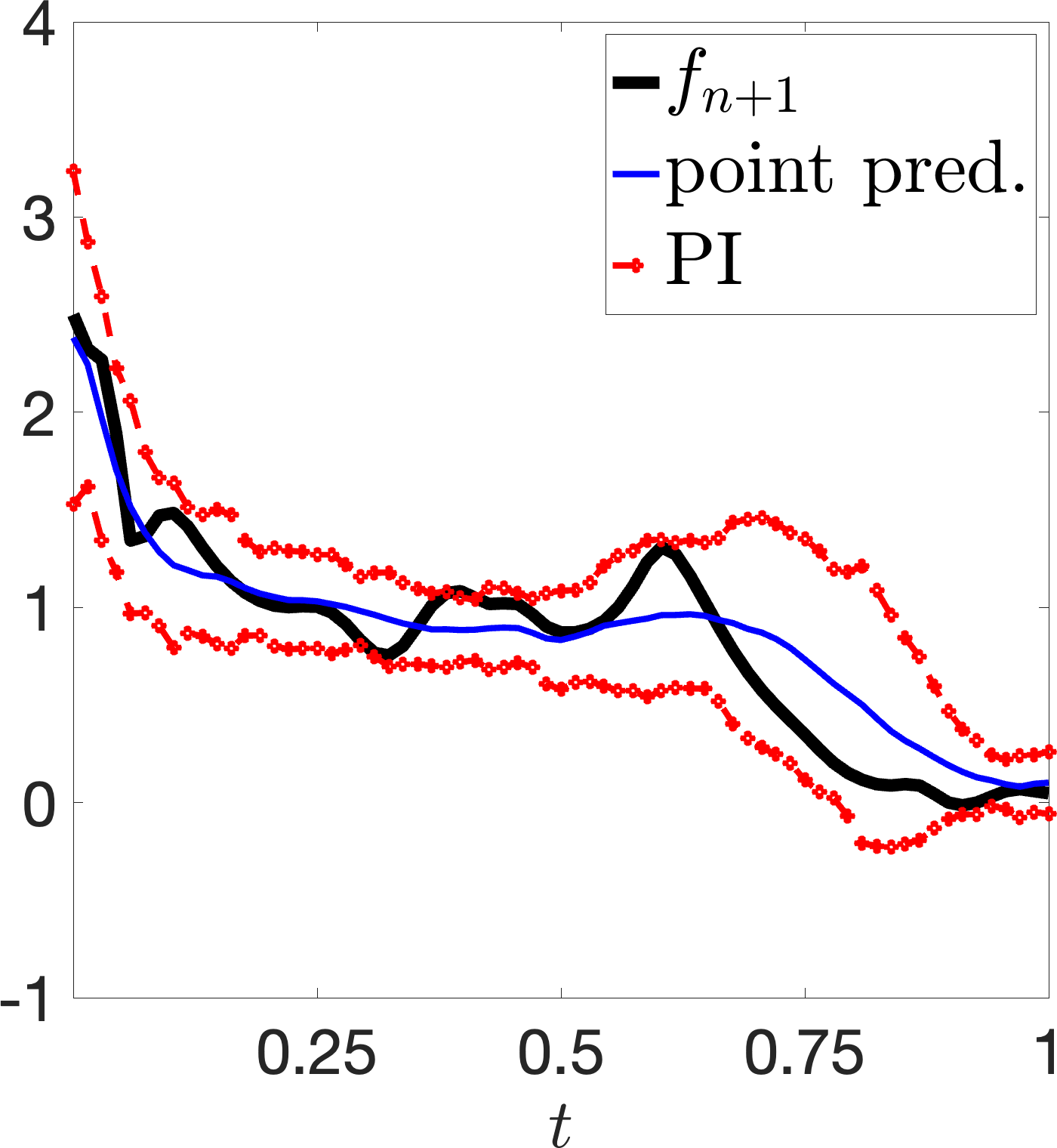} &
        \includegraphics[width = 0.16\textwidth]{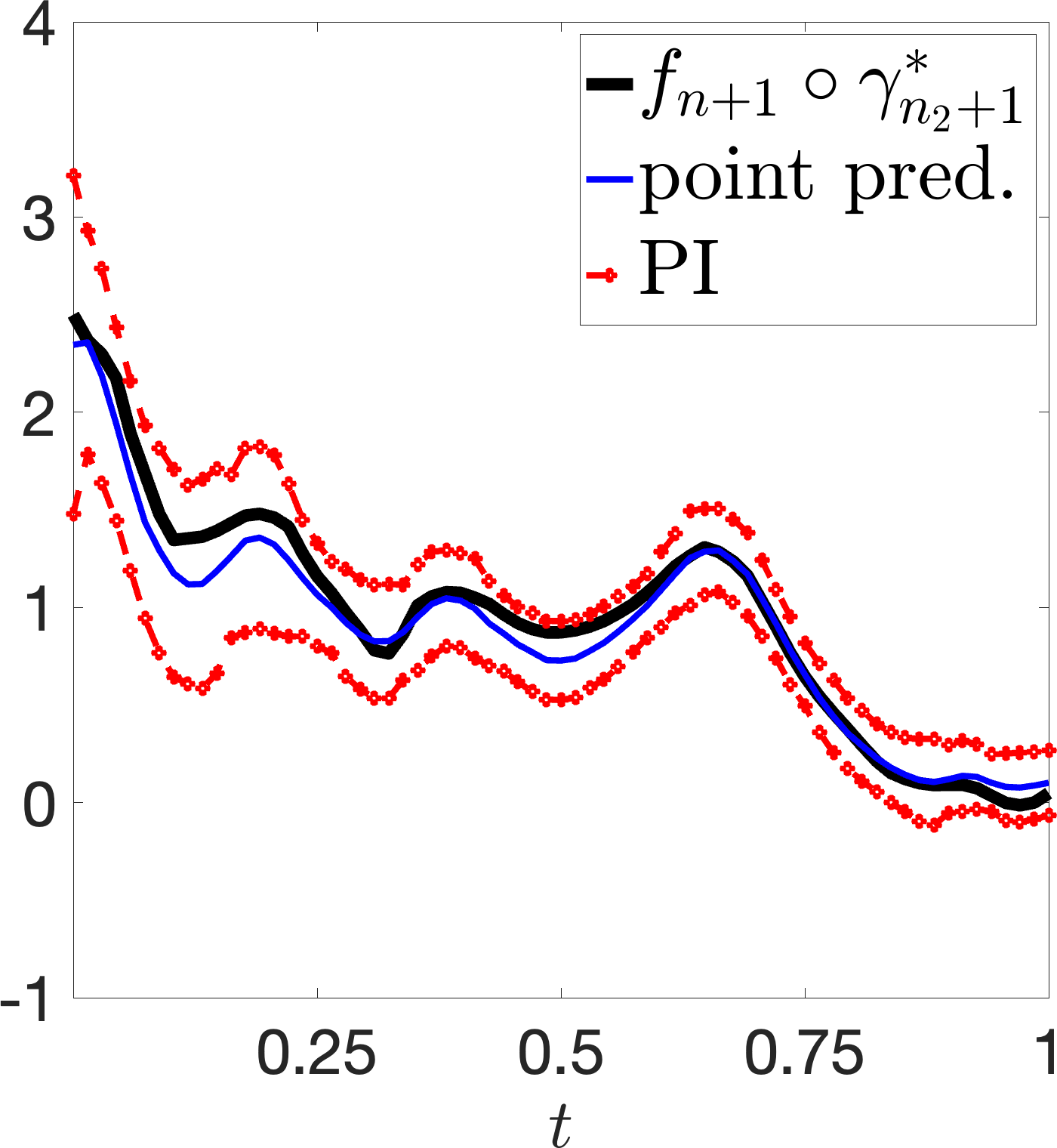}
        \\
        \includegraphics[width = 0.16\textwidth]{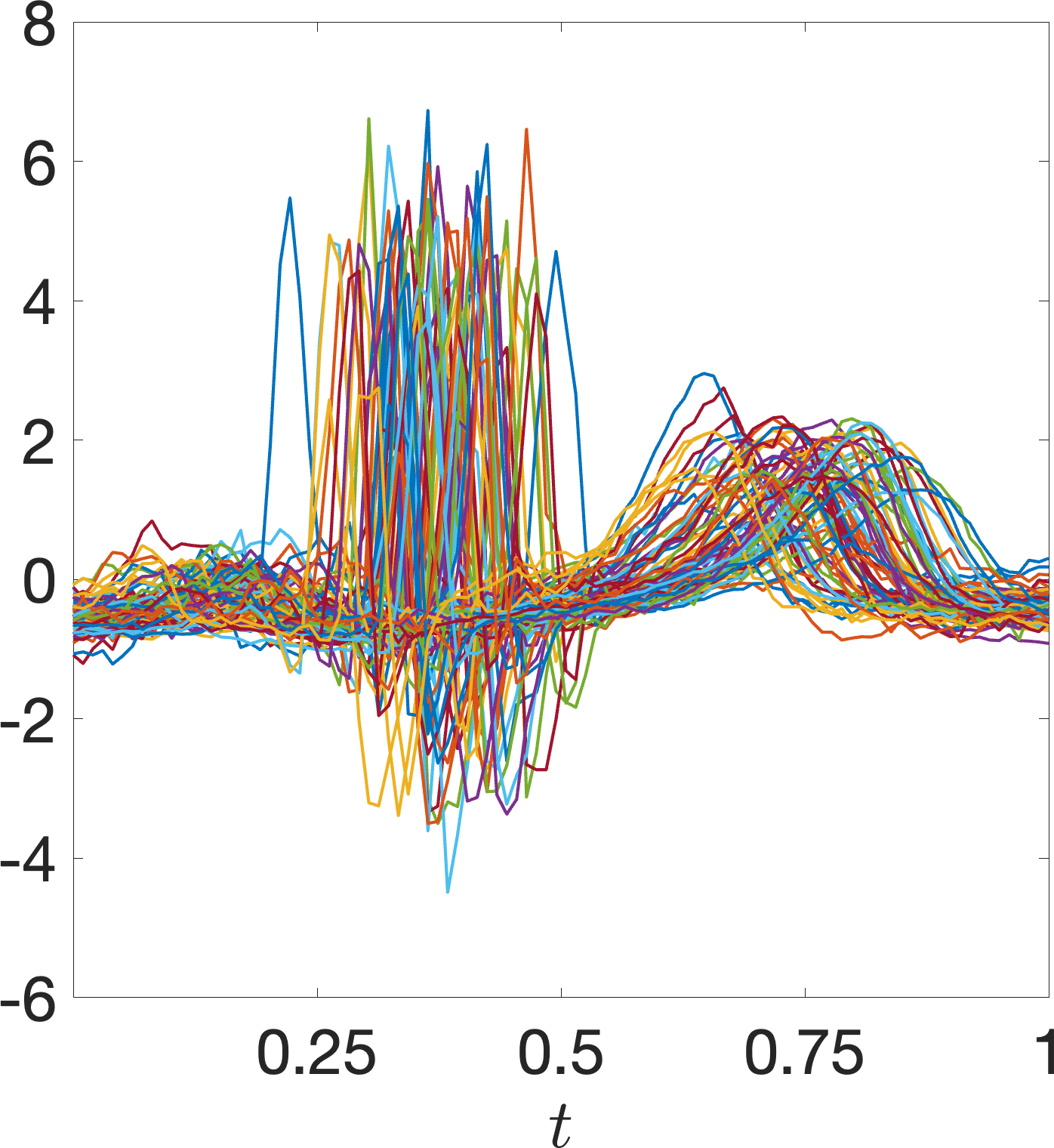} &
        \includegraphics[width = 0.16\textwidth]{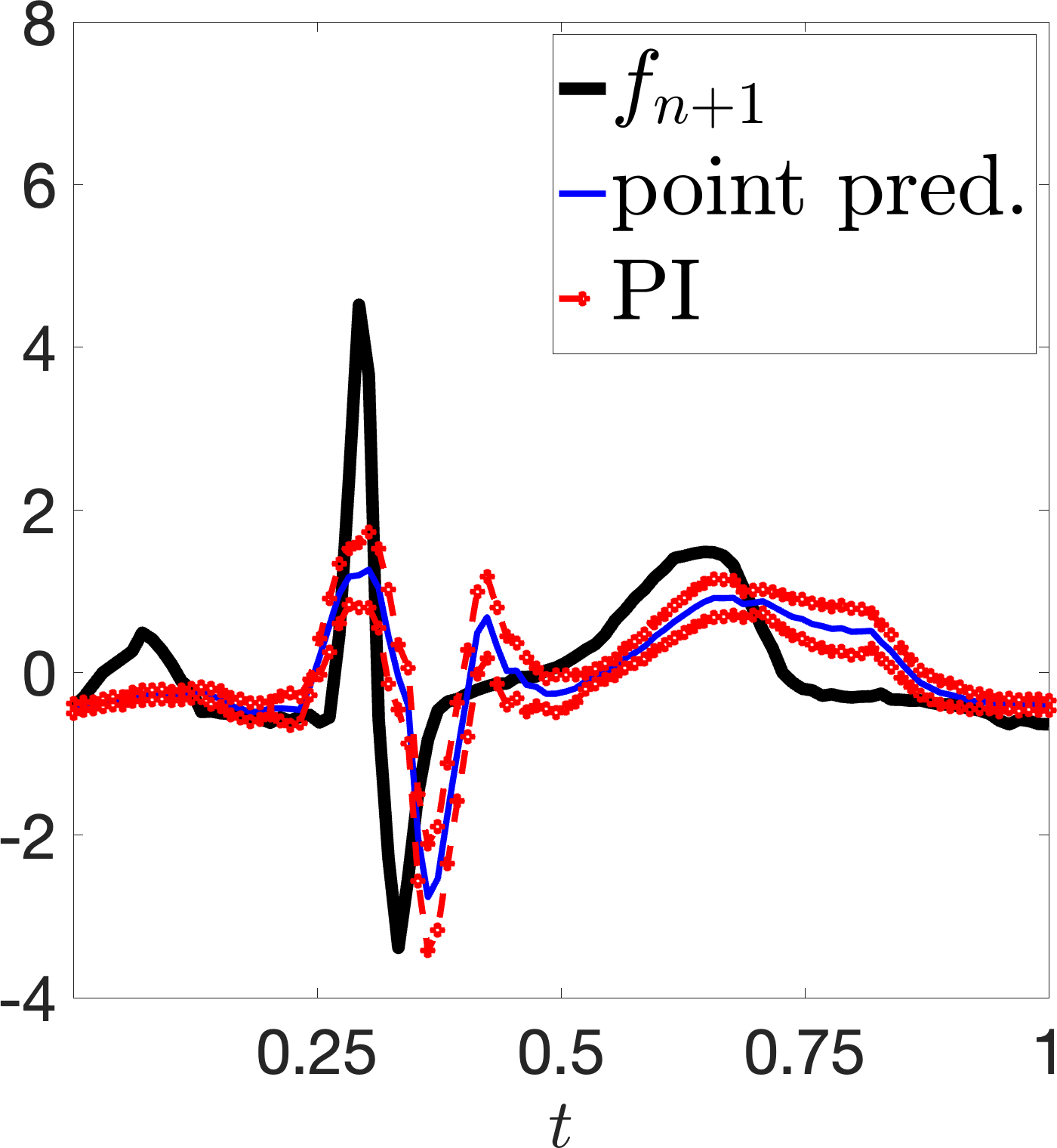} &
        \includegraphics[width = 0.16\textwidth]{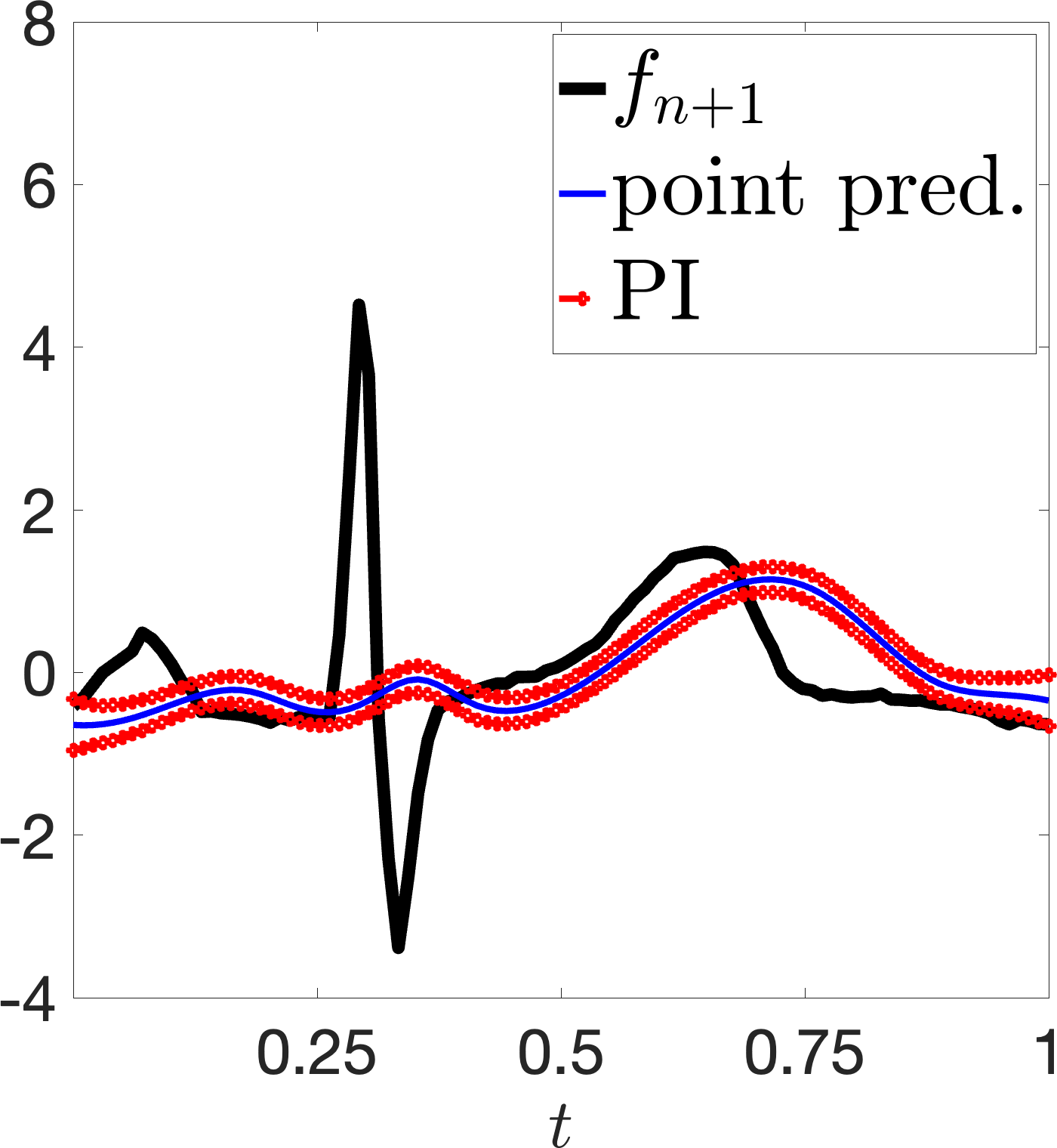} &
        \includegraphics[width = 0.16\textwidth]{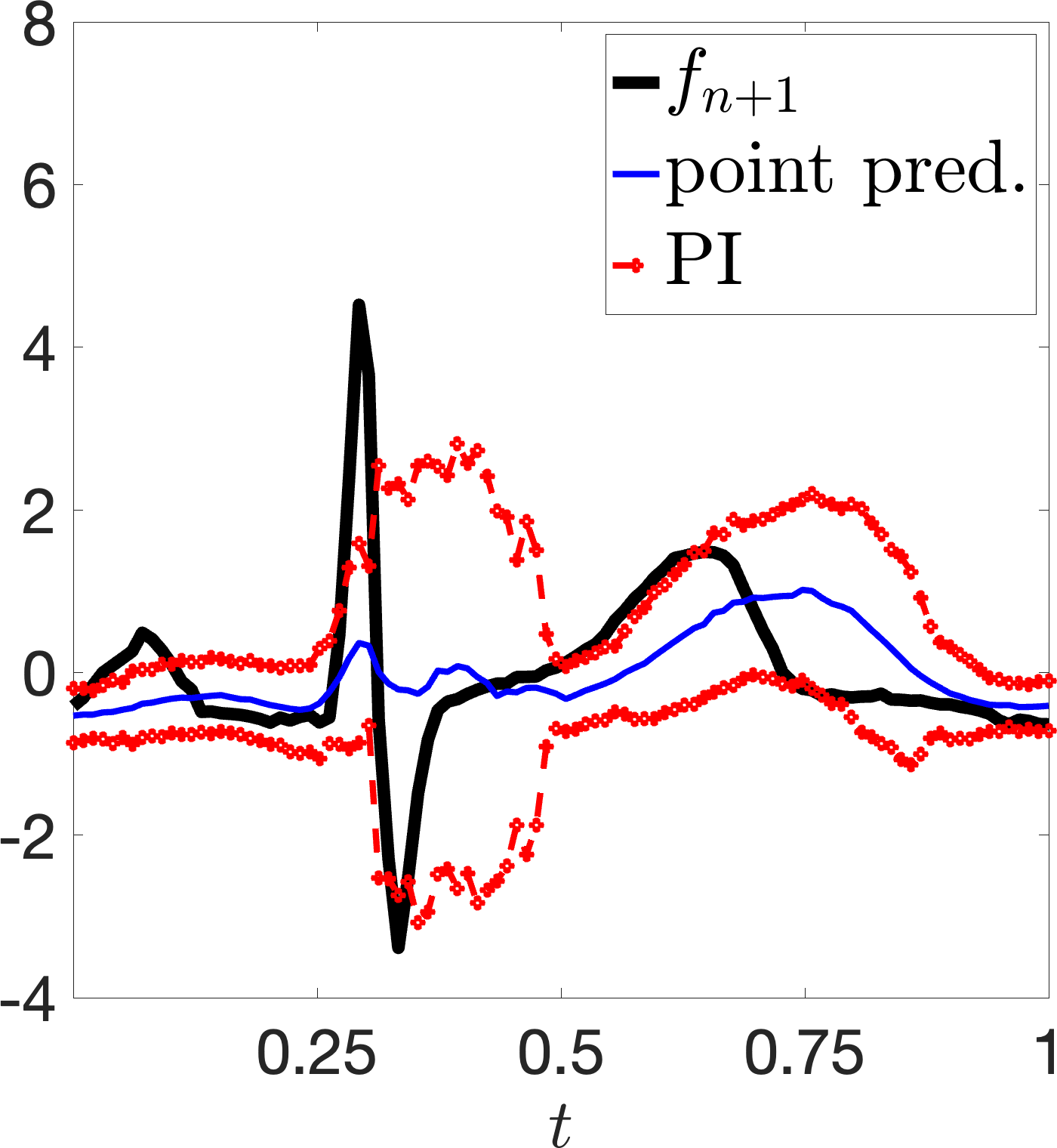} &
        \includegraphics[width = 0.16\textwidth]{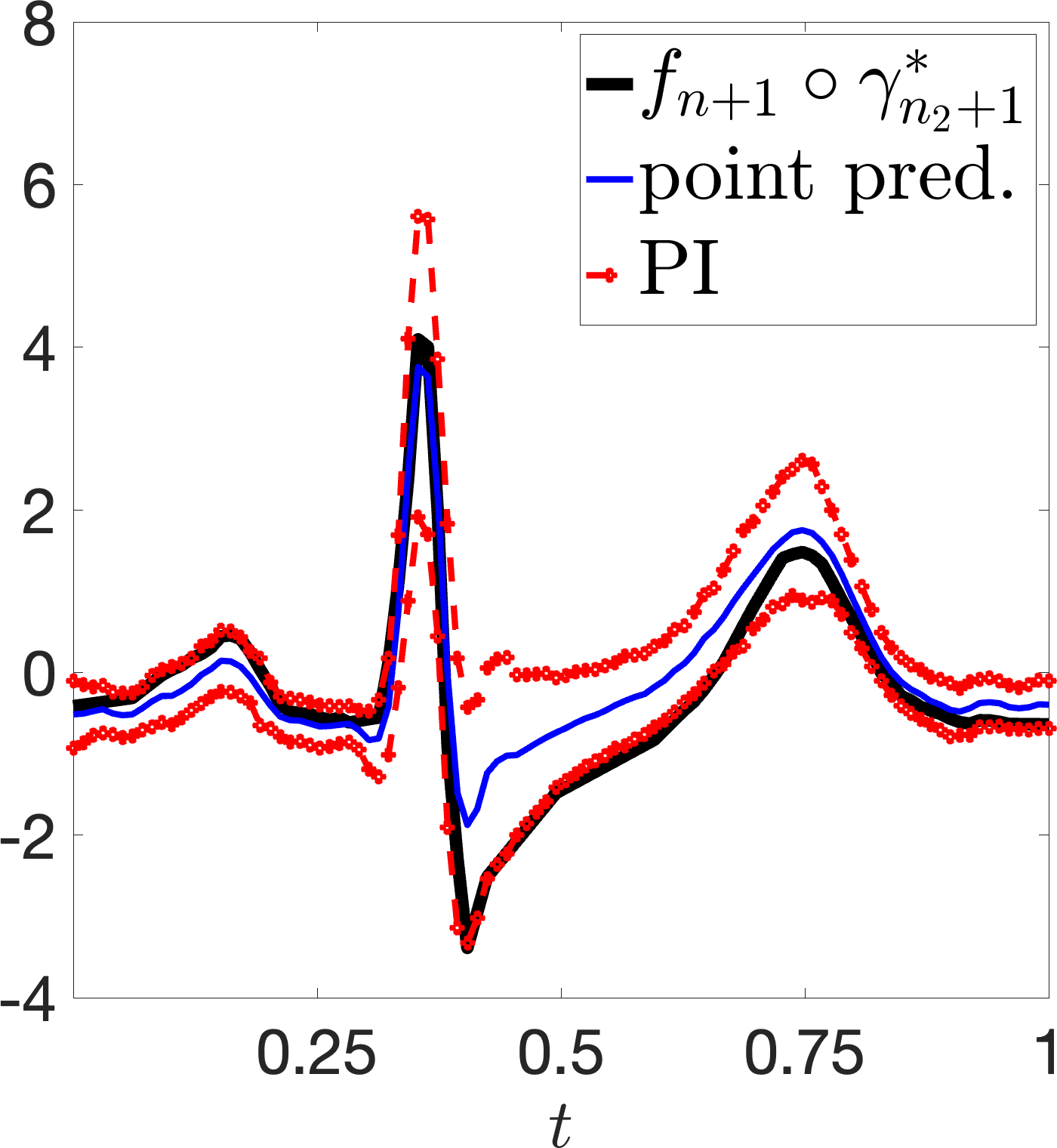}
        \\
        \includegraphics[width = 0.16\textwidth]{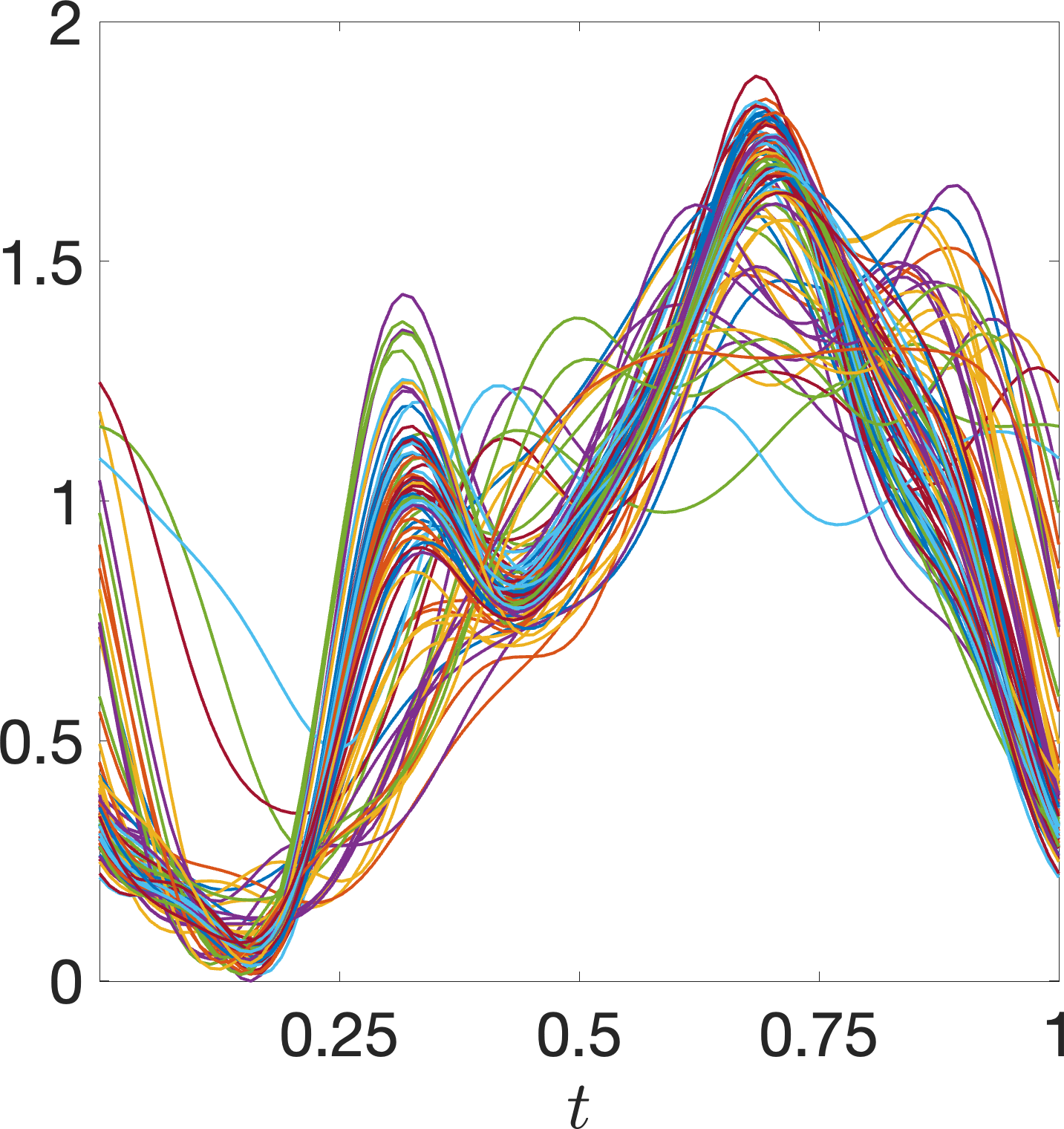} &
        \includegraphics[width = 0.16\textwidth]{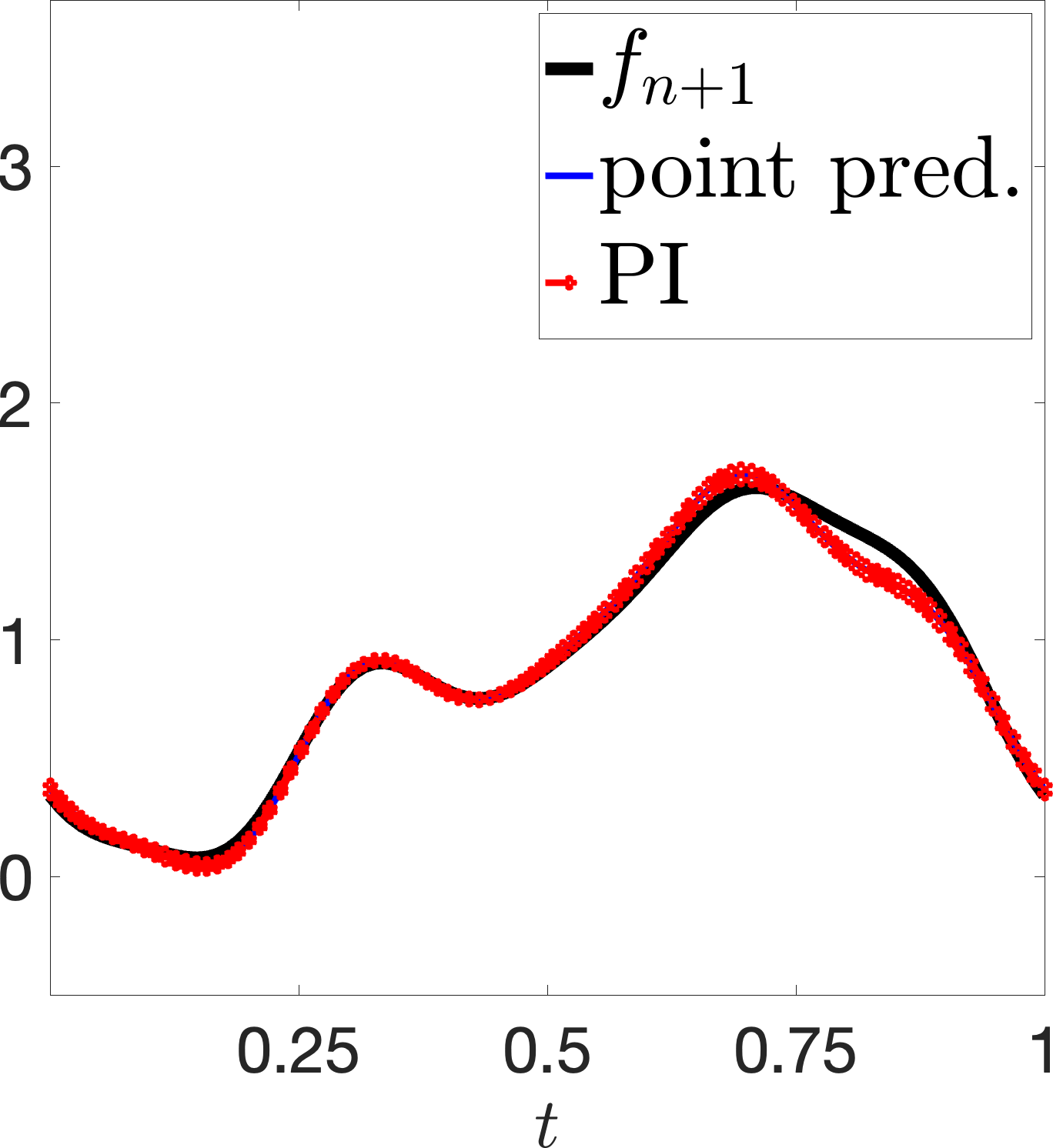} &
        \includegraphics[width = 0.16\textwidth]{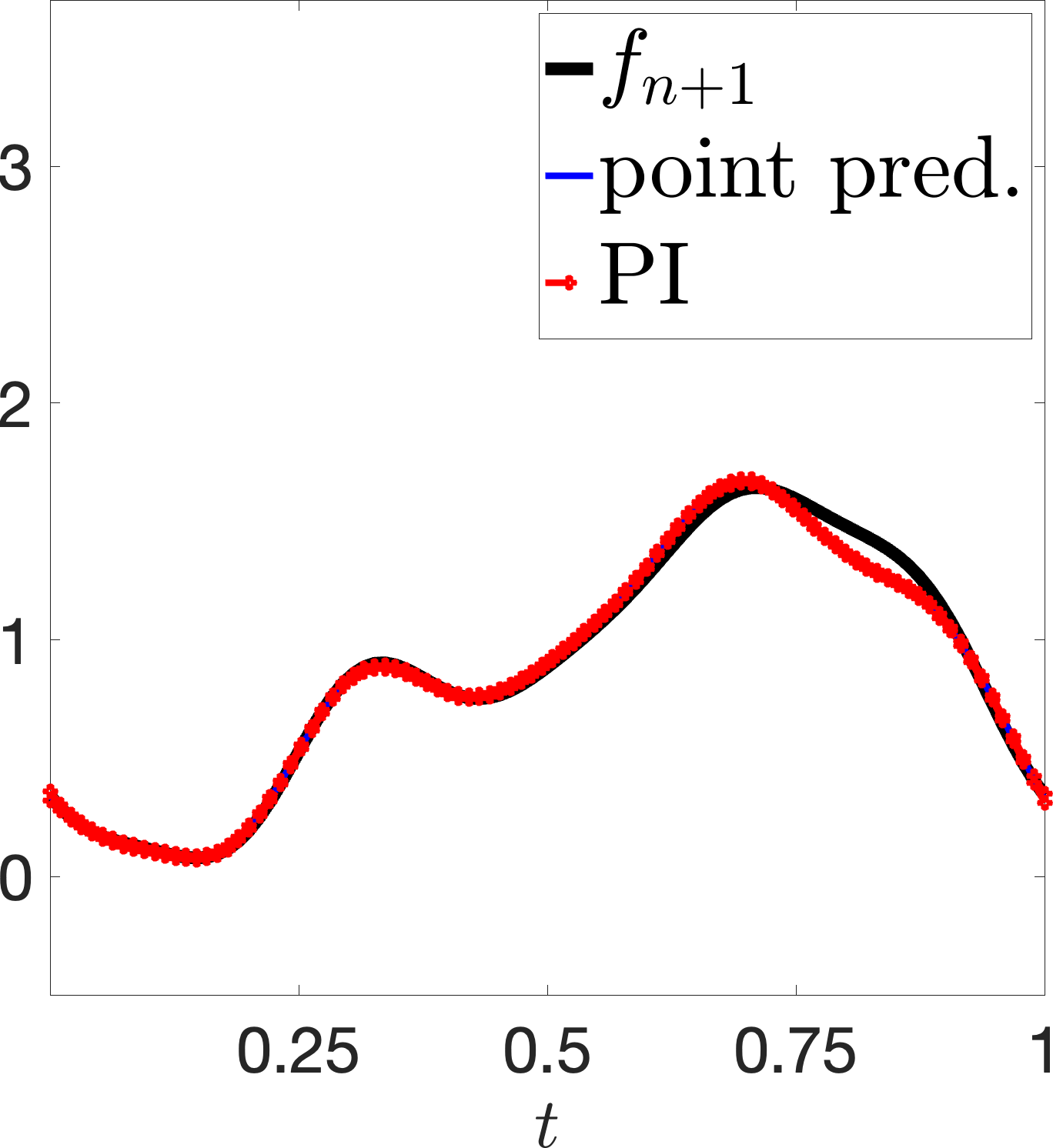} &
        \includegraphics[width = 0.16\textwidth]{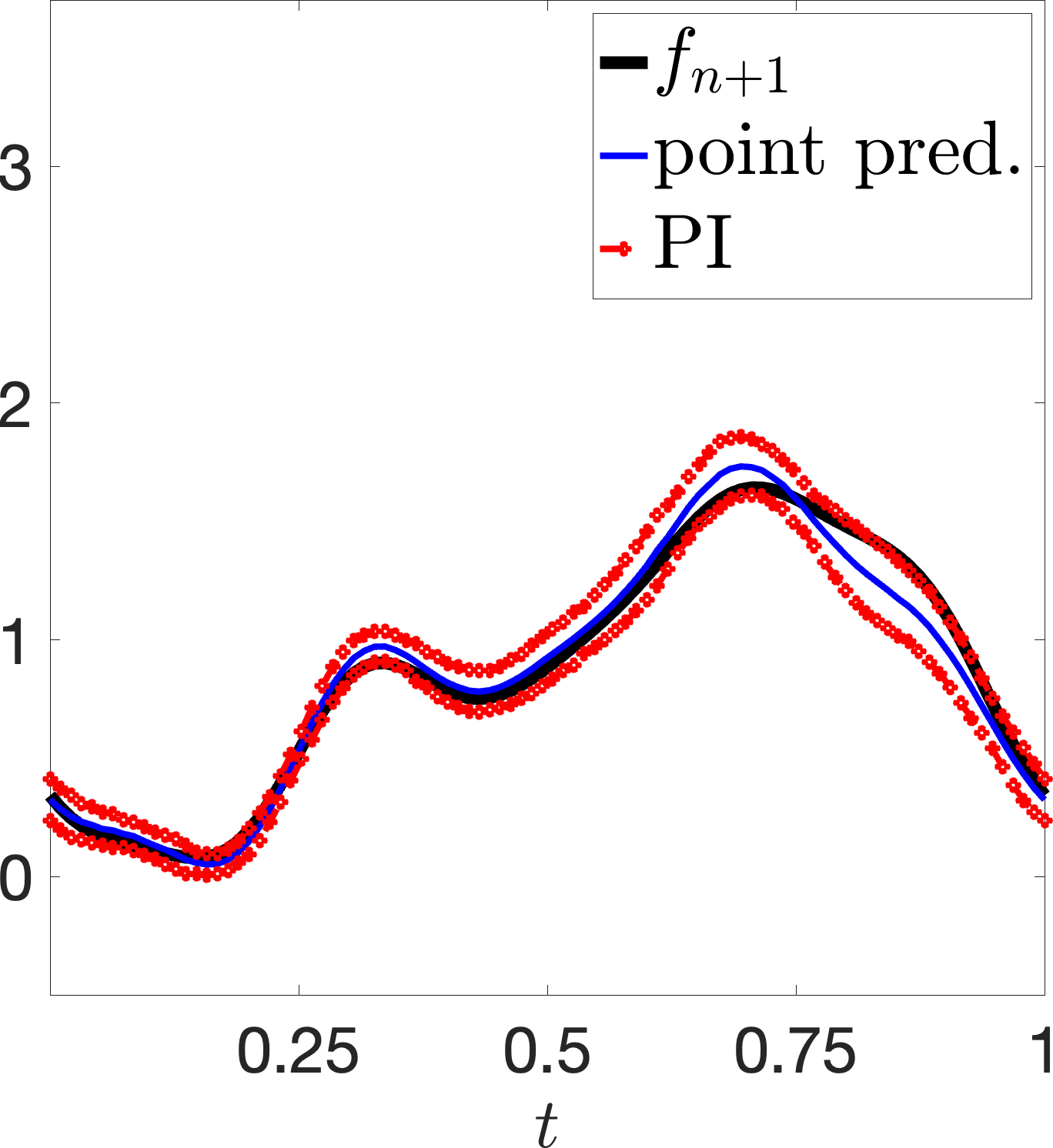} &
        \includegraphics[width = 0.16\textwidth]{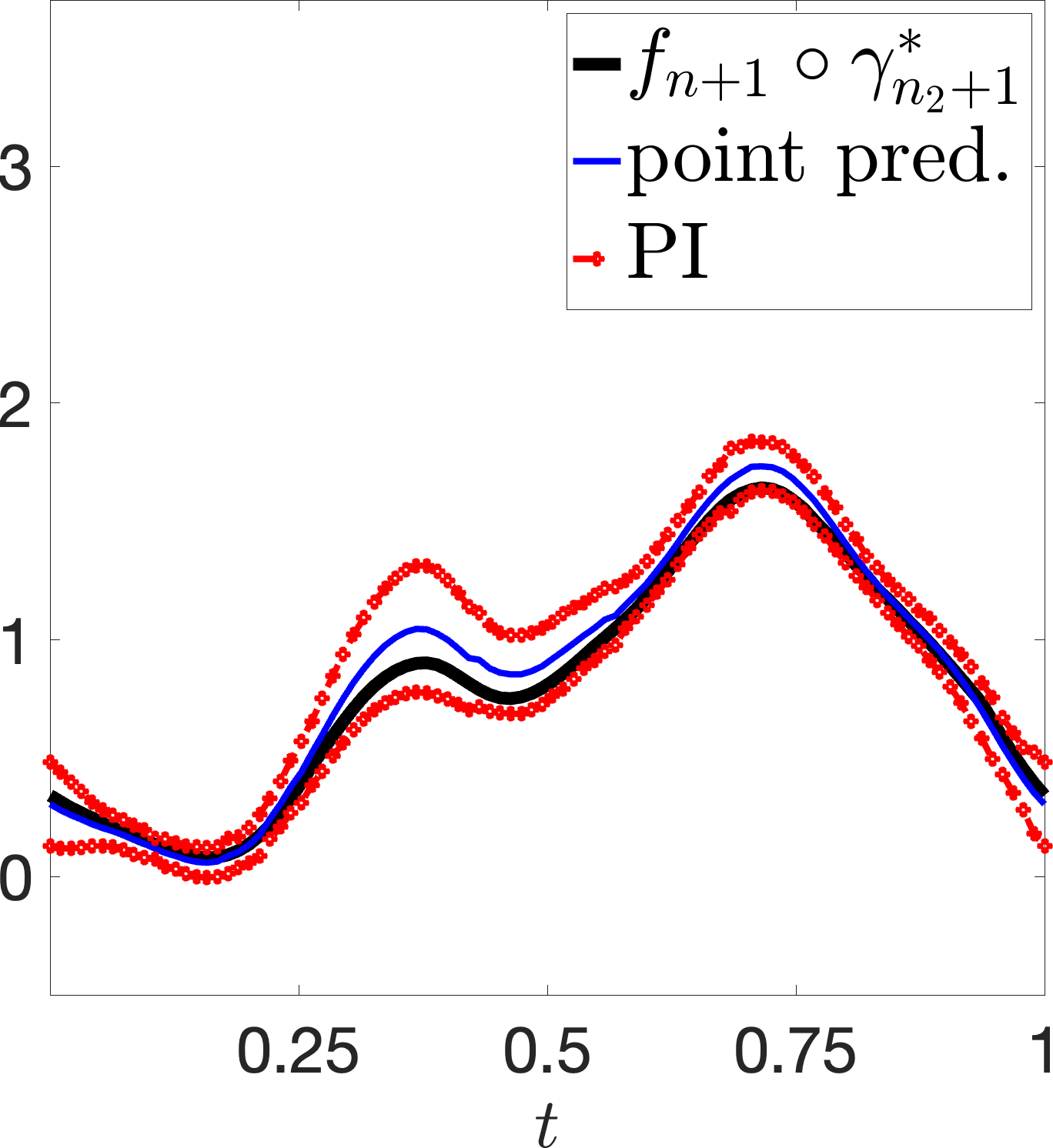}
        \end{tabular}
    \vspace{-10pt}
    \caption{\small Rows 1-3: Berkeley growth rate functions, PQRST complexes and traffic flow rate functions. (a) Data. (b)-(e) Prediction results, with target function (black), point prediction (blue) and pointwise PIs (red), for {\tt SoF}, {\tt FoF}, {\tt FFCP} and {\tt SFCP}, respectively.}
    \label{fig::real-data-simulated-fragmentation}
    \vspace{-10pt}
\end{figure}

\noindent\textbf{Example 1: real-world data with simulated truncation time point.} We compare prediction performance of {\tt SoF}, {\tt FoF}, {\tt FFCP} and {\tt SFCP} on three functional datasets with a simulated truncation point. Rows 1-3 in Figure \ref{fig::real-data-simulated-fragmentation}(a) show the following three datasets, respectively:
    (i) {\bf Berkeley growth rate functions}, first derivative of measurements on heights in centimeters for $n=93$ boys and girls from age 1-18 \citep{RamsaySilverman2005};
    (ii) {\bf PQRST complexes}, $n=80$ segmented PQRST complexes from a long electrocardiogram (ECG) signal; \citep{Kurtek2013SegmentationAA};
    (iii) {\bf traffic flow rate functions}, pre-smoothed traffic flow rate on National Highway 5 in Taiwan for $n=84$ days \citep{chiou2012dynamical, jiao2023functional}.
In each case, we rescaled the time axis to $[0,1]$ and normalized all functions to have unit $\mathbb{L}^2$ norm. For each dataset, $f_{n+1}$ is chosen at random and truncated using $U=0.5$. Panels (b)-(e) show PIs (red) and point predictions (blue) from {\tt SoF}, {\tt FoF}, {\tt FFCP} and {\tt SFCP}, respectively; the target function is in black. Compared to the other approaches, {\tt SFCP} yields much more accurate point predictions and PIs, which capture the main geometric features of the target functions. In most cases, {\tt SoF} and {\tt FoF} generate PIs that are overly smooth and fail to capture the true target function. {\tt FFCP} generates much wider PIs to maintain its coverage guarantee, but the resulting point predictions are not accurate.

\setlength{\tabcolsep}{0pt}

\begin{figure}[!t]
    \centering
    \begin{tabular}{ccccc}
     (a)&(b)&(c)&(d)&(e)\\
        \includegraphics[width = 0.2\textwidth]{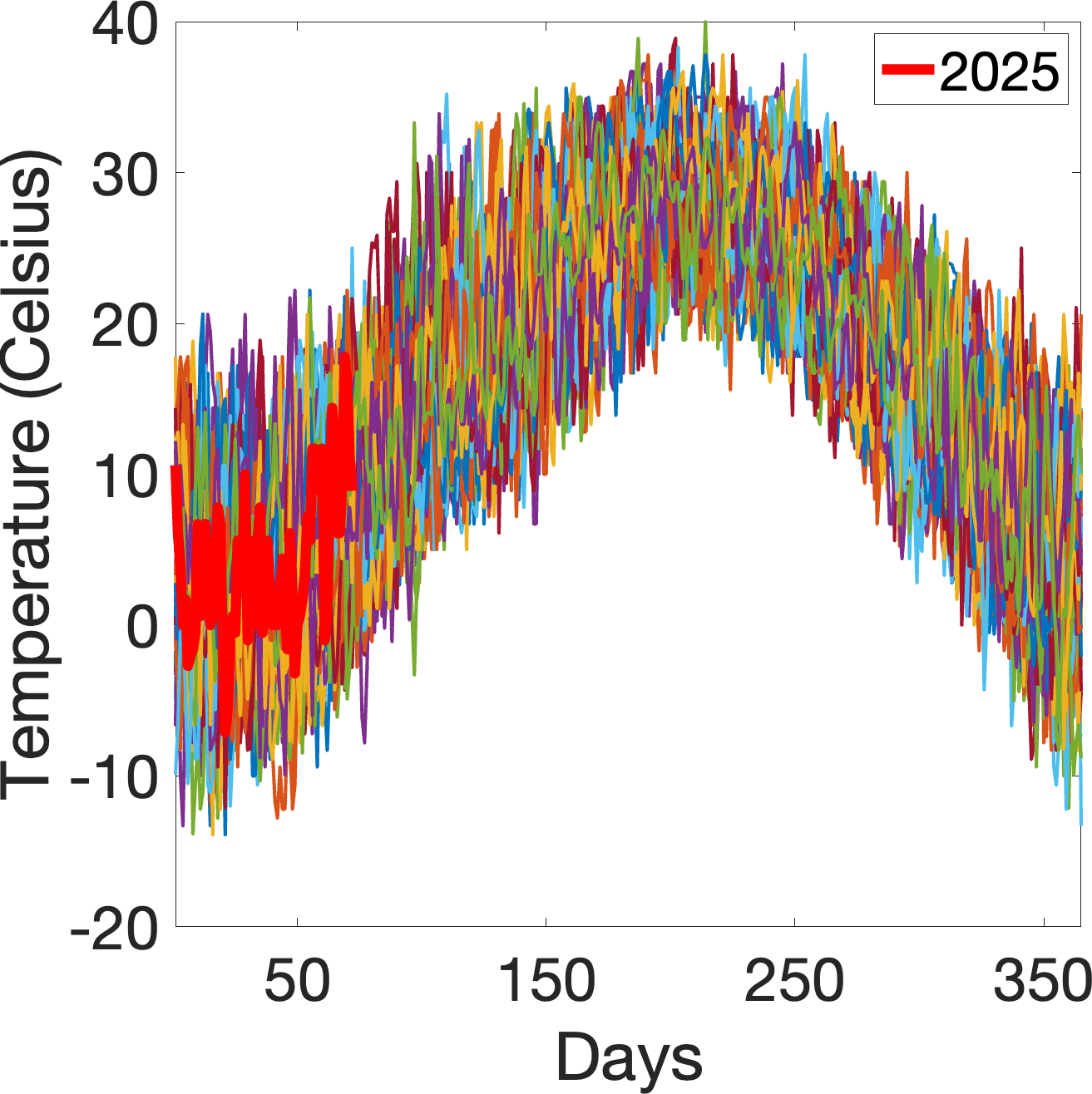} &
        \includegraphics[width = 0.2\textwidth]{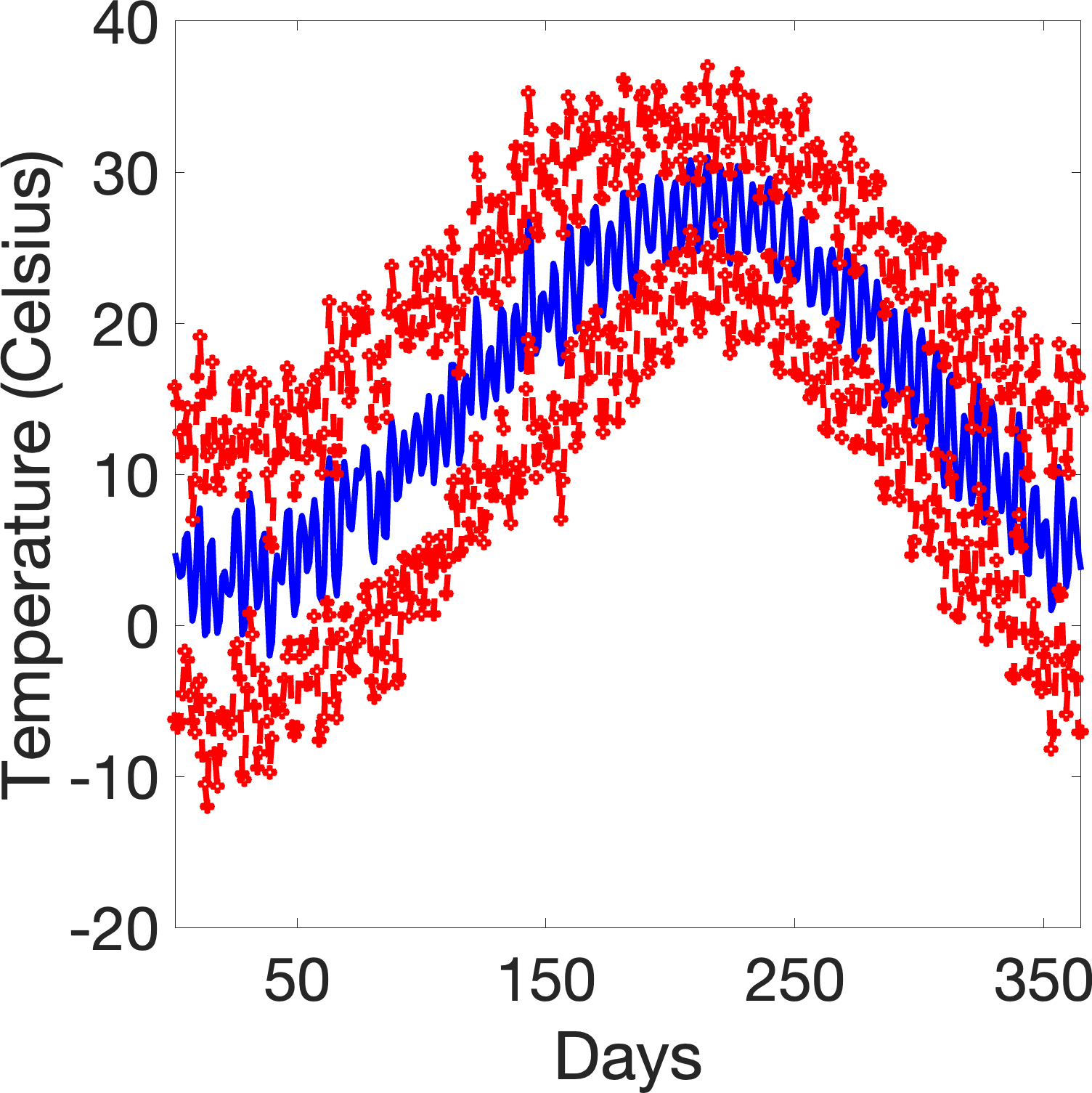} &
        \includegraphics[width = 0.2\textwidth]{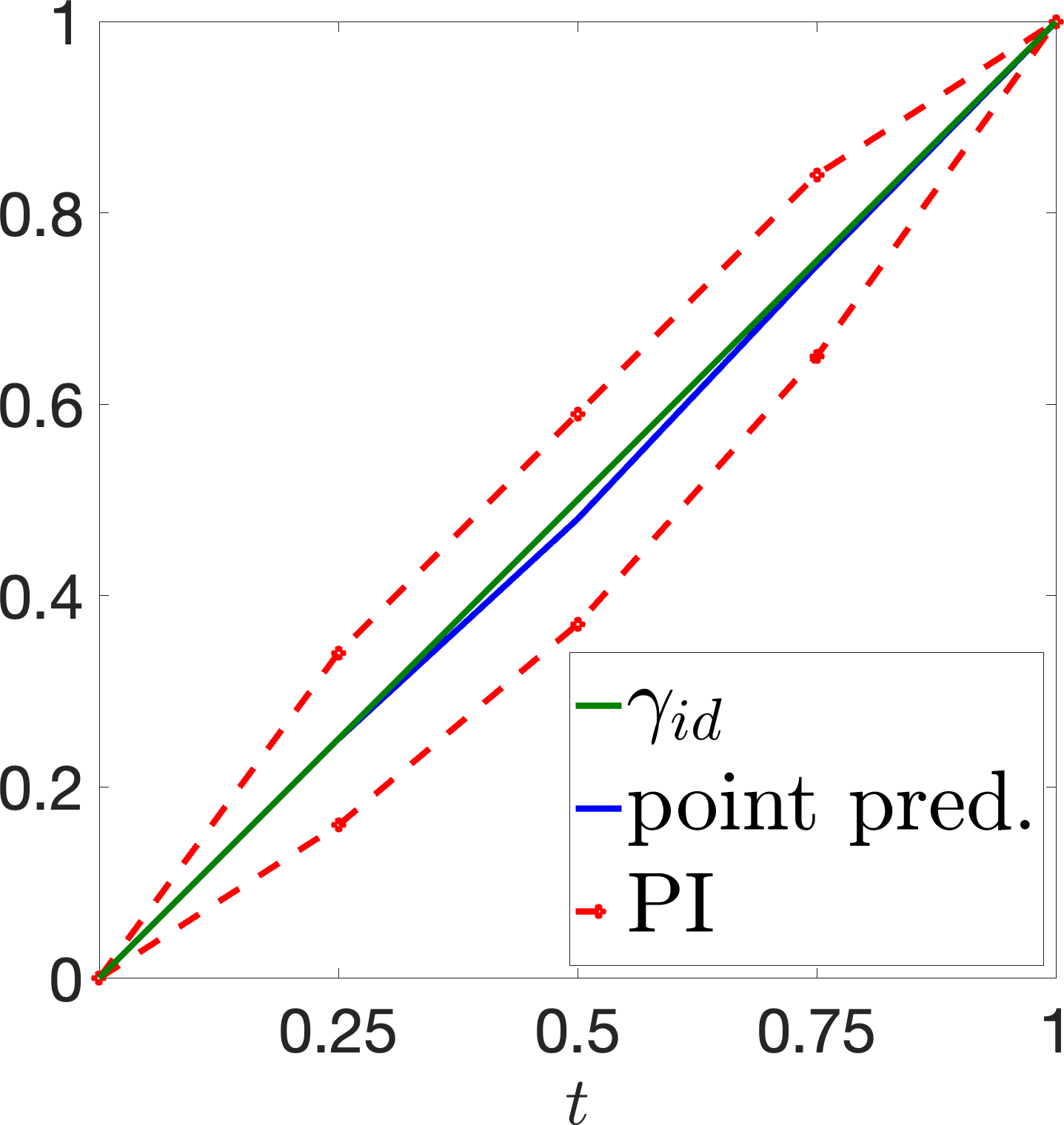}&
        \includegraphics[width = 0.2\textwidth]{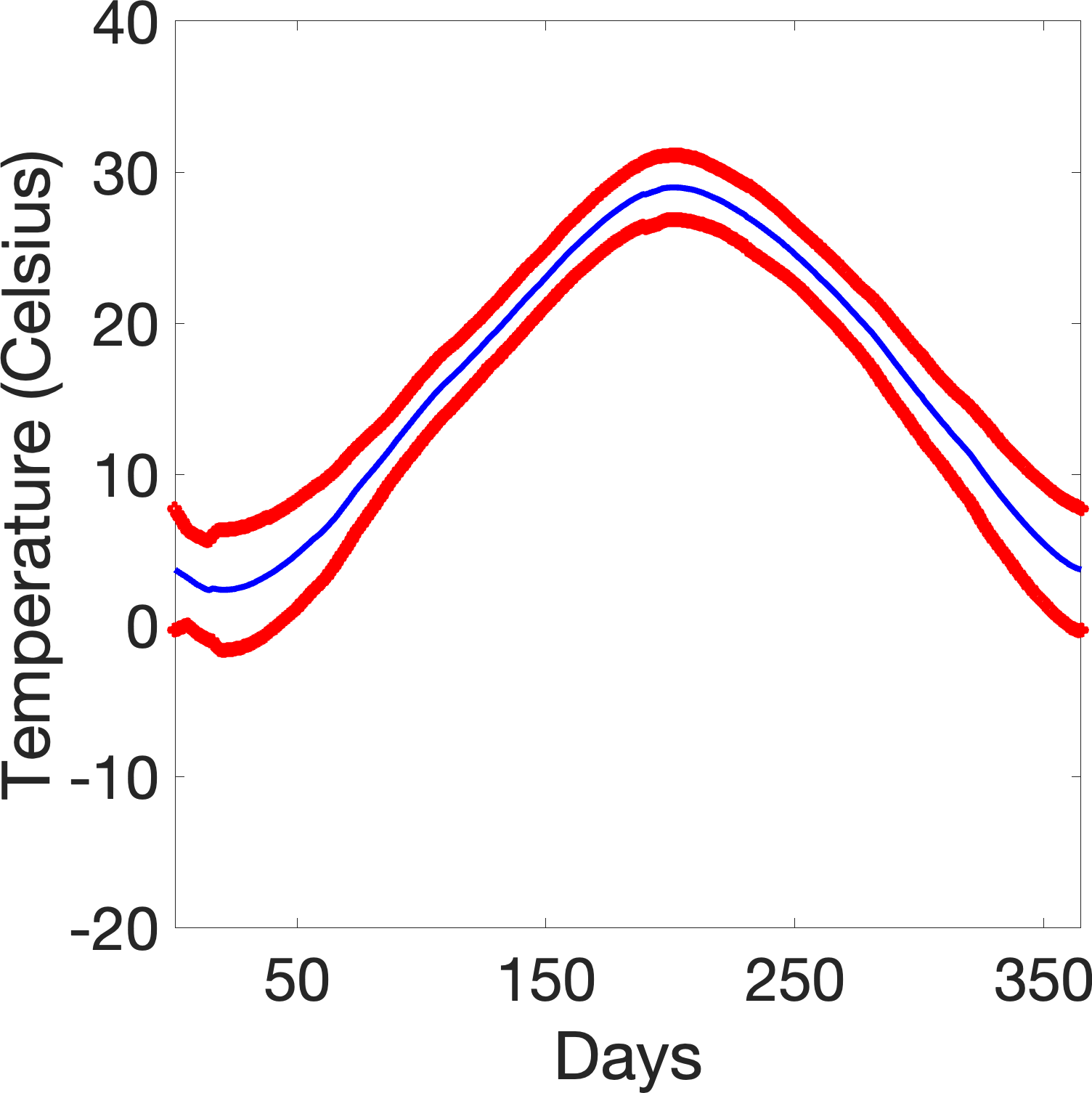}&
        \includegraphics[width = 0.2\textwidth]{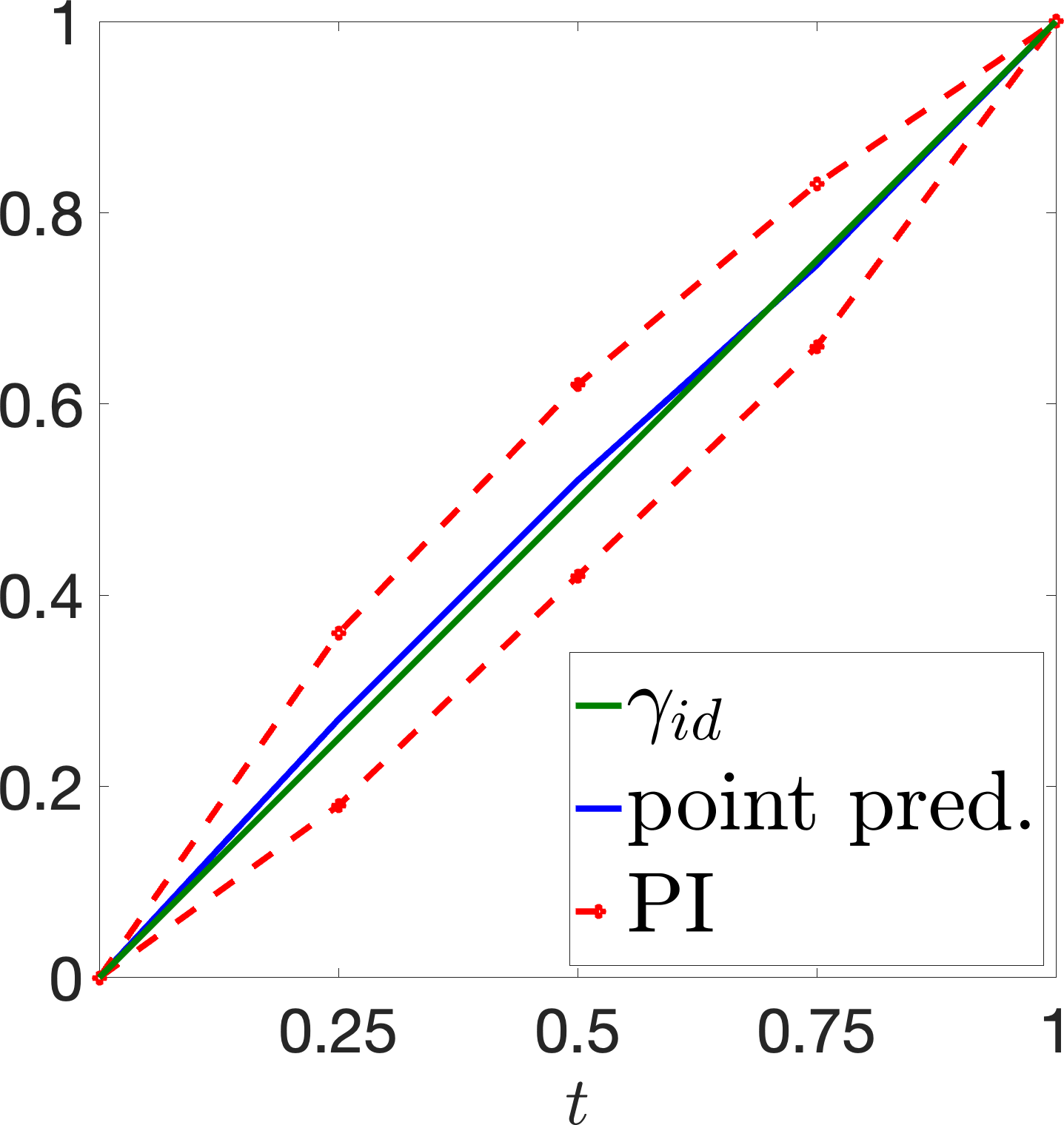}
        \end{tabular}
    \vspace{-10pt}
    \caption{\small Prediction of MDT in Rhode Island. (a) Complete historical data with partial observation for 2025 in red. (b) \& (d) Amplitude prediction. PIs (red) and point prediction (blue) for raw and pre-smoothed data, respectively. (c) \& (e) Relative phase prediction. PIs (red), point prediction (blue) and identity warping (green) for raw and pre-smoothed data, respectively. }
    \label{fig::real-data-weather-forecasting}
    \vspace{-10pt}
\end{figure}

\noindent\textbf{Example 2: prediction of maximum daily temperature.} Finally, we focus on forecasting maximum daily temperatures (MDTs) in Rhode Island. We use complete observations of MDT from 1950 to 2024 to predict the amplitude and relative phase of MDT for 2025, based on a partial observation from January 1 to March 13 \citep{nceiData2024}. The data is shown in Figure \ref{fig::real-data-weather-forecasting}(a) with the partial observation highlighted in red. We apply {\tt SFCP} and {\tt SFCPP} to both, raw data and data after smoothing using a Fourier basis projection. Panels (b) and (d) in Figure \ref{fig::real-data-weather-forecasting} show the PIs (red) and point predictions (blue) for amplitude generated by {\tt SFCP} based on raw and smoothed data, respectively. Panels (c) and (e) show the corresponding results for relative phase generated by {\tt SFCPP}, with point predictions in blue, PIs in red and the identity warping function in green. Overall, the proposed approaches perform very well. Based on raw data, {\tt SFCP} generates PIs that capture the overall MDT trend and (local) daily fluctuations. Based on smoothed data, the prediction band for amplitude captures the global MDT trend with reasonable pointwise PI lengths. The 2025 MDT trend in Rhode Island has one maximum in mid-July and one minimum in late January. Relative phase predictions based on both raw and smoothed data show a small relative timing shift of 2025 MDT with respect to the historical mean, indicated by point predictions that are close to the identity warping and tight PIs.

\setlength{\tabcolsep}{0pt}

\vspace{-.3in}
\section{Discussion}
\label{sec::discussion}
\vspace{-.1in}
We introduced a novel conformal prediction framework for partial functional data that incorporates registration. Results based on simulations and real-world data examples validate the proposed method's finite-sample coverage, high prediction accuracy and computational efficiency. Despite these advantages over competing approaches, several challenges remain that we leave as future work. First, {\tt SFCP} may be less effective for heterogeneous populations with significant amplitude variation across subpopulations. In such cases, inaccurate estimation of the Karcher mean, a key step in our framework, may impair prediction accuracy. Addressing this issue requires strategies that account for such heterogeneity, e.g., a group conditional approach given observed labels for subpopulations. Second, {\tt SFCP} generates pointwise prediction intervals with marginal coverage validity. We will explore alternative formulations that utilize basis expansions with a global coverage guarantee. Finally, {\tt SFCP} relies on a grid search over trial values when constructing prediction intervals, which can be computationally expensive for a dense grid of time points. This issue is further amplified in {\tt SFCPP}, where joint prediction over multiple time points is required. For reference, the approximate {\tt SFCPP} computing times for $T=5,6,7,8,9$ (number of time points along $x$-axis) with 50 trial values at each time point, using a fixed bandwidth, are $0.12, 0.78, 7.26, 61.5$ and $>1440$ minutes, respectively. While the use of a coarse time grid often results in a good approximation for the underlying warping function and enforces regularity in the PIs, exploring alternative conformal prediction methods that avoid an exhaustive grid search can significantly improve computational efficiency and broaden the utility of both {\tt SFCP} and {\tt SFCPP}.

\nolinenumbers

\if1\blind
{
\noindent\textbf{Acknowledgments:} This research was partially supported by grants NSF DMS-2311109 (to YZ), and NIH R37-CA214955, NSF DMS-2015226 and NSF DMS-2413747 (to SK).
} \fi

\if0\blind
{
} \fi
\vspace{-0.1 in}
\bibliographystyle{plainnat}
\spacingset{1.4}
\bibliography{ref}

\end{document}